\documentstyle[a4,epsfig,12pt]{article}
\begin{document}

\newdimen\picraise
\newcommand\picbox[1]
{
  \setbox0=\hbox{\input{#1}}
  \picraise=-0.5\ht0 
  \advance\picraise by 0.5\dp0
  \advance\picraise by 3pt      
  \hbox{\raise\picraise \box0}
}
\def\del{\partial}

\begin{tabbing}
\` Bonn TK 97-08 
\end{tabbing}
\baselineskip 10mm
\renewcommand{\thefootnote}{\fnsymbol{footnote}}
\begin{center}{\Large\bf 
QCD at large and short distances\footnote{This work was
completed in Germany under the Humboldt Research Award Program}
}
\end{center}
\begin{center}{\large 
Vladimir Gribov
}\end{center}
\baselineskip 7mm
\begin{center}{\it 
Landau Institute for Theoretical Physics, Moscow \\
and \\
Research Institute for Particle and Nuclear Physics, Budapest \\
and \\
Institut f\"ur Theoretische Kernphysik der Universit\"at Bonn 
}
\end{center}
\begin{center}
July 28, 1997
\end{center}
\renewcommand{\thefootnote}{\arabic{footnote}}
\vspace{3mm}
\begin{abstract}
\baselineskip 5mm
The formulation of QCD which contains no divergences and no
renormalization procedure is presented. It contains both perturbative
and non-perturbative phenomena. It is shown that, due to its
asymptotically free nature, the theory is not defined uniquely. The
chiral symmetry breaking and the nature of the octet of pseudo-scalar
particles as quasi-Goldstone states are analysed in the theory with
massless and massive quarks. The $U(1)$ problem is discussed.
\end{abstract}

\tiny
\normalsize



\section{Introduction}

In this paper I show how to formulate an asymptotically free theory
in such a way that it includes perturbative and non-perturbative
phenomena simultaneously.

The idea is the following. Contrary to an infrared free theory, in
an asymptotically free theory the divergences prevent us from
writing even perturbation expansions in a unique, well-defined
way. We can, however, make use of the fact that divergences in the theory
enter only the Green's functions and the vertices. On the other hand,
knowing the Green's functions and the vertices we can express all
the other amplitudes through them perturbatively in a unique way.
Thus, we have to formulate equations for Green's functions and
vertices in a form which does not contain any divergences. If
this is done, the solutions of these equations will contain both
perturbative and non-perturbative phenomena.

To avoid technical complications, in the first section of the paper
we will derive these equations in an Abelian theory in which usual
perturbation theory is also well defined. In the second section we
generalize the equations for a non-Abelian theory and discuss the
connection between perturbative and non-perturbative effects. The
equations have an integro-differential structure in which the
asymptotic behaviour of the Green's functions is defined by the
boundary conditions. The main conclusion in this section is that in
the region of large momenta the Green's functions of quarks and gluons
contain additional parameters compared to normal perturbation theory
which can be associated with different types of "condensates". These
non-perturbative parameters can be defined by solving the equations
and finding non-singular solutions in the infrared region. A priori
it is not clear what type of additional conditions have to be
imposed on this system of equations in order to fix a non-singular
solution. It can be, for example, the conservation of the axial
current or of other currents which is formally satisfied from the
point of view of the Lagrangian but is not ensured because of
the divergences.

In the third section we will show that these equations make it
possible, for the first time, to analyse the problem of spontaneous
symmetry breaking in an asymptotically free theory and
to find approximately the value of the critical coupling at which
symmetry breaking occures. Also, they allow us to answer the
fundamental question for asymptotically free theories, namely:
how the bound states -- the hadrons -- have to be treated in these
theories and how these bound states influence the equations for the
Green's functions.

The analysis leads to the following conclusion.
The conditions for axial current conservation of flavour non-singlet
currents (in the limit of zero bare quark masses) require that eight
Goldstone bosons (the pseudo-scalar octet) have to be regarded
as elementary objects with couplings defined by Ward identities.
This is so in spite of the fact that the couplings of these states
to fermions decrease at large fermion virtualities.

The same analysis provides a new possibility for the solution of the
$U(1)$ problem. In this solution the flavour singlet pseudoscalar
$\eta'$ is a normal bound state of $q\bar{q}$ without a point-like
structure. The mass of this bound state is different from zero and
can be calculated in the limit of massless quarks. For massive quarks
the pseudoscalar octet becomes massive. Their masses, however,
are not calculable in terms of bare quark masses
because of logarithmic divergences and have to be
regarded as unknown parameters which in the real case of confined
quarks are defined by the self-consistence of the solution of the
infrared problem. These states have an essential influence on the
equations for the Green's functions, which, as it will be shown
in the next paper, can be used constructively in solving the
confinement problem if the effective coupling in the infrared
region is not too large. In this case the integro-differential
equations can be reduced to a system of non-linear differential
equations and the theory looks like a theory of particle propagation
in self-consistent fields defined by the Green's functions themselves
(as it is the case in Landau's Fermi liquid theory). The
self-consistent fields are fields of gluons and $\pi$-mesons.

\section{Equations for Green's functions in QED}\label{I}

In QED we have two Green's functions: that of the photon $D_{\mu\nu}(k)$
and of the electron $G(q)$, and a vertex function $\Gamma_{\mu}(k,q)$.
The photon Green's function is defined by the vacuum polarization
operator $\Pi_{\mu\nu}(k)$ which can be expressed symbolically by
a sum of Feynman diagrams
\begin{equation}
\label{1.1}
\Pi_{\mu\nu}(k) = e_0^2\{ \gamma_{\mu} \begin{picture}(0,0)%
\epsfig{file=fig1.pstex}%
\end{picture}%
\setlength{\unitlength}{0.00083300in}%
\begingroup\makeatletter\ifx\SetFigFont\undefined
\def\x#1#2#3#4#5#6#7\relax{\def\x{#1#2#3#4#5#6}}%
\expandafter\x\fmtname xxxxxx\relax \def\y{splain}%
\ifx\x\y   
\gdef\SetFigFont#1#2#3{%
  \ifnum #1<17\tiny\else \ifnum #1<20\small\else
  \ifnum #1<24\normalsize\else \ifnum #1<29\large\else
  \ifnum #1<34\Large\else \ifnum #1<41\LARGE\else
     \huge\fi\fi\fi\fi\fi\fi
  \csname #3\endcsname}%
\else
\gdef\SetFigFont#1#2#3{\begingroup
  \count@#1\relax \ifnum 25<\count@\count@25\fi
  \def\x{\endgroup\@setsize\SetFigFont{#2pt}}%
  \expandafter\x
    \csname \romannumeral\the\count@ pt\expandafter\endcsname
    \csname @\romannumeral\the\count@ pt\endcsname
  \csname #3\endcsname}%
\fi
\fi\endgroup
\begin{picture}(616,251)(593,-486)
\end{picture}
 \gamma_{\nu} +
 \gamma_{\mu} \begin{picture}(0,0)%
\epsfig{file=fig2.pstex}%
\end{picture}%
\setlength{\unitlength}{0.00083300in}%
\begingroup\makeatletter\ifx\SetFigFont\undefined
\def\x#1#2#3#4#5#6#7\relax{\def\x{#1#2#3#4#5#6}}%
\expandafter\x\fmtname xxxxxx\relax \def\y{splain}%
\ifx\x\y   
\gdef\SetFigFont#1#2#3{%
  \ifnum #1<17\tiny\else \ifnum #1<20\small\else
  \ifnum #1<24\normalsize\else \ifnum #1<29\large\else
  \ifnum #1<34\Large\else \ifnum #1<41\LARGE\else
     \huge\fi\fi\fi\fi\fi\fi
  \csname #3\endcsname}%
\else
\gdef\SetFigFont#1#2#3{\begingroup
  \count@#1\relax \ifnum 25<\count@\count@25\fi
  \def\x{\endgroup\@setsize\SetFigFont{#2pt}}%
  \expandafter\x
    \csname \romannumeral\the\count@ pt\expandafter\endcsname
    \csname @\romannumeral\the\count@ pt\endcsname
  \csname #3\endcsname}%
\fi
\fi\endgroup
\begin{picture}(616,251)(593,-486)
\end{picture}
 \gamma_{\nu} + \cdots \}
\end{equation}
In order to obtain series not containing divergences we will try to
consider the derivatives of $\Pi_{\mu\nu}(k)$ as functions of momenta.
Due to current conservation, we have
\begin{equation}
\label{1.2}
\Pi_{\mu\nu}(k) = (\delta_{\mu\nu}k^2-k_{\mu}k_{\nu})\Pi(k^2)
\end{equation}
where $\Pi(k^2)$ contains only logarithmic divergences.

Differentiating (\ref{1.2}) twice we will have
\begin{equation}
\label{1.3}
\partial^2 \Pi_{\mu\nu}(k) = 6\Pi(k^2) + \left(\delta_{\mu\nu} -
 \frac{k_{\mu}k_{\nu}}{k^2}\right)(\partial_{\xi}+6)
 \partial_{\xi}\Pi(k^2) ;
\end{equation}
we here use
\[ \partial_{\xi} \equiv q_{\mu}\frac{\partial}{\partial q_{\mu}} .\]
The second term in (\ref{1.3}) does not contain divergences, the
first one does. In order to obtain a finite expression, consider
\begin{equation}
\label{1.4}
\partial_{\mu}\partial_{\sigma} \Pi_{\sigma\nu}(k) = - 3\delta_{\mu\nu}
 \Pi(k^2) - 3 \frac{k_{\mu}k_{\nu}}{k^2} \partial_{\xi}\Pi(k^2) .
\end{equation}
Because of this, the quantity
\begin{equation}
\label{1.5}
\partial^2 \Pi_{\mu\nu}(k) + 2\partial_{\mu}\partial_{\sigma}
 \Pi_{\sigma\nu}(k) = -6\frac{k_{\mu}k_{\nu}}{k^2}\partial_{\xi}\Pi(k^2)
 + \left(\delta_{\mu\nu} -
 \frac{k_{\mu}k_{\nu}}{k^2}\right)(\partial_{\xi}+6)
 \partial_{\xi}\Pi(k^2)
\end{equation}
must not contain divergences.

In Feynman gauge $D_{\mu\nu}(k^2)$ is equal
\begin{equation}
\label{1.6}
D_{\mu\nu}(k) = \frac{1}{k^2(1-\Pi(k^2))} \equiv \frac{1}{k^2}
 \frac{e^2(k^2)}{e_0^2(k^2)}
\end{equation}
The diagrams contain only the product $e_0^2 D_{\mu\nu}(k)$ and therefore
\begin{eqnarray}
\label{1.7}
e_0^2 D_{\mu\nu}(k) & = & \frac{1}{k^2} e^2(k^2) \nonumber\\
 \partial_{\xi}\Pi(k^2) & = & -\partial_{\xi}\frac{e_0^2}{e^2} .
\end{eqnarray}
In first order this means
\begin{eqnarray}
\label{1.8}
\lefteqn{\partial_{\xi}\frac{1}{e^2} = } \nonumber\\
 & = & \frac{1}{3}\frac{k_{\mu}k_{\nu}}{k^2}\Bigg\{ \; \begin{picture}(0,0)%
\epsfig{file=fig3.pstex}%
\end{picture}%
\setlength{\unitlength}{0.00083300in}%
\begingroup\makeatletter\ifx\SetFigFont\undefined
\def\x#1#2#3#4#5#6#7\relax{\def\x{#1#2#3#4#5#6}}%
\expandafter\x\fmtname xxxxxx\relax \def\y{splain}%
\ifx\x\y   
\gdef\SetFigFont#1#2#3{%
  \ifnum #1<17\tiny\else \ifnum #1<20\small\else
  \ifnum #1<24\normalsize\else \ifnum #1<29\large\else
  \ifnum #1<34\Large\else \ifnum #1<41\LARGE\else
     \huge\fi\fi\fi\fi\fi\fi
  \csname #3\endcsname}%
\else
\gdef\SetFigFont#1#2#3{\begingroup
  \count@#1\relax \ifnum 25<\count@\count@25\fi
  \def\x{\endgroup\@setsize\SetFigFont{#2pt}}%
  \expandafter\x
    \csname \romannumeral\the\count@ pt\expandafter\endcsname
    \csname @\romannumeral\the\count@ pt\endcsname
  \csname #3\endcsname}%
\fi
\fi\endgroup
\begin{picture}(1137,775)(451,-704)
\put(1351,-61){\makebox(0,0)[lb]{\smash{\SetFigFont{10}{12.0}{rm}$\gamma_\sigma$}}}
\put(1501,-511){\makebox(0,0)[lb]{\smash{\SetFigFont{10}{12.0}{rm}$\gamma_\nu$}}}
\put(451,-511){\makebox(0,0)[lb]{\smash{\SetFigFont{10}{12.0}{rm}$\gamma_\mu$}}}
\put(601,-61){\makebox(0,0)[lb]{\smash{\SetFigFont{10}{12.0}{rm}$\gamma_\sigma$}}}
\end{picture}
 +
 \begin{picture}(0,0)%
\epsfig{file=fig4.pstex}%
\end{picture}%
\setlength{\unitlength}{0.00083300in}%
\begingroup\makeatletter\ifx\SetFigFont\undefined
\def\x#1#2#3#4#5#6#7\relax{\def\x{#1#2#3#4#5#6}}%
\expandafter\x\fmtname xxxxxx\relax \def\y{splain}%
\ifx\x\y   
\gdef\SetFigFont#1#2#3{%
  \ifnum #1<17\tiny\else \ifnum #1<20\small\else
  \ifnum #1<24\normalsize\else \ifnum #1<29\large\else
  \ifnum #1<34\Large\else \ifnum #1<41\LARGE\else
     \huge\fi\fi\fi\fi\fi\fi
  \csname #3\endcsname}%
\else
\gdef\SetFigFont#1#2#3{\begingroup
  \count@#1\relax \ifnum 25<\count@\count@25\fi
  \def\x{\endgroup\@setsize\SetFigFont{#2pt}}%
  \expandafter\x
    \csname \romannumeral\the\count@ pt\expandafter\endcsname
    \csname @\romannumeral\the\count@ pt\endcsname
  \csname #3\endcsname}%
\fi
\fi\endgroup
\begin{picture}(1137,775)(451,-704)
\put(601,-61){\makebox(0,0)[lb]{\smash{\SetFigFont{10}{12.0}{rm}$\gamma_\sigma$}}}
\put(451,-511){\makebox(0,0)[lb]{\smash{\SetFigFont{10}{12.0}{rm}$\gamma_\sigma$}}}
\put(1501,-511){\makebox(0,0)[lb]{\smash{\SetFigFont{10}{12.0}{rm}$\gamma_\nu$}}}
\put(1351,-61){\makebox(0,0)[lb]{\smash{\SetFigFont{10}{12.0}{rm}$\gamma_\mu$}}}
\end{picture}
  + \begin{picture}(0,0)%
\epsfig{file=fig5.pstex}%
\end{picture}%
\setlength{\unitlength}{0.00083300in}%
\begingroup\makeatletter\ifx\SetFigFont\undefined
\def\x#1#2#3#4#5#6#7\relax{\def\x{#1#2#3#4#5#6}}%
\expandafter\x\fmtname xxxxxx\relax \def\y{splain}%
\ifx\x\y   
\gdef\SetFigFont#1#2#3{%
  \ifnum #1<17\tiny\else \ifnum #1<20\small\else
  \ifnum #1<24\normalsize\else \ifnum #1<29\large\else
  \ifnum #1<34\Large\else \ifnum #1<41\LARGE\else
     \huge\fi\fi\fi\fi\fi\fi
  \csname #3\endcsname}%
\else
\gdef\SetFigFont#1#2#3{\begingroup
  \count@#1\relax \ifnum 25<\count@\count@25\fi
  \def\x{\endgroup\@setsize\SetFigFont{#2pt}}%
  \expandafter\x
    \csname \romannumeral\the\count@ pt\expandafter\endcsname
    \csname @\romannumeral\the\count@ pt\endcsname
  \csname #3\endcsname}%
\fi
\fi\endgroup
\begin{picture}(1137,775)(451,-704)
\put(1501,-511){\makebox(0,0)[lb]{\smash{\SetFigFont{10}{12.0}{rm}$\gamma_\nu$}}}
\put(601,-61){\makebox(0,0)[lb]{\smash{\SetFigFont{10}{12.0}{rm}$\gamma_\mu$}}}
\put(1351,-61){\makebox(0,0)[lb]{\smash{\SetFigFont{10}{12.0}{rm}$\gamma_\sigma$}}}
\put(451,-511){\makebox(0,0)[lb]{\smash{\SetFigFont{10}{12.0}{rm}$\gamma_\sigma$}}}
\end{picture}
\; \Bigg\} ;
\end{eqnarray}
the integrals on the right-hand-side are convergent. This fact
has to hold in any order. Hence, we can include in (\ref{1.8}) the
exact electron Green's functions and the exact vertex functions and
add all the corresponding diagrams which were not included. As a
result, we can write
\begin{eqnarray}
\label{1.9}
\lefteqn{k_{\mu}\partial_{\mu}\frac{1}{e^2} = } \nonumber\\
 & = & \frac{1}{3}\frac{k_{\mu}k_{\nu}}{k^2} \sum_p \Bigg\{
 \; \begin{picture}(0,0)%
\epsfig{file=fig6.pstex}%
\end{picture}%
\setlength{\unitlength}{0.00083300in}%
\begingroup\makeatletter\ifx\SetFigFont\undefined
\def\x#1#2#3#4#5#6#7\relax{\def\x{#1#2#3#4#5#6}}%
\expandafter\x\fmtname xxxxxx\relax \def\y{splain}%
\ifx\x\y   
\gdef\SetFigFont#1#2#3{%
  \ifnum #1<17\tiny\else \ifnum #1<20\small\else
  \ifnum #1<24\normalsize\else \ifnum #1<29\large\else
  \ifnum #1<34\Large\else \ifnum #1<41\LARGE\else
     \huge\fi\fi\fi\fi\fi\fi
  \csname #3\endcsname}%
\else
\gdef\SetFigFont#1#2#3{\begingroup
  \count@#1\relax \ifnum 25<\count@\count@25\fi
  \def\x{\endgroup\@setsize\SetFigFont{#2pt}}%
  \expandafter\x
    \csname \romannumeral\the\count@ pt\expandafter\endcsname
    \csname @\romannumeral\the\count@ pt\endcsname
  \csname #3\endcsname}%
\fi
\fi\endgroup
\begin{picture}(1058,960)(451,-838)
\put(1201,-811){\makebox(0,0)[lb]{\smash{\SetFigFont{10}{12.0}{rm}$G$}}}
\put(451,-286){\makebox(0,0)[lb]{\smash{\SetFigFont{10}{12.0}{rm}$G$}}}
\put(1501,-511){\makebox(0,0)[lb]{\smash{\SetFigFont{10}{12.0}{rm}$\Gamma_\nu$}}}
\put(526,-61){\makebox(0,0)[lb]{\smash{\SetFigFont{10}{12.0}{rm}$\del_\sigma G^{-1}$}}}
\put(1351,-61){\makebox(0,0)[lb]{\smash{\SetFigFont{10}{12.0}{rm}$\del_\sigma G^{-1}$}}}
\put(1126, 14){\makebox(0,0)[lb]{\smash{\SetFigFont{10}{12.0}{rm}$G$}}}
\put(1501,-286){\makebox(0,0)[lb]{\smash{\SetFigFont{10}{12.0}{rm}$G$}}}
\put(451,-511){\makebox(0,0)[lb]{\smash{\SetFigFont{10}{12.0}{rm}$\Gamma_\mu$}}}
\end{picture}
 \;\; + \;\; \begin{picture}(0,0)%
\epsfig{file=fig7.pstex}%
\end{picture}%
\setlength{\unitlength}{0.00083300in}%
\begingroup\makeatletter\ifx\SetFigFont\undefined
\def\x#1#2#3#4#5#6#7\relax{\def\x{#1#2#3#4#5#6}}%
\expandafter\x\fmtname xxxxxx\relax \def\y{splain}%
\ifx\x\y   
\gdef\SetFigFont#1#2#3{%
  \ifnum #1<17\tiny\else \ifnum #1<20\small\else
  \ifnum #1<24\normalsize\else \ifnum #1<29\large\else
  \ifnum #1<34\Large\else \ifnum #1<41\LARGE\else
     \huge\fi\fi\fi\fi\fi\fi
  \csname #3\endcsname}%
\else
\gdef\SetFigFont#1#2#3{\begingroup
  \count@#1\relax \ifnum 25<\count@\count@25\fi
  \def\x{\endgroup\@setsize\SetFigFont{#2pt}}%
  \expandafter\x
    \csname \romannumeral\the\count@ pt\expandafter\endcsname
    \csname @\romannumeral\the\count@ pt\endcsname
  \csname #3\endcsname}%
\fi
\fi\endgroup
\begin{picture}(1058,990)(451,-844)
\put(1351,-61){\makebox(0,0)[lb]{\smash{\SetFigFont{10}{12.0}{rm}$\del_\sigma G^{-1}$}}}
\put(451,-511){\makebox(0,0)[lb]{\smash{\SetFigFont{10}{12.0}{rm}$\Gamma_\mu$}}}
\put(1501,-511){\makebox(0,0)[lb]{\smash{\SetFigFont{10}{12.0}{rm}$\Gamma_\nu$}}}
\put(1051, 14){\makebox(0,0)[lb]{\smash{\SetFigFont{10}{12.0}{rm}$\Gamma_\rho$}}}
\put(451,-61){\makebox(0,0)[lb]{\smash{\SetFigFont{10}{12.0}{rm}$\del_\sigma G^{-1}$}}}
\put(1126,-811){\makebox(0,0)[lb]{\smash{\SetFigFont{10}{12.0}{rm}$\Gamma_\rho$}}}
\end{picture}
 \;\; + \cdots \Bigg\}
\end{eqnarray}
where $\sum_p$ denotes the sum over the permutations of the indices
similarly to (\ref{1.8}); a quantity $\frac{e^2}{k^2}$ corresponds
to each photon line. (\ref{1.9}) is an equation for $e^2$ (in the
form of series in $e^2$) provided that $G(k)$ and $\Gamma_{\mu}(k,q)$
are known.

In order to obtain equations for electron Green's functions and
vertices we have to remember that these function can change
rapidly even if $e^2$ is small because they can have infrared
singularities of the type $e^2\ln\frac{q^2-m^2}{m^2}$ or ultraviolet
singularities of the type $\left(\frac{\alpha}{\alpha_0}
\right)^{\gamma}$. It is proven to be possible to arrange the
differentiation in such a way that there will be an expansion
only in $e^2$.

To write an equation for the fermion Green's function not containing
any divergences, we have to differentiate twice the self-energy of
the fermion or its inverse Green's function. Let us consider
$\partial_{\mu\mu}G^{-1}(q)$; the diagrammatic expression for
$G^{-1}$ will be the following. The simplest diagram is
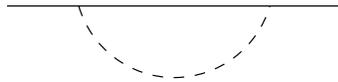
\vspace{.3cm}
\[ \begin{picture}(0,0)%
\epsfig{file=fig8.pstex}%
\end{picture}%
\setlength{\unitlength}{0.00083300in}%
\begingroup\makeatletter\ifx\SetFigFont\undefined
\def\x#1#2#3#4#5#6#7\relax{\def\x{#1#2#3#4#5#6}}%
\expandafter\x\fmtname xxxxxx\relax \def\y{splain}%
\ifx\x\y   
\gdef\SetFigFont#1#2#3{%
  \ifnum #1<17\tiny\else \ifnum #1<20\small\else
  \ifnum #1<24\normalsize\else \ifnum #1<29\large\else
  \ifnum #1<34\Large\else \ifnum #1<41\LARGE\else
     \huge\fi\fi\fi\fi\fi\fi
  \csname #3\endcsname}%
\else
\gdef\SetFigFont#1#2#3{\begingroup
  \count@#1\relax \ifnum 25<\count@\count@25\fi
  \def\x{\endgroup\@setsize\SetFigFont{#2pt}}%
  \expandafter\x
    \csname \romannumeral\the\count@ pt\expandafter\endcsname
    \csname @\romannumeral\the\count@ pt\endcsname
  \csname #3\endcsname}%
\fi
\fi\endgroup
\begin{picture}(2124,471)(439,-520)
\end{picture}
 \]
Diagrams of the next order are
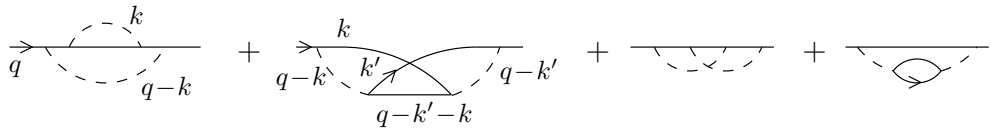
\[ \begin{picture}(0,0)%
\epsfig{file=combine1.pstex}%
\end{picture}%
\setlength{\unitlength}{0.00083300in}%
\begingroup\makeatletter\ifx\SetFigFont\undefined
\def\x#1#2#3#4#5#6#7\relax{\def\x{#1#2#3#4#5#6}}%
\expandafter\x\fmtname xxxxxx\relax \def\y{splain}%
\ifx\x\y   
\gdef\SetFigFont#1#2#3{%
  \ifnum #1<17\tiny\else \ifnum #1<20\small\else
  \ifnum #1<24\normalsize\else \ifnum #1<29\large\else
  \ifnum #1<34\Large\else \ifnum #1<41\LARGE\else
     \huge\fi\fi\fi\fi\fi\fi
  \csname #3\endcsname}%
\else
\gdef\SetFigFont#1#2#3{\begingroup
  \count@#1\relax \ifnum 25<\count@\count@25\fi
  \def\x{\endgroup\@setsize\SetFigFont{#2pt}}%
  \expandafter\x
    \csname \romannumeral\the\count@ pt\expandafter\endcsname
    \csname @\romannumeral\the\count@ pt\endcsname
  \csname #3\endcsname}%
\fi
\fi\endgroup
\begin{picture}(6174,753)(439,-856)
\put(1201,-211){\makebox(0,0)[lb]{\smash{\SetFigFont{10}{12.0}{rm}$k$}}}
\put(451,-511){\makebox(0,0)[lb]{\smash{\SetFigFont{10}{12.0}{rm}$q$}}}
\put(1276,-661){\makebox(0,0)[lb]{\smash{\SetFigFont{10}{12.0}{rm}$q\!-\!k$}}}
\put(2495,-297){\makebox(0,0)[lb]{\smash{\SetFigFont{10}{12.0}{rm}$k$}}}
\put(5401,-436){\makebox(0,0)[lb]{\smash{\SetFigFont{12}{14.4}{rm}$+$}}}
\put(2742,-826){\makebox(0,0)[lb]{\smash{\SetFigFont{10}{12.0}{rm}$q\!-\!k'\!-\!k$}}}
\put(2638,-541){\makebox(0,0)[lb]{\smash{\SetFigFont{10}{12.0}{rm}$k'$}}}
\put(3515,-556){\makebox(0,0)[lb]{\smash{\SetFigFont{10}{12.0}{rm}$q\!-\!k'$}}}
\put(2116,-594){\makebox(0,0)[lb]{\smash{\SetFigFont{10}{12.0}{rm}$q\!-\!k$}}}
\put(1876,-436){\makebox(0,0)[lb]{\smash{\SetFigFont{12}{14.4}{rm}$+$}}}
\put(4051,-436){\makebox(0,0)[lb]{\smash{\SetFigFont{12}{14.4}{rm}$+$}}}
\end{picture}
 \]
It can be easily shown that
\begin{equation}
\label{1.10}
 \partial^2 \frac{1}{k^2+i\varepsilon} = - 4\pi^2 i\delta^4(k)
\end{equation}
Using this equality, we have
\[ \partial^2  \, \begin{picture}(0,0)%
\epsfig{file=fig13.pstex}%
\end{picture}%
\setlength{\unitlength}{0.00083300in}%
\begingroup\makeatletter\ifx\SetFigFont\undefined
\def\x#1#2#3#4#5#6#7\relax{\def\x{#1#2#3#4#5#6}}%
\expandafter\x\fmtname xxxxxx\relax \def\y{splain}%
\ifx\x\y   
\gdef\SetFigFont#1#2#3{%
  \ifnum #1<17\tiny\else \ifnum #1<20\small\else
  \ifnum #1<24\normalsize\else \ifnum #1<29\large\else
  \ifnum #1<34\Large\else \ifnum #1<41\LARGE\else
     \huge\fi\fi\fi\fi\fi\fi
  \csname #3\endcsname}%
\else
\gdef\SetFigFont#1#2#3{\begingroup
  \count@#1\relax \ifnum 25<\count@\count@25\fi
  \def\x{\endgroup\@setsize\SetFigFont{#2pt}}%
  \expandafter\x
    \csname \romannumeral\the\count@ pt\expandafter\endcsname
    \csname @\romannumeral\the\count@ pt\endcsname
  \csname #3\endcsname}%
\fi
\fi\endgroup
\begin{picture}(924,246)(589,-295)
\end{picture}
 = -\frac{e_0^2}{4\pi^2}\gamma_{\mu}G_0
 \gamma_{\mu} = -g\gamma_{\mu}G_0\gamma_{\mu} ,\]
\vspace{.3cm}
\[ \partial^2 \, \begin{picture}(0,0)%
\epsfig{file=fig14.pstex}%
\end{picture}%
\setlength{\unitlength}{0.00083300in}%
\begingroup\makeatletter\ifx\SetFigFont\undefined
\def\x#1#2#3#4#5#6#7\relax{\def\x{#1#2#3#4#5#6}}%
\expandafter\x\fmtname xxxxxx\relax \def\y{splain}%
\ifx\x\y   
\gdef\SetFigFont#1#2#3{%
  \ifnum #1<17\tiny\else \ifnum #1<20\small\else
  \ifnum #1<24\normalsize\else \ifnum #1<29\large\else
  \ifnum #1<34\Large\else \ifnum #1<41\LARGE\else
     \huge\fi\fi\fi\fi\fi\fi
  \csname #3\endcsname}%
\else
\gdef\SetFigFont#1#2#3{\begingroup
  \count@#1\relax \ifnum 25<\count@\count@25\fi
  \def\x{\endgroup\@setsize\SetFigFont{#2pt}}%
  \expandafter\x
    \csname \romannumeral\the\count@ pt\expandafter\endcsname
    \csname @\romannumeral\the\count@ pt\endcsname
  \csname #3\endcsname}%
\fi
\fi\endgroup
\begin{picture}(924,393)(589,-295)
\end{picture}
 = -g\gamma_{\mu}G_0\gamma_{\mu} ,\]
\vspace{.3cm}
\[ \partial^2 \, \begin{picture}(0,0)%
\epsfig{file=fig15.pstex}%
\end{picture}%
\setlength{\unitlength}{0.00083300in}%
\begingroup\makeatletter\ifx\SetFigFont\undefined
\def\x#1#2#3#4#5#6#7\relax{\def\x{#1#2#3#4#5#6}}%
\expandafter\x\fmtname xxxxxx\relax \def\y{splain}%
\ifx\x\y   
\gdef\SetFigFont#1#2#3{%
  \ifnum #1<17\tiny\else \ifnum #1<20\small\else
  \ifnum #1<24\normalsize\else \ifnum #1<29\large\else
  \ifnum #1<34\Large\else \ifnum #1<41\LARGE\else
     \huge\fi\fi\fi\fi\fi\fi
  \csname #3\endcsname}%
\else
\gdef\SetFigFont#1#2#3{\begingroup
  \count@#1\relax \ifnum 25<\count@\count@25\fi
  \def\x{\endgroup\@setsize\SetFigFont{#2pt}}%
  \expandafter\x
    \csname \romannumeral\the\count@ pt\expandafter\endcsname
    \csname @\romannumeral\the\count@ pt\endcsname
  \csname #3\endcsname}%
\fi
\fi\endgroup
\begin{picture}(1224,451)(289,-459)
\end{picture}
 = 
  - g \,\,\,\begin{picture}(0,0)%
\epsfig{file=fig16.pstex}%
\end{picture}%
\setlength{\unitlength}{0.00083300in}%
\begingroup\makeatletter\ifx\SetFigFont\undefined
\def\x#1#2#3#4#5#6#7\relax{\def\x{#1#2#3#4#5#6}}%
\expandafter\x\fmtname xxxxxx\relax \def\y{splain}%
\ifx\x\y   
\gdef\SetFigFont#1#2#3{%
  \ifnum #1<17\tiny\else \ifnum #1<20\small\else
  \ifnum #1<24\normalsize\else \ifnum #1<29\large\else
  \ifnum #1<34\Large\else \ifnum #1<41\LARGE\else
     \huge\fi\fi\fi\fi\fi\fi
  \csname #3\endcsname}%
\else
\gdef\SetFigFont#1#2#3{\begingroup
  \count@#1\relax \ifnum 25<\count@\count@25\fi
  \def\x{\endgroup\@setsize\SetFigFont{#2pt}}%
  \expandafter\x
    \csname \romannumeral\the\count@ pt\expandafter\endcsname
    \csname @\romannumeral\the\count@ pt\endcsname
  \csname #3\endcsname}%
\fi
\fi\endgroup
\begin{picture}(938,342)(575,-544)
\put(1201,-511){\makebox(0,0)[lb]{\smash{\SetFigFont{10}{12.0}{rm}$\gamma_\mu$}}}
\put(601,-511){\makebox(0,0)[lb]{\smash{\SetFigFont{10}{12.0}{rm}$\gamma_\mu$}}}
\end{picture}
 
  - g \,\,\, \begin{picture}(0,0)%
\epsfig{file=fig17.pstex}%
\end{picture}%
\setlength{\unitlength}{0.00083300in}%
\begingroup\makeatletter\ifx\SetFigFont\undefined
\def\x#1#2#3#4#5#6#7\relax{\def\x{#1#2#3#4#5#6}}%
\expandafter\x\fmtname xxxxxx\relax \def\y{splain}%
\ifx\x\y   
\gdef\SetFigFont#1#2#3{%
  \ifnum #1<17\tiny\else \ifnum #1<20\small\else
  \ifnum #1<24\normalsize\else \ifnum #1<29\large\else
  \ifnum #1<34\Large\else \ifnum #1<41\LARGE\else
     \huge\fi\fi\fi\fi\fi\fi
  \csname #3\endcsname}%
\else
\gdef\SetFigFont#1#2#3{\begingroup
  \count@#1\relax \ifnum 25<\count@\count@25\fi
  \def\x{\endgroup\@setsize\SetFigFont{#2pt}}%
  \expandafter\x
    \csname \romannumeral\the\count@ pt\expandafter\endcsname
    \csname @\romannumeral\the\count@ pt\endcsname
  \csname #3\endcsname}%
\fi
\fi\endgroup
\begin{picture}(934,342)(593,-1144)
\put(1501,-1111){\makebox(0,0)[lb]{\smash{\SetFigFont{10}{12.0}{rm}$\gamma_\mu$}}}
\put(901,-1111){\makebox(0,0)[lb]{\smash{\SetFigFont{10}{12.0}{rm}$\gamma_\mu$}}}
\end{picture}
 + \delta_2 , \]
\vspace{.3cm}
\[ \partial^2 \, \begin{picture}(0,0)%
\epsfig{file=fig18.pstex}%
\end{picture}%
\setlength{\unitlength}{0.00083300in}%
\begingroup\makeatletter\ifx\SetFigFont\undefined
\def\x#1#2#3#4#5#6#7\relax{\def\x{#1#2#3#4#5#6}}%
\expandafter\x\fmtname xxxxxx\relax \def\y{splain}%
\ifx\x\y   
\gdef\SetFigFont#1#2#3{%
  \ifnum #1<17\tiny\else \ifnum #1<20\small\else
  \ifnum #1<24\normalsize\else \ifnum #1<29\large\else
  \ifnum #1<34\Large\else \ifnum #1<41\LARGE\else
     \huge\fi\fi\fi\fi\fi\fi
  \csname #3\endcsname}%
\else
\gdef\SetFigFont#1#2#3{\begingroup
  \count@#1\relax \ifnum 25<\count@\count@25\fi
  \def\x{\endgroup\@setsize\SetFigFont{#2pt}}%
  \expandafter\x
    \csname \romannumeral\the\count@ pt\expandafter\endcsname
    \csname @\romannumeral\the\count@ pt\endcsname
  \csname #3\endcsname}%
\fi
\fi\endgroup
\begin{picture}(924,246)(589,-595)
\end{picture}
 = 
  \gamma_{\mu}\int\frac{d^4 k}{4\pi^2 i}
 G(q')\partial^2 \frac{g_1(k^2)}{k^2} \gamma_{\mu} . \]
Restricting ourselves to $\partial^2\frac{1}{k^2}$, we can write
\[ \partial^2 \sigma_1 = -g_0\gamma_{\mu}G_0\gamma_{\mu} ,\]
\[ \partial^2 \sigma_2 = -g_0\gamma_{\mu}G_0\partial_{\mu}\sigma_1 +
 g_0\partial_{\mu}\sigma_1 G_0\gamma_{\mu} -
 g_0\gamma_{\mu}G_1\gamma_{\mu} - g_1\gamma_{\mu}G_0\gamma_{\mu} \]
and, consequently,
\begin{equation}
\label{1.11}
\partial^2 G^{-1} = g\partial_{\mu}G^{-1}G\partial_{\mu}G^{-1},
 \quad\mbox{}\quad G_1^{-1}+G_2^{-1} = G^{-1} .
\end{equation}
The term $\delta_2$ has a contribution to $\,\,\begin{picture}(0,0)%
\epsfig{file=fig19.pstex}%
\end{picture}%
\setlength{\unitlength}{0.00083300in}%
\begingroup\makeatletter\ifx\SetFigFont\undefined
\def\x#1#2#3#4#5#6#7\relax{\def\x{#1#2#3#4#5#6}}%
\expandafter\x\fmtname xxxxxx\relax \def\y{splain}%
\ifx\x\y   
\gdef\SetFigFont#1#2#3{%
  \ifnum #1<17\tiny\else \ifnum #1<20\small\else
  \ifnum #1<24\normalsize\else \ifnum #1<29\large\else
  \ifnum #1<34\Large\else \ifnum #1<41\LARGE\else
     \huge\fi\fi\fi\fi\fi\fi
  \csname #3\endcsname}%
\else
\gdef\SetFigFont#1#2#3{\begingroup
  \count@#1\relax \ifnum 25<\count@\count@25\fi
  \def\x{\endgroup\@setsize\SetFigFont{#2pt}}%
  \expandafter\x
    \csname \romannumeral\the\count@ pt\expandafter\endcsname
    \csname @\romannumeral\the\count@ pt\endcsname
  \csname #3\endcsname}%
\fi
\fi\endgroup
\begin{picture}(924,174)(589,-373)
\end{picture}
$ not
containing overlapping divergences and is of the form
\begin{eqnarray}
\delta_2 &=& \,\,\begin{picture}(0,0)%
\epsfig{file=combine2.pstex}%
\end{picture}%
\setlength{\unitlength}{0.00083300in}%
\begingroup\makeatletter\ifx\SetFigFont\undefined
\def\x#1#2#3#4#5#6#7\relax{\def\x{#1#2#3#4#5#6}}%
\expandafter\x\fmtname xxxxxx\relax \def\y{splain}%
\ifx\x\y   
\gdef\SetFigFont#1#2#3{%
  \ifnum #1<17\tiny\else \ifnum #1<20\small\else
  \ifnum #1<24\normalsize\else \ifnum #1<29\large\else
  \ifnum #1<34\Large\else \ifnum #1<41\LARGE\else
     \huge\fi\fi\fi\fi\fi\fi
  \csname #3\endcsname}%
\else
\gdef\SetFigFont#1#2#3{\begingroup
  \count@#1\relax \ifnum 25<\count@\count@25\fi
  \def\x{\endgroup\@setsize\SetFigFont{#2pt}}%
  \expandafter\x
    \csname \romannumeral\the\count@ pt\expandafter\endcsname
    \csname @\romannumeral\the\count@ pt\endcsname
  \csname #3\endcsname}%
\fi
\fi\endgroup
\begin{picture}(3624,690)(289,-544)
\put(3301, 14){\makebox(0,0)[lb]{\smash{\SetFigFont{11}{13.2}{rm}$-\frac{2k'_\sigma}{k'^4}$}}}
\put(1126,-511){\makebox(0,0)[lb]{\smash{\SetFigFont{11}{13.2}{rm}$\gamma_\sigma$}}}
\put(901,-511){\makebox(0,0)[lb]{\smash{\SetFigFont{11}{13.2}{rm}$\gamma_\sigma$}}}
\put(2026,-286){\makebox(0,0)[lb]{\smash{\SetFigFont{12}{14.4}{rm}$+$}}}
\put(2776, 14){\makebox(0,0)[lb]{\smash{\SetFigFont{11}{13.2}{rm}$-\frac{2k_\sigma}{k^4}$}}}
\end{picture}
 \nonumber\\[0.3cm]
         & & + \,\, \begin{picture}(0,0)%
\epsfig{file=fig22.pstex}%
\end{picture}%
\setlength{\unitlength}{0.00083300in}%
\begingroup\makeatletter\ifx\SetFigFont\undefined
\def\x#1#2#3#4#5#6#7\relax{\def\x{#1#2#3#4#5#6}}%
\expandafter\x\fmtname xxxxxx\relax \def\y{splain}%
\ifx\x\y   
\gdef\SetFigFont#1#2#3{%
  \ifnum #1<17\tiny\else \ifnum #1<20\small\else
  \ifnum #1<24\normalsize\else \ifnum #1<29\large\else
  \ifnum #1<34\Large\else \ifnum #1<41\LARGE\else
     \huge\fi\fi\fi\fi\fi\fi
  \csname #3\endcsname}%
\else
\gdef\SetFigFont#1#2#3{\begingroup
  \count@#1\relax \ifnum 25<\count@\count@25\fi
  \def\x{\endgroup\@setsize\SetFigFont{#2pt}}%
  \expandafter\x
    \csname \romannumeral\the\count@ pt\expandafter\endcsname
    \csname @\romannumeral\the\count@ pt\endcsname
  \csname #3\endcsname}%
\fi
\fi\endgroup
\begin{picture}(1524,690)(289,-544)
\put(1051,-511){\makebox(0,0)[lb]{\smash{\SetFigFont{11}{13.2}{rm}$\gamma_\sigma$}}}
\put(676, 14){\makebox(0,0)[lb]{\smash{\SetFigFont{11}{13.2}{rm}$-\frac{2k_\sigma}{k^4}$}}}
\end{picture}
 \,\, + \,\, \begin{picture}(0,0)%
\epsfig{file=fig23.pstex}%
\end{picture}%
\setlength{\unitlength}{0.00083300in}%
\begingroup\makeatletter\ifx\SetFigFont\undefined
\def\x#1#2#3#4#5#6#7\relax{\def\x{#1#2#3#4#5#6}}%
\expandafter\x\fmtname xxxxxx\relax \def\y{splain}%
\ifx\x\y   
\gdef\SetFigFont#1#2#3{%
  \ifnum #1<17\tiny\else \ifnum #1<20\small\else
  \ifnum #1<24\normalsize\else \ifnum #1<29\large\else
  \ifnum #1<34\Large\else \ifnum #1<41\LARGE\else
     \huge\fi\fi\fi\fi\fi\fi
  \csname #3\endcsname}%
\else
\gdef\SetFigFont#1#2#3{\begingroup
  \count@#1\relax \ifnum 25<\count@\count@25\fi
  \def\x{\endgroup\@setsize\SetFigFont{#2pt}}%
  \expandafter\x
    \csname \romannumeral\the\count@ pt\expandafter\endcsname
    \csname @\romannumeral\the\count@ pt\endcsname
  \csname #3\endcsname}%
\fi
\fi\endgroup
\begin{picture}(1524,690)(289,-544)
\put(1201, 14){\makebox(0,0)[lb]{\smash{\SetFigFont{11}{13.2}{rm}$-\frac{2k'_\sigma}{k'^4}$}}}
\put(1051,-511){\makebox(0,0)[lb]{\smash{\SetFigFont{11}{13.2}{rm}$\gamma_\sigma$}}}
\end{picture}
  
\label{1.12}
\end{eqnarray}
For the calculation of higher order diagrams it is convenient to
adopt the following principle. Beginning from the first point of
interaction we shall relate the external momentum to the photon
line:
\begin{equation}
\begin{picture}(0,0)%
\epsfig{file=fig24.pstex}%
\end{picture}%
\setlength{\unitlength}{0.00083300in}%
\begingroup\makeatletter\ifx\SetFigFont\undefined
\def\x#1#2#3#4#5#6#7\relax{\def\x{#1#2#3#4#5#6}}%
\expandafter\x\fmtname xxxxxx\relax \def\y{splain}%
\ifx\x\y   
\gdef\SetFigFont#1#2#3{%
  \ifnum #1<17\tiny\else \ifnum #1<20\small\else
  \ifnum #1<24\normalsize\else \ifnum #1<29\large\else
  \ifnum #1<34\Large\else \ifnum #1<41\LARGE\else
     \huge\fi\fi\fi\fi\fi\fi
  \csname #3\endcsname}%
\else
\gdef\SetFigFont#1#2#3{\begingroup
  \count@#1\relax \ifnum 25<\count@\count@25\fi
  \def\x{\endgroup\@setsize\SetFigFont{#2pt}}%
  \expandafter\x
    \csname \romannumeral\the\count@ pt\expandafter\endcsname
    \csname @\romannumeral\the\count@ pt\endcsname
  \csname #3\endcsname}%
\fi
\fi\endgroup
\begin{picture}(1224,699)(589,-598)
\put(1651,-436){\makebox(0,0)[lb]{\smash{\SetFigFont{10}{12.0}{rm}$q\!-\!q'$}}}
\put(818,-533){\makebox(0,0)[lb]{\smash{\SetFigFont{10}{12.0}{rm}$q$}}}
\put(1201,-136){\makebox(0,0)[lb]{\smash{\SetFigFont{10}{12.0}{rm}$q'$}}}
\end{picture}
\hspace{1.5cm} .
\label{1.13}
\end{equation}
The next photon interaction can be expressed as
\begin{equation}
\begin{picture}(0,0)%
\epsfig{file=fig25.pstex}%
\end{picture}%
\setlength{\unitlength}{0.00083300in}%
\begingroup\makeatletter\ifx\SetFigFont\undefined
\def\x#1#2#3#4#5#6#7\relax{\def\x{#1#2#3#4#5#6}}%
\expandafter\x\fmtname xxxxxx\relax \def\y{splain}%
\ifx\x\y   
\gdef\SetFigFont#1#2#3{%
  \ifnum #1<17\tiny\else \ifnum #1<20\small\else
  \ifnum #1<24\normalsize\else \ifnum #1<29\large\else
  \ifnum #1<34\Large\else \ifnum #1<41\LARGE\else
     \huge\fi\fi\fi\fi\fi\fi
  \csname #3\endcsname}%
\else
\gdef\SetFigFont#1#2#3{\begingroup
  \count@#1\relax \ifnum 25<\count@\count@25\fi
  \def\x{\endgroup\@setsize\SetFigFont{#2pt}}%
  \expandafter\x
    \csname \romannumeral\the\count@ pt\expandafter\endcsname
    \csname @\romannumeral\the\count@ pt\endcsname
  \csname #3\endcsname}%
\fi
\fi\endgroup
\begin{picture}(1374,774)(589,-748)
\put(1726,-511){\makebox(0,0)[lb]{\smash{\SetFigFont{10}{12.0}{rm}$q\!-\!q'\!-\!q''$}}}
\put(743,-533){\makebox(0,0)[lb]{\smash{\SetFigFont{10}{12.0}{rm}$q\!-\!q'$}}}
\put(1351,-136){\makebox(0,0)[lb]{\smash{\SetFigFont{10}{12.0}{rm}$q''$}}}
\end{picture}
 \hspace{1.5cm} ;
\label{1.14}
\end{equation}
now we relate the external momentum to the positron line. If the
positron is emitting a photon, we keep sending the external momentum
along the positron line up to its annihilation. As a result, we get
a tree structure
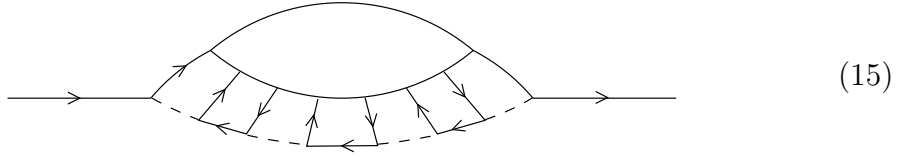
\begin{equation}
\begin{picture}(0,0)%
\epsfig{file=fig26.pstex}%
\end{picture}%
\setlength{\unitlength}{0.00083300in}%
\begingroup\makeatletter\ifx\SetFigFont\undefined
\def\x#1#2#3#4#5#6#7\relax{\def\x{#1#2#3#4#5#6}}%
\expandafter\x\fmtname xxxxxx\relax \def\y{splain}%
\ifx\x\y   
\gdef\SetFigFont#1#2#3{%
  \ifnum #1<17\tiny\else \ifnum #1<20\small\else
  \ifnum #1<24\normalsize\else \ifnum #1<29\large\else
  \ifnum #1<34\Large\else \ifnum #1<41\LARGE\else
     \huge\fi\fi\fi\fi\fi\fi
  \csname #3\endcsname}%
\else
\gdef\SetFigFont#1#2#3{\begingroup
  \count@#1\relax \ifnum 25<\count@\count@25\fi
  \def\x{\endgroup\@setsize\SetFigFont{#2pt}}%
  \expandafter\x
    \csname \romannumeral\the\count@ pt\expandafter\endcsname
    \csname @\romannumeral\the\count@ pt\endcsname
  \csname #3\endcsname}%
\fi
\fi\endgroup
\begin{picture}(4224,958)(589,-1310)
\end{picture}

\label{1.15}
\end{equation}
in which the initial photon line can end only at the final electron.
It can be shown that every line in the diagram will be passed
only once if the photons included in the fermion self energy
are not taken into account. This means that in this approach
exact electron Green's functions have to be used with bare vertices
since the idea is basically the exclusion of overlapping divergences.
Taking the second derivative of one of the photon propagators and
restricting ourselves to the contribution $4\pi^4\delta^4(k)g(0)$
we obtain both on the left-hand side and the right-hand side
photon emission amplitudes with zero momentum. Due to gauge
invariance, however, a zero momentum photon cannot change the state
of the system (it is emitted from the external end). The photon
emission amplitude equals $\Gamma_{\mu}(q,0)=\partial_{\mu}G^{-1}(q)$.
The final contribution of the differentiation has the structure
(\ref{1.11}). The above statement, which essentially means that the
emission of a zero momentum photon is determined by the total charge,
can be proven by the Ward identity.

Thus, we have
\begin{equation}
\label{1.16}
\partial^2 G^{-1} = g(0)\partial_{\mu}G^{-1}G \partial_{\mu}G^{-1}
 + \,\, \begin{picture}(0,0)%
\epsfig{file=fig27.pstex}%
\end{picture}%
\setlength{\unitlength}{0.00083300in}%
\begingroup\makeatletter\ifx\SetFigFont\undefined
\def\x#1#2#3#4#5#6#7\relax{\def\x{#1#2#3#4#5#6}}%
\expandafter\x\fmtname xxxxxx\relax \def\y{splain}%
\ifx\x\y   
\gdef\SetFigFont#1#2#3{%
  \ifnum #1<17\tiny\else \ifnum #1<20\small\else
  \ifnum #1<24\normalsize\else \ifnum #1<29\large\else
  \ifnum #1<34\Large\else \ifnum #1<41\LARGE\else
     \huge\fi\fi\fi\fi\fi\fi
  \csname #3\endcsname}%
\else
\gdef\SetFigFont#1#2#3{\begingroup
  \count@#1\relax \ifnum 25<\count@\count@25\fi
  \def\x{\endgroup\@setsize\SetFigFont{#2pt}}%
  \expandafter\x
    \csname \romannumeral\the\count@ pt\expandafter\endcsname
    \csname @\romannumeral\the\count@ pt\endcsname
  \csname #3\endcsname}%
\fi
\fi\endgroup
\begin{picture}(1374,332)(1114,-759)
\end{picture}

\end{equation}
The remaining diagrams contain first derivatives of photon and
positron lines and second derivatives of positron lines.
All these diagrams can be expressed in terms of the exact $\Gamma_{\mu}$,
$G$ and $D$ functions. They do not contain divergences except those
graphs which correspond to the photon self energy.

The first term in (\ref{1.16}) has the structure
\[ -\frac{1}{2}g(0)\tilde{M}_{\nu\nu}(q,k=0) \]
where $\tilde{M}_{\nu\nu}(q,k=0)$ is a quantity close to the Compton
scattering amplitude in the sense that they would be equal if we
differentiated all photon propagators including those inside the
electron Green's function. (The factor $\frac{1}{2}$ is due to the
fact that the Compton amplitude contains the sum of diagrams with
momenta $k$ and $-k$). The Compton scattering amplitude $M_{\nu\nu}
(q,k=0)$ satisfies the Ward identity
\[ M_{\nu\nu}(q) = G^{-1}\partial^2 GG^{-1} = 2\partial_{\mu}G^{-1}G
 \partial_{\mu}G^{-1} = \partial^2 G^{-1} \]
which has the simple diagrammatical meaning
\begin{equation}
\label{1.17}
 \begin{picture}(0,0)%
\epsfig{file=fig28.pstex}%
\end{picture}%
\setlength{\unitlength}{0.00083300in}%
\begingroup\makeatletter\ifx\SetFigFont\undefined
\def\x#1#2#3#4#5#6#7\relax{\def\x{#1#2#3#4#5#6}}%
\expandafter\x\fmtname xxxxxx\relax \def\y{splain}%
\ifx\x\y   
\gdef\SetFigFont#1#2#3{%
  \ifnum #1<17\tiny\else \ifnum #1<20\small\else
  \ifnum #1<24\normalsize\else \ifnum #1<29\large\else
  \ifnum #1<34\Large\else \ifnum #1<41\LARGE\else
     \huge\fi\fi\fi\fi\fi\fi
  \csname #3\endcsname}%
\else
\gdef\SetFigFont#1#2#3{\begingroup
  \count@#1\relax \ifnum 25<\count@\count@25\fi
  \def\x{\endgroup\@setsize\SetFigFont{#2pt}}%
  \expandafter\x
    \csname \romannumeral\the\count@ pt\expandafter\endcsname
    \csname @\romannumeral\the\count@ pt\endcsname
  \csname #3\endcsname}%
\fi
\fi\endgroup
\begin{picture}(1824,846)(1339,-1195)
\put(1426,-736){\makebox(0,0)[lb]{\smash{\SetFigFont{10}{12.0}{rm}$k\!=\!0$}}}
\put(2776,-736){\makebox(0,0)[lb]{\smash{\SetFigFont{10}{12.0}{rm}$k'\!=\!0$}}}
\end{picture}
 = \begin{picture}(0,0)%
\epsfig{file=fig29.pstex}%
\end{picture}%
\setlength{\unitlength}{0.00083300in}%
\begingroup\makeatletter\ifx\SetFigFont\undefined
\def\x#1#2#3#4#5#6#7\relax{\def\x{#1#2#3#4#5#6}}%
\expandafter\x\fmtname xxxxxx\relax \def\y{splain}%
\ifx\x\y   
\gdef\SetFigFont#1#2#3{%
  \ifnum #1<17\tiny\else \ifnum #1<20\small\else
  \ifnum #1<24\normalsize\else \ifnum #1<29\large\else
  \ifnum #1<34\Large\else \ifnum #1<41\LARGE\else
     \huge\fi\fi\fi\fi\fi\fi
  \csname #3\endcsname}%
\else
\gdef\SetFigFont#1#2#3{\begingroup
  \count@#1\relax \ifnum 25<\count@\count@25\fi
  \def\x{\endgroup\@setsize\SetFigFont{#2pt}}%
  \expandafter\x
    \csname \romannumeral\the\count@ pt\expandafter\endcsname
    \csname @\romannumeral\the\count@ pt\endcsname
  \csname #3\endcsname}%
\fi
\fi\endgroup
\begin{picture}(1674,913)(1414,-1262)
\end{picture}
 +  
   \begin{picture}(0,0)%
\epsfig{file=fig30.pstex}%
\end{picture}%
\setlength{\unitlength}{0.00083300in}%
\begingroup\makeatletter\ifx\SetFigFont\undefined
\def\x#1#2#3#4#5#6#7\relax{\def\x{#1#2#3#4#5#6}}%
\expandafter\x\fmtname xxxxxx\relax \def\y{splain}%
\ifx\x\y   
\gdef\SetFigFont#1#2#3{%
  \ifnum #1<17\tiny\else \ifnum #1<20\small\else
  \ifnum #1<24\normalsize\else \ifnum #1<29\large\else
  \ifnum #1<34\Large\else \ifnum #1<41\LARGE\else
     \huge\fi\fi\fi\fi\fi\fi
  \csname #3\endcsname}%
\else
\gdef\SetFigFont#1#2#3{\begingroup
  \count@#1\relax \ifnum 25<\count@\count@25\fi
  \def\x{\endgroup\@setsize\SetFigFont{#2pt}}%
  \expandafter\x
    \csname \romannumeral\the\count@ pt\expandafter\endcsname
    \csname @\romannumeral\the\count@ pt\endcsname
  \csname #3\endcsname}%
\fi
\fi\endgroup
\begin{picture}(2124,924)(1189,-1273)
\end{picture}

\end{equation}
The second term on the right-hand side of (\ref{1.17}) represents
the contribution of photons which, if we carried out the mentioned
differentiation, would correspond to photons inside the Green's
function and which are not present in $\tilde{M}_{\nu\nu}$. Hence,
\[ \tilde{M}_{\nu\nu}(q,0)=2\partial_{\mu}G^{-1}G\partial_{\mu}G^{-1} .\]

The remaining diagrams in (\ref{1.16}) contain first derivatives of
photon and posi\-tron lines and second derivatives of positron lines.
All these diagrams can be expressed in terms of the exact $\Gamma_{\mu}$,
$G$ and $D$ functions. They do not contain divergences except those
graphs which correspond to the photon self energy. All these diagrams
are of the form
\begin{equation}
\begin{picture}(0,0)%
\epsfig{file=fig31.pstex}%
\end{picture}%
\setlength{\unitlength}{0.00083300in}%
\begingroup\makeatletter\ifx\SetFigFont\undefined
\def\x#1#2#3#4#5#6#7\relax{\def\x{#1#2#3#4#5#6}}%
\expandafter\x\fmtname xxxxxx\relax \def\y{splain}%
\ifx\x\y   
\gdef\SetFigFont#1#2#3{%
  \ifnum #1<17\tiny\else \ifnum #1<20\small\else
  \ifnum #1<24\normalsize\else \ifnum #1<29\large\else
  \ifnum #1<34\Large\else \ifnum #1<41\LARGE\else
     \huge\fi\fi\fi\fi\fi\fi
  \csname #3\endcsname}%
\else
\gdef\SetFigFont#1#2#3{\begingroup
  \count@#1\relax \ifnum 25<\count@\count@25\fi
  \def\x{\endgroup\@setsize\SetFigFont{#2pt}}%
  \expandafter\x
    \csname \romannumeral\the\count@ pt\expandafter\endcsname
    \csname @\romannumeral\the\count@ pt\endcsname
  \csname #3\endcsname}%
\fi
\fi\endgroup
\begin{picture}(3624,901)(589,-1188)
\end{picture}

\label{1.18}
\end{equation}
and correspond to the photon line with the exact Green's function
(\ref{1.6}) which equals $4\pi^2 g(k)\frac{1}{k^2}$.
By differentiating the photon lines we have calculated the
contribution of $\partial^2\frac{1}{k^2}$. So, there remains
\begin{equation}
\label{1.19}
 {\cal I} = \int \frac{d^4 k}{4\pi^2 i}\tilde{M}_{\nu\nu}(q,k)
 \partial^2 \big( g(k)-g(0) \big) \frac{1}{k^2}
\end{equation}
where we have introduced $g(0)$ in order to avoid a contribution
from $\delta^4(k)$. This expression contains logarithmic corrections
coming from the ultraviolet region ($k>q$). Replacing $g(k)-g(0)$ by
$g(q)-g(0)+g(k)-g(q)$ and performing an integration by parts, we get
\begin{eqnarray}
\label{1.20}
\lefteqn{{\cal I} = - \big(g(q)-g(0)\big)\tilde{M}_{\nu\nu}(q,0) + }
\nonumber\\
& + & \int \frac{d^4 k}{4\pi^2 i}\big( g(k)-g(q) \big)
 \frac{1}{k^2} \partial^2 \tilde{M}_{\nu\nu}(q,k).
\end{eqnarray}
The first term in (\ref{1.20}) can be explicitly calculated while
the second one does not contain any logarithms because of the
presence of the difference $g(k)-g(q)$. We can rewrite (\ref{1.20})
in the form
\[ {\cal I} \equiv \big( g(q)-g(k) \big)\partial_{\mu}G^{-1}G
  \partial_{\mu}G^{-1} + \delta_1 ; \]
consequently, $\partial^2 G^{-1}$ can be expressed as
\begin{equation}
\label{1.21}
\partial^2 G^{-1}(q) = g(q)\partial_{\mu}G^{-1}G\partial_{\mu}G^{-1}
  + \delta_1 + \delta_2
\end{equation}
where $\delta_1$ is defined by (\ref{1.20}) and $\delta_2$ contains
first order derivatives of the photon and positron lines and second order
derivatives of the positron lines, as it has been explained above. The
first term contains all singularities of the types $\alpha\ln
\frac{q^2-m^2}{m^2}$ and $\left(\frac{\alpha}{\alpha_0}\right)^{\gamma}$.

\section{Equations for the vertex function and for the amplitudes of
interaction with the external field}\label{II}

To obtain the equation for the vertex function, let us express
$\Gamma_{\mu}(p,q)$ as a set of diagrams containing exact Green's
functions:
\begin{equation}
\label{2.1}
 \Gamma_{\mu}(p,q) =  
   \hspace{-.5cm} \begin{picture}(0,0)%
\epsfig{file=fig32.pstex}%
\end{picture}%
\setlength{\unitlength}{0.00083300in}%
\begingroup\makeatletter\ifx\SetFigFont\undefined
\def\x#1#2#3#4#5#6#7\relax{\def\x{#1#2#3#4#5#6}}%
\expandafter\x\fmtname xxxxxx\relax \def\y{splain}%
\ifx\x\y   
\gdef\SetFigFont#1#2#3{%
  \ifnum #1<17\tiny\else \ifnum #1<20\small\else
  \ifnum #1<24\normalsize\else \ifnum #1<29\large\else
  \ifnum #1<34\Large\else \ifnum #1<41\LARGE\else
     \huge\fi\fi\fi\fi\fi\fi
  \csname #3\endcsname}%
\else
\gdef\SetFigFont#1#2#3{\begingroup
  \count@#1\relax \ifnum 25<\count@\count@25\fi
  \def\x{\endgroup\@setsize\SetFigFont{#2pt}}%
  \expandafter\x
    \csname \romannumeral\the\count@ pt\expandafter\endcsname
    \csname @\romannumeral\the\count@ pt\endcsname
  \csname #3\endcsname}%
\fi
\fi\endgroup
\begin{picture}(1569,1224)(544,-673)
\put(1576,164){\makebox(0,0)[lb]{\smash{\SetFigFont{10}{12.0}{rm}
$p$
}}}
\put(544,-597){\makebox(0,0)[lb]{\smash{\SetFigFont{10}{12.0}{rm}
$q\!-\!\frac{p}{2}$
}}}
\put(1534,-597){\makebox(0,0)[lb]{\smash{\SetFigFont{10}{12.0}{rm}
$q\!+\!\frac{p}{2}$
}}}
\end{picture}
 \hspace{-0.3cm}
   =  \hspace{0.3cm} \begin{picture}(0,0)%
\epsfig{file=fig33.pstex}%
\end{picture}%
\setlength{\unitlength}{0.00083300in}%
\begingroup\makeatletter\ifx\SetFigFont\undefined
\def\x#1#2#3#4#5#6#7\relax{\def\x{#1#2#3#4#5#6}}%
\expandafter\x\fmtname xxxxxx\relax \def\y{splain}%
\ifx\x\y   
\gdef\SetFigFont#1#2#3{%
  \ifnum #1<17\tiny\else \ifnum #1<20\small\else
  \ifnum #1<24\normalsize\else \ifnum #1<29\large\else
  \ifnum #1<34\Large\else \ifnum #1<41\LARGE\else
     \huge\fi\fi\fi\fi\fi\fi
  \csname #3\endcsname}%
\else
\gdef\SetFigFont#1#2#3{\begingroup
  \count@#1\relax \ifnum 25<\count@\count@25\fi
  \def\x{\endgroup\@setsize\SetFigFont{#2pt}}%
  \expandafter\x
    \csname \romannumeral\the\count@ pt\expandafter\endcsname
    \csname @\romannumeral\the\count@ pt\endcsname
  \csname #3\endcsname}%
\fi
\fi\endgroup
\begin{picture}(1224,1317)(889,-916)
\put(976,-886){\makebox(0,0)[lb]{\smash{\SetFigFont{10}{12.0}{rm}
$q\!-\!\frac{p}{2}$
}}}
\put(1576,164){\makebox(0,0)[lb]{\smash{\SetFigFont{10}{12.0}{rm}
$p$
}}}
\put(2101,-661){\makebox(0,0)[lb]{\smash{\SetFigFont{10}{12.0}{rm}
$q\!+\!\frac{p}{2}$
}}}
\put(1726,-286){\makebox(0,0)[lb]{\smash{\SetFigFont{10}{12.0}{rm}
$q'\!+\!\frac{p}{2}$
}}}
\put(901,-286){\makebox(0,0)[lb]{\smash{\SetFigFont{10}{12.0}{rm}
$q'\!-\!\frac{p}{2}$
}}}
\end{picture}
 \hspace{0.3cm} + 
      \hspace{.3cm} \begin{picture}(0,0)%
\epsfig{file=fig34.pstex}%
\end{picture}%
\setlength{\unitlength}{0.00083300in}%
\begingroup\makeatletter\ifx\SetFigFont\undefined
\def\x#1#2#3#4#5#6#7\relax{\def\x{#1#2#3#4#5#6}}%
\expandafter\x\fmtname xxxxxx\relax \def\y{splain}%
\ifx\x\y   
\gdef\SetFigFont#1#2#3{%
  \ifnum #1<17\tiny\else \ifnum #1<20\small\else
  \ifnum #1<24\normalsize\else \ifnum #1<29\large\else
  \ifnum #1<34\Large\else \ifnum #1<41\LARGE\else
     \huge\fi\fi\fi\fi\fi\fi
  \csname #3\endcsname}%
\else
\gdef\SetFigFont#1#2#3{\begingroup
  \count@#1\relax \ifnum 25<\count@\count@25\fi
  \def\x{\endgroup\@setsize\SetFigFont{#2pt}}%
  \expandafter\x
    \csname \romannumeral\the\count@ pt\expandafter\endcsname
    \csname @\romannumeral\the\count@ pt\endcsname
  \csname #3\endcsname}%
\fi
\fi\endgroup
\begin{picture}(1224,1224)(889,-823)
\end{picture}
 
 + \cdots
\end{equation}
We shall relate the external momenta to the lines in the diagrams
in the same way as we did when we derived the equation for the Green's
function. Calculating the second order derivative in $q$ we obtain
\begin{equation}
\label{2.2}
\partial_q^2 \Gamma_{\mu}(p,q) = g(0)\tilde{M}_{\nu\nu}^{\mu}(p,q,k)
 |_{k=0} + \cdots
\end{equation}
where $\tilde{M}_{\nu\nu}(p,q,k)|_{k=0}$ is the contribution to the
amplitude of the process
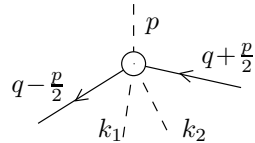
\[ \begin{picture}(0,0)%
\epsfig{file=fig35.pstex}%
\end{picture}%
\setlength{\unitlength}{0.00083300in}%
\begingroup\makeatletter\ifx\SetFigFont\undefined
\def\x#1#2#3#4#5#6#7\relax{\def\x{#1#2#3#4#5#6}}%
\expandafter\x\fmtname xxxxxx\relax \def\y{splain}%
\ifx\x\y   
\gdef\SetFigFont#1#2#3{%
  \ifnum #1<17\tiny\else \ifnum #1<20\small\else
  \ifnum #1<24\normalsize\else \ifnum #1<29\large\else
  \ifnum #1<34\Large\else \ifnum #1<41\LARGE\else
     \huge\fi\fi\fi\fi\fi\fi
  \csname #3\endcsname}%
\else
\gdef\SetFigFont#1#2#3{\begingroup
  \count@#1\relax \ifnum 25<\count@\count@25\fi
  \def\x{\endgroup\@setsize\SetFigFont{#2pt}}%
  \expandafter\x
    \csname \romannumeral\the\count@ pt\expandafter\endcsname
    \csname @\romannumeral\the\count@ pt\endcsname
  \csname #3\endcsname}%
\fi
\fi\endgroup
\begin{picture}(1449,924)(739,-598)
\put(1576,164){\makebox(0,0)[lb]{\smash{\SetFigFont{10}{12.0}{rm}
$p$
}}}
\put(1924,-46){\makebox(0,0)[lb]{\smash{\SetFigFont{10}{12.0}{rm}
$q\!+\!\frac{p}{2}$
}}}
\put(1276,-511){\makebox(0,0)[lb]{\smash{\SetFigFont{10}{12.0}{rm}$k_1$}}}
\put(1801,-511){\makebox(0,0)[lb]{\smash{\SetFigFont{10}{12.0}{rm}$k_2$}}}
\put(739,-230){\makebox(0,0)[lb]{\smash{\SetFigFont{10}{12.0}{rm}
$q\!-\!\frac{p}{2}$
}}}
\end{picture}
 \]
which corresponds to the emission of photons with momenta $k_1,k_2=0$
from the external legs:
\begin{equation}
\label{2.3}
\tilde{M}_{\nu\nu}^{\mu}(p,q) =  \begin{picture}(0,0)%
\epsfig{file=fig36.pstex}%
\end{picture}%
\setlength{\unitlength}{0.00083300in}%
\begingroup\makeatletter\ifx\SetFigFont\undefined
\def\x#1#2#3#4#5#6#7\relax{\def\x{#1#2#3#4#5#6}}%
\expandafter\x\fmtname xxxxxx\relax \def\y{splain}%
\ifx\x\y   
\gdef\SetFigFont#1#2#3{%
  \ifnum #1<17\tiny\else \ifnum #1<20\small\else
  \ifnum #1<24\normalsize\else \ifnum #1<29\large\else
  \ifnum #1<34\Large\else \ifnum #1<41\LARGE\else
     \huge\fi\fi\fi\fi\fi\fi
  \csname #3\endcsname}%
\else
\gdef\SetFigFont#1#2#3{\begingroup
  \count@#1\relax \ifnum 25<\count@\count@25\fi
  \def\x{\endgroup\@setsize\SetFigFont{#2pt}}%
  \expandafter\x
    \csname \romannumeral\the\count@ pt\expandafter\endcsname
    \csname @\romannumeral\the\count@ pt\endcsname
  \csname #3\endcsname}%
\fi
\fi\endgroup
\begin{picture}(1212,720)(451,-619)
\put(451,-586){\makebox(0,0)[lb]{\smash{\SetFigFont{10}{12.0}{rm}$\del_{\nu}G^{-1}$}}}
\put(1051,-586){\makebox(0,0)[lb]{\smash{\SetFigFont{10}{12.0}{rm}$\Gamma^\mu$}}}
\put(1501,-586){\makebox(0,0)[lb]{\smash{\SetFigFont{10}{12.0}{rm}$\del_{\nu}G^{-1}$}}}
\end{picture}
  \hspace{0.3cm}
  - \hspace{0.3cm} \begin{picture}(0,0)%
\epsfig{file=fig37.pstex}%
\end{picture}%
\setlength{\unitlength}{0.00083300in}%
\begingroup\makeatletter\ifx\SetFigFont\undefined
\def\x#1#2#3#4#5#6#7\relax{\def\x{#1#2#3#4#5#6}}%
\expandafter\x\fmtname xxxxxx\relax \def\y{splain}%
\ifx\x\y   
\gdef\SetFigFont#1#2#3{%
  \ifnum #1<17\tiny\else \ifnum #1<20\small\else
  \ifnum #1<24\normalsize\else \ifnum #1<29\large\else
  \ifnum #1<34\Large\else \ifnum #1<41\LARGE\else
     \huge\fi\fi\fi\fi\fi\fi
  \csname #3\endcsname}%
\else
\gdef\SetFigFont#1#2#3{\begingroup
  \count@#1\relax \ifnum 25<\count@\count@25\fi
  \def\x{\endgroup\@setsize\SetFigFont{#2pt}}%
  \expandafter\x
    \csname \romannumeral\the\count@ pt\expandafter\endcsname
    \csname @\romannumeral\the\count@ pt\endcsname
  \csname #3\endcsname}%
\fi
\fi\endgroup
\begin{picture}(1233,720)(451,-619)
\put(451,-586){\makebox(0,0)[lb]{\smash{\SetFigFont{10}{12.0}{rm}$\del_{\nu}G^{-1}$}}}
\put(1501,-586){\makebox(0,0)[lb]{\smash{\SetFigFont{10}{12.0}{rm}$\del_{\nu}\Gamma^{\mu}$}}}
\end{picture}
  \hspace{0.3cm}
  - \hspace{0.3cm} \begin{picture}(0,0)%
\epsfig{file=fig38.pstex}%
\end{picture}%
\setlength{\unitlength}{0.00083300in}%
\begingroup\makeatletter\ifx\SetFigFont\undefined
\def\x#1#2#3#4#5#6#7\relax{\def\x{#1#2#3#4#5#6}}%
\expandafter\x\fmtname xxxxxx\relax \def\y{splain}%
\ifx\x\y   
\gdef\SetFigFont#1#2#3{%
  \ifnum #1<17\tiny\else \ifnum #1<20\small\else
  \ifnum #1<24\normalsize\else \ifnum #1<29\large\else
  \ifnum #1<34\Large\else \ifnum #1<41\LARGE\else
     \huge\fi\fi\fi\fi\fi\fi
  \csname #3\endcsname}%
\else
\gdef\SetFigFont#1#2#3{\begingroup
  \count@#1\relax \ifnum 25<\count@\count@25\fi
  \def\x{\endgroup\@setsize\SetFigFont{#2pt}}%
  \expandafter\x
    \csname \romannumeral\the\count@ pt\expandafter\endcsname
    \csname @\romannumeral\the\count@ pt\endcsname
  \csname #3\endcsname}%
\fi
\fi\endgroup
\begin{picture}(1212,720)(451,-619)
\put(1501,-586){\makebox(0,0)[lb]{\smash{\SetFigFont{10}{12.0}{rm}$\del_{\nu}G^{-1}$}}}
\put(451,-586){\makebox(0,0)[lb]{\smash{\SetFigFont{10}{12.0}{rm}$\del_{\nu}\Gamma^{\mu}$}}}
\end{picture}

\end{equation}
Inserting (\ref{2.3}) into (\ref{2.2}) and replacing $g(0)$ by $g(k)$
we get
\begin{eqnarray}
\label{2.4}
\lefteqn{ \partial^2 \Gamma^{\mu}(p,q) = } \nonumber\\
 & = & g(q)\{ A_{\nu}(q_2) \partial_{\nu} \Gamma^{\mu}(p,q) +
  \partial_{\nu} \Gamma^{\mu}(p,q)\tilde{A}_{\nu}(q_1) - \nonumber\\
 & - & A_{\nu}(q_2)\Gamma^{\mu}(p,q) A_{\nu}(q_1)\} + \partial^2
 \tilde{\Gamma}^{\mu}(p,q) .
\end{eqnarray}
Here we have introced the notations
\begin{equation}
\label{2.5}
q_{1,2}=q\pm \frac{p}{2}, \quad A_{\mu}(q)=\partial_{\mu}G^{-1}(q)G(q)
\quad\mbox{and}\quad \tilde{A}_{\mu}(q)=G(q)\partial_{\mu}G^{-1}(q).
\end{equation}
The correction terms $\partial^2 \tilde{\Gamma}^{\mu}$ are defined as
a set of diagrams with exact Green's functions and vertices. They
contain first order derivatives of both the photon lines and the
positron lines, second order derivatives of the
positron lines and, in the same way as in the case of the Green's
function, corrections due to the replacement of $g(0)$ by $g(q)$.

The same equation is valid for the interaction amplitude of fermions
with the external field provided this interaction does not depend on
the relative momentum $q$ of the fermions.

The equations for the interaction with external fields are essential.
Indeed, if these equations have solutions that decrease at
large virtualities of the fermions -- i.e. solutions which do not
require driving terms -- this means the existence of bound states.

For the sake of convenience we rewrite the equation (\ref{2.4}) in
the form
\begin{equation}
\label{2.6}
 \partial^2 \phi(p,q) =
 g(q)\{ A_{\nu}(q_2) \partial_{\nu} \phi(p,q) +
  \partial_{\nu} \phi(p,q)\tilde{A}_{\nu}(q_1) -
 A_{\nu}(q_2)\phi(p,q) A_{\nu}(q_1)\}
\end{equation}
which will be understood as the equation for the bound state with a
spin which is defined by the invariant structure of the matrix $\phi$.
It is important to note that the accuracy of the equations for
the vertex $\Gamma_{\mu}(p,q)$ and for the bound states
differs from the accuracy of the equation for the Green's function.
The functions $\Gamma_{\mu}(p,q)$ and $\phi(p,q)$ depend, among others,
on the ratio $p^4/q_1^2 q_2^2$. If this parameter becomes large,
then $\Gamma_{\mu}$ and $\phi(p,q)$ contain, in general, the
so-called Sudakov logarithms $\alpha\ln\frac{p^4}{q_1^2 q_2^2}$ which
are not included in the equations (\ref{2.4}) and (\ref{2.6}).
Hence, the equations are valid only if
\begin{equation}
\label{2.7}
 \alpha\ln\left(\frac{p^4}{q_1^2 q_2^2}\right)<1
\end{equation}

\section{Equations for Green's functions in QCD}\label{III}
QCD is the theory of interacting quarks and gluons. The description
of quarks is more or less the same as in QED. Gluons, however, are
very different. Even the fact that being a spin $1$ particle the
gluon has to have a multi-component wave function is not seen
explicitly. In the usual approach it is described by a Green's
function $D_{\mu\nu}=\langle A_{\mu}(x),A_{\nu}(y) \rangle$ which
in momentum space (in Feynman gauge) always can be written in the
form
\begin{equation}
\label{3.1}
D_{\mu\nu} = \frac{\delta_{\mu\nu}}{k^2} C(k^2) .
\end{equation}
It contains only one unknown function. All spin properties of the
gluon are included in the momentum dependence of its interaction.
In order to introduce multi-component Green's functions, we have to
formulate the theory in a form in which the interaction is
momentum independent: we have to replace the usual description
of the gluons and their interactions by a Duffin-Kemmer type
formulation. In the framework of this description the gluon
Lagrangian is
\begin{equation}
\label{3.2}
{\cal L}(x) = \{\partial_{\mu}A_{\nu}-\partial_{\nu}A_{\mu} +
g[A_{\nu},A_{\mu}]\}F_{\mu\nu} + \frac{1}{2}F_{\mu\nu}F_{\mu\nu}.
\end{equation}
The potential $A_{\mu}$ and the field strength
$F_{\mu\nu}$ are independent quantities. The interaction is
momentum independent and equals $[A_{\mu},A_{\nu}]F_{\mu\nu}$.
In this formulation we have three independent Green's functions
\begin{equation}
\label{3.3}
\langle A_{\mu},A_{\nu}\rangle,\qquad \langle F_{\mu\nu},A_{\rho}\rangle
,\qquad \langle F_{\mu\nu},F_{\rho\sigma}\rangle .
\end{equation}
In order to fix the gauge in a covariant way, it is, of course,
necessary to introduce ghosts by adding a gauge-fixing term
\begin{equation}
\label{3.4}
\Delta{\cal L} = \frac{\zeta}{2}(\partial_{\mu}A_{\mu})^2 .
\end{equation}
The three Green's functions can be combined into one by introducing
the ten-component state
\begin{equation}
\label{3.5}
\Psi = {A_{\mu} \choose \frac{1}{m} F_{\mu\nu} } .
\end{equation}
We use the parameter $m$, which has the dimension of mass, to convert the
lower component of $\Psi$ into the same dimension as the upper component.
The equations for the fields corresponding to the Lagrangian (\ref{3.2}),
(\ref{3.4}) are
\[ \Delta_{\mu}F_{\mu\nu} + \zeta\partial_{\mu}\partial_{\nu}
 A_{\nu} = 0  , \]
\begin{equation}
\label{3.6}
\partial_{\mu}A_{\nu}-\partial_{\nu}A_{\mu} + g[A_{\mu},A_{\nu}]
 - F_{\mu\nu} = 0 .
\end{equation}
In terms of the state $\Psi$ we have
\begin{equation}
\label{3.7}
\beta_{\mu}(-i\partial_{\mu}+gA_{\mu})\Psi - m\gamma_{-}\Psi -
 \frac{\gamma_+}{m}\zeta(\hat{p}^2-p^2)\Psi = 0 ;
\end{equation}
\[ \hat{p}^2 = -\beta_{\mu}\beta_{\nu}\partial_{\mu}\partial_{\nu}
\quad\mbox{,}\quad p^2 = \partial_{\mu}\partial_{\mu} . \]
$\beta_{\mu}$ are Duffin-Kemmer matrices satisfying the
commutation relation
\begin{equation}
\label{3.8}
\beta_i\beta_k\beta_l + \beta_l\beta_k\beta_i =
 \delta_{ik}\beta_l + \delta_{kl}\beta_i ;
\end{equation}
they are connecting the upper $\Psi^{\rho}$ and lower
$\Psi_{\sigma\gamma}$ components of $\Psi$ and have a simple
representation
\begin{equation}
\label{3.8'}
(\beta^{\mu})^{\rho}_{\sigma\gamma} = \frac{1}{\sqrt{2}}
 (\delta_{\mu\sigma}\delta_{\rho\gamma} - \delta_{\mu\gamma}
 \delta_{\rho\sigma}).
\end{equation}
The quantities $\gamma_{\pm}$ are projector operators for the
upper and lower components of $\Psi$ .

If we want to preserve the dimension of the
$\langle A_{\mu},A_{\nu} \rangle$ component of the free Green's
function, $D$ has to satisfy the equation
\begin{equation}
\label{3.9}
[\frac{\hat{p}}{m} -\gamma_{-} - \frac{\gamma_{+}}{m^2}\zeta
 (\hat{p}^2-p^2)]D =\frac{1}{m^2}.
\end{equation}
The solution of this equation is
\begin{equation}
\label{3.10}
D = \frac{1}{p^2}\left[ \frac{\hat{p}}{m} + C_1 \gamma_{+} +
 \frac{\gamma_{-}}{m^2}(\hat{p}^2-p^2)\right] .
\end{equation}
In Feynman gauge $\zeta = 1$ and $C_1=1$, in Landau gauge $C_1=
\frac{\hat{p}^2}{p^2}$. The three terms in (\ref{3.10}) correspond
to the three independent Green's functions (\ref{3.3}).

The vertex for the interaction of three gluons with momenta $p_1,
p_2,p_3$, colours $a,b,c$ and Duffin-Kemmer indices $\alpha,\beta,
\gamma$ has the structure
\[ \Gamma_{a,\alpha,p_1;b,\beta,p_2;c,\gamma,p_3} = \hspace{8cm}  \]
\begin{equation}
\label{3.11}
 = \begin{picture}(0,0)%
\epsfig{file=fig39.pstex}%
\end{picture}%
\setlength{\unitlength}{0.00083300in}%
\begingroup\makeatletter\ifx\SetFigFont\undefined
\def\x#1#2#3#4#5#6#7\relax{\def\x{#1#2#3#4#5#6}}%
\expandafter\x\fmtname xxxxxx\relax \def\y{splain}%
\ifx\x\y   
\gdef\SetFigFont#1#2#3{%
  \ifnum #1<17\tiny\else \ifnum #1<20\small\else
  \ifnum #1<24\normalsize\else \ifnum #1<29\large\else
  \ifnum #1<34\Large\else \ifnum #1<41\LARGE\else
     \huge\fi\fi\fi\fi\fi\fi
  \csname #3\endcsname}%
\else
\gdef\SetFigFont#1#2#3{\begingroup
  \count@#1\relax \ifnum 25<\count@\count@25\fi
  \def\x{\endgroup\@setsize\SetFigFont{#2pt}}%
  \expandafter\x
    \csname \romannumeral\the\count@ pt\expandafter\endcsname
    \csname @\romannumeral\the\count@ pt\endcsname
  \csname #3\endcsname}%
\fi
\fi\endgroup
\begin{picture}(1287,669)(589,-673)
\put(1876,-586){\makebox(0,0)[lb]{\smash{\SetFigFont{10}{12.0}{rm}$p_3$}}}
\put(676,-511){\makebox(0,0)[lb]{\smash{\SetFigFont{10}{12.0}{rm}$p_1$}}}
\put(1876,-136){\makebox(0,0)[lb]{\smash{\SetFigFont{10}{12.0}{rm}$p_2$}}}
\end{picture}
 \,\,\,+\,\,\,  \begin{picture}(0,0)%
\epsfig{file=fig40.pstex}%
\end{picture}%
\setlength{\unitlength}{0.00083300in}%
\begingroup\makeatletter\ifx\SetFigFont\undefined
\def\x#1#2#3#4#5#6#7\relax{\def\x{#1#2#3#4#5#6}}%
\expandafter\x\fmtname xxxxxx\relax \def\y{splain}%
\ifx\x\y   
\gdef\SetFigFont#1#2#3{%
  \ifnum #1<17\tiny\else \ifnum #1<20\small\else
  \ifnum #1<24\normalsize\else \ifnum #1<29\large\else
  \ifnum #1<34\Large\else \ifnum #1<41\LARGE\else
     \huge\fi\fi\fi\fi\fi\fi
  \csname #3\endcsname}%
\else
\gdef\SetFigFont#1#2#3{\begingroup
  \count@#1\relax \ifnum 25<\count@\count@25\fi
  \def\x{\endgroup\@setsize\SetFigFont{#2pt}}%
  \expandafter\x
    \csname \romannumeral\the\count@ pt\expandafter\endcsname
    \csname @\romannumeral\the\count@ pt\endcsname
  \csname #3\endcsname}%
\fi
\fi\endgroup
\begin{picture}(1287,669)(589,-673)
\put(1876,-586){\makebox(0,0)[lb]{\smash{\SetFigFont{10}{12.0}{rm}$p_3$}}}
\put(676,-511){\makebox(0,0)[lb]{\smash{\SetFigFont{10}{12.0}{rm}$p_1$}}}
\put(1876,-136){\makebox(0,0)[lb]{\smash{\SetFigFont{10}{12.0}{rm}$p_2$}}}
\end{picture}
 
   \,\,\, + \,\,\, \begin{picture}(0,0)%
\epsfig{file=fig41.pstex}%
\end{picture}%
\setlength{\unitlength}{0.00083300in}%
\begingroup\makeatletter\ifx\SetFigFont\undefined
\def\x#1#2#3#4#5#6#7\relax{\def\x{#1#2#3#4#5#6}}%
\expandafter\x\fmtname xxxxxx\relax \def\y{splain}%
\ifx\x\y   
\gdef\SetFigFont#1#2#3{%
  \ifnum #1<17\tiny\else \ifnum #1<20\small\else
  \ifnum #1<24\normalsize\else \ifnum #1<29\large\else
  \ifnum #1<34\Large\else \ifnum #1<41\LARGE\else
     \huge\fi\fi\fi\fi\fi\fi
  \csname #3\endcsname}%
\else
\gdef\SetFigFont#1#2#3{\begingroup
  \count@#1\relax \ifnum 25<\count@\count@25\fi
  \def\x{\endgroup\@setsize\SetFigFont{#2pt}}%
  \expandafter\x
    \csname \romannumeral\the\count@ pt\expandafter\endcsname
    \csname @\romannumeral\the\count@ pt\endcsname
  \csname #3\endcsname}%
\fi
\fi\endgroup
\begin{picture}(1287,669)(589,-673)
\put(1876,-586){\makebox(0,0)[lb]{\smash{\SetFigFont{10}{12.0}{rm}$p_3$}}}
\put(676,-511){\makebox(0,0)[lb]{\smash{\SetFigFont{10}{12.0}{rm}$p_1$}}}
\put(1876,-136){\makebox(0,0)[lb]{\smash{\SetFigFont{10}{12.0}{rm}$p_2$}}}
\end{picture}

\end{equation}
\vspace{.3cm}
\[ \begin{picture}(0,0)%
\epsfig{file=fig42.pstex}%
\end{picture}%
\setlength{\unitlength}{0.00083300in}%
\begingroup\makeatletter\ifx\SetFigFont\undefined
\def\x#1#2#3#4#5#6#7\relax{\def\x{#1#2#3#4#5#6}}%
\expandafter\x\fmtname xxxxxx\relax \def\y{splain}%
\ifx\x\y   
\gdef\SetFigFont#1#2#3{%
  \ifnum #1<17\tiny\else \ifnum #1<20\small\else
  \ifnum #1<24\normalsize\else \ifnum #1<29\large\else
  \ifnum #1<34\Large\else \ifnum #1<41\LARGE\else
     \huge\fi\fi\fi\fi\fi\fi
  \csname #3\endcsname}%
\else
\gdef\SetFigFont#1#2#3{\begingroup
  \count@#1\relax \ifnum 25<\count@\count@25\fi
  \def\x{\endgroup\@setsize\SetFigFont{#2pt}}%
  \expandafter\x
    \csname \romannumeral\the\count@ pt\expandafter\endcsname
    \csname @\romannumeral\the\count@ pt\endcsname
  \csname #3\endcsname}%
\fi
\fi\endgroup
\begin{picture}(624,567)(589,-541)
\put(976,-211){\makebox(0,0)[lb]{\smash{\SetFigFont{10}{12.0}{rm}$\mu$}}}
\put(1051,-511){\makebox(0,0)[lb]{\smash{\SetFigFont{10}{12.0}{rm}$\rho \sigma$}}}
\put(601,-511){\makebox(0,0)[lb]{\smash{\SetFigFont{10}{12.0}{rm}$\nu$}}}
\end{picture}
 \,\,\, +  \,\,\, \begin{picture}(0,0)%
\epsfig{file=fig43.pstex}%
\end{picture}%
\setlength{\unitlength}{0.00083300in}%
\begingroup\makeatletter\ifx\SetFigFont\undefined
\def\x#1#2#3#4#5#6#7\relax{\def\x{#1#2#3#4#5#6}}%
\expandafter\x\fmtname xxxxxx\relax \def\y{splain}%
\ifx\x\y   
\gdef\SetFigFont#1#2#3{%
  \ifnum #1<17\tiny\else \ifnum #1<20\small\else
  \ifnum #1<24\normalsize\else \ifnum #1<29\large\else
  \ifnum #1<34\Large\else \ifnum #1<41\LARGE\else
     \huge\fi\fi\fi\fi\fi\fi
  \csname #3\endcsname}%
\else
\gdef\SetFigFont#1#2#3{\begingroup
  \count@#1\relax \ifnum 25<\count@\count@25\fi
  \def\x{\endgroup\@setsize\SetFigFont{#2pt}}%
  \expandafter\x
    \csname \romannumeral\the\count@ pt\expandafter\endcsname
    \csname @\romannumeral\the\count@ pt\endcsname
  \csname #3\endcsname}%
\fi
\fi\endgroup
\begin{picture}(624,567)(589,-541)
\put(976,-211){\makebox(0,0)[lb]{\smash{\SetFigFont{10}{12.0}{rm}$\mu$}}}
\put(1051,-511){\makebox(0,0)[lb]{\smash{\SetFigFont{10}{12.0}{rm}$\nu$}}}
\put(601,-511){\makebox(0,0)[lb]{\smash{\SetFigFont{10}{12.0}{rm}$\rho \sigma$}}}
\end{picture}
 
  = \beta^{\mu} .\]
The coupling constant remains in our notation $g$.

Let us consider in this approach the properties of the exact Green's
function
\begin{equation}
\label{3.12}
D^{-1} = \hat{k} - \gamma_- -\zeta(\hat{k}^2-k^2) - \Sigma - \delta\zeta
(\hat{k}^2-k^2) ,
\end{equation}
\[ \hat{k}=\frac{\hat{p}}{m} , \]
where $\Sigma$ is defined digrammatically:
\begin{equation}
\label{3.13}
\Sigma = \,\, \begin{picture}(0,0)%
\epsfig{file=fig44.pstex}%
\end{picture}%
\setlength{\unitlength}{0.00083300in}%
\begingroup\makeatletter\ifx\SetFigFont\undefined
\def\x#1#2#3#4#5#6#7\relax{\def\x{#1#2#3#4#5#6}}%
\expandafter\x\fmtname xxxxxx\relax \def\y{splain}%
\ifx\x\y   
\gdef\SetFigFont#1#2#3{%
  \ifnum #1<17\tiny\else \ifnum #1<20\small\else
  \ifnum #1<24\normalsize\else \ifnum #1<29\large\else
  \ifnum #1<34\Large\else \ifnum #1<41\LARGE\else
     \huge\fi\fi\fi\fi\fi\fi
  \csname #3\endcsname}%
\else
\gdef\SetFigFont#1#2#3{\begingroup
  \count@#1\relax \ifnum 25<\count@\count@25\fi
  \def\x{\endgroup\@setsize\SetFigFont{#2pt}}%
  \expandafter\x
    \csname \romannumeral\the\count@ pt\expandafter\endcsname
    \csname @\romannumeral\the\count@ pt\endcsname
  \csname #3\endcsname}%
\fi
\fi\endgroup
\begin{picture}(806,468)(873,-295)
\end{picture}
 \,\, + 
  \,\, \begin{picture}(0,0)%
\epsfig{file=fig45.pstex}%
\end{picture}%
\setlength{\unitlength}{0.00083300in}%
\begingroup\makeatletter\ifx\SetFigFont\undefined
\def\x#1#2#3#4#5#6#7\relax{\def\x{#1#2#3#4#5#6}}%
\expandafter\x\fmtname xxxxxx\relax \def\y{splain}%
\ifx\x\y   
\gdef\SetFigFont#1#2#3{%
  \ifnum #1<17\tiny\else \ifnum #1<20\small\else
  \ifnum #1<24\normalsize\else \ifnum #1<29\large\else
  \ifnum #1<34\Large\else \ifnum #1<41\LARGE\else
     \huge\fi\fi\fi\fi\fi\fi
  \csname #3\endcsname}%
\else
\gdef\SetFigFont#1#2#3{\begingroup
  \count@#1\relax \ifnum 25<\count@\count@25\fi
  \def\x{\endgroup\@setsize\SetFigFont{#2pt}}%
  \expandafter\x
    \csname \romannumeral\the\count@ pt\expandafter\endcsname
    \csname @\romannumeral\the\count@ pt\endcsname
  \csname #3\endcsname}%
\fi
\fi\endgroup
\begin{picture}(806,468)(873,-295)
\end{picture}
 \,\, + \cdots
\end{equation}
It contains four matrix elements $\Sigma_{++}$, $\Sigma_{-+}$,
$\Sigma_{+-}$, $\Sigma_{--}$. It can be also represented in the form
\begin{equation}
\label{3.14}
\Sigma = \hat{k}\Sigma_1 + \gamma_-\Sigma_2 + \gamma_+\hat{k}^2\Sigma_3 .
\end{equation}
The factor $\hat{k}^2$ in the third term is necessary to preserve the current
conservation; in first order, the ghost contributes only to $\Sigma_3$.

We added the term $\delta\zeta$ in (\ref{3.12}) to be able to fix the
gauge for the exact Green's function. Instead of (\ref{3.12}) we can
write
\begin{equation}
\label{3.15}
D^{-1} = Z_1^{-1}\hat{k} - Z_2^{-1}\gamma_- - Z_3^{-1}\hat{k}^2 -
\zeta(\hat{k}^2-k^2) .
\end{equation}
The Green's function then will be equal
\begin{equation}
\label{3.16}
D = \frac{1}{(Z_1^{-2} - Z_2^{-1}Z_3^{-1})p^2}\{Z_1^{-1}\hat{k} +
 Z_2^{-1}C\gamma_+ + Z_2 Z_1^{-2}\hat{k}\gamma_-^2 \} 
 + \frac{Z_2 \gamma_-}{m^2}
\end{equation}
where
\begin{equation}
\label{3.17}
C = 1 + \left(\frac{\hat{k}^2}{k^2}-1 \right)\left[\frac{Z_2 Z_1^{-2}
 -Z_3^{-1}}{\zeta} - 1 \right] .
\end{equation}
In Feynman gauge
\begin{equation}
\label{3.18}
\zeta = Z_2 Z_1^{-2} - Z_3^{-1} .
\end{equation}
In order to understand the meaning of the denominator in (\ref{3.16}),
let us consider the renormalization properties of simple diagrams, for
example $\Sigma_{+-} + \Sigma_{-+}$.
\begin{equation}
\label{3.19}
\Sigma_{+-} + \Sigma_{-+} = g_0^2 \Gamma_{+-+} \begin{picture}(0,0)%
\epsfig{file=fig46.pstex}%
\end{picture}%
\setlength{\unitlength}{0.00083300in}%
\begingroup\makeatletter\ifx\SetFigFont\undefined
\def\x#1#2#3#4#5#6#7\relax{\def\x{#1#2#3#4#5#6}}%
\expandafter\x\fmtname xxxxxx\relax \def\y{splain}%
\ifx\x\y   
\gdef\SetFigFont#1#2#3{%
  \ifnum #1<17\tiny\else \ifnum #1<20\small\else
  \ifnum #1<24\normalsize\else \ifnum #1<29\large\else
  \ifnum #1<34\Large\else \ifnum #1<41\LARGE\else
     \huge\fi\fi\fi\fi\fi\fi
  \csname #3\endcsname}%
\else
\gdef\SetFigFont#1#2#3{\begingroup
  \count@#1\relax \ifnum 25<\count@\count@25\fi
  \def\x{\endgroup\@setsize\SetFigFont{#2pt}}%
  \expandafter\x
    \csname \romannumeral\the\count@ pt\expandafter\endcsname
    \csname @\romannumeral\the\count@ pt\endcsname
  \csname #3\endcsname}%
\fi
\fi\endgroup
\begin{picture}(806,690)(873,-394)
\put(1201,-361){\makebox(0,0)[lb]{\smash{\SetFigFont{10}{12.0}{rm}$D_{-+}$}}}
\put(1201,164){\makebox(0,0)[lb]{\smash{\SetFigFont{10}{12.0}{rm}$D_{++}$}}}
\end{picture}

 \Gamma_{++-} + (+\rightarrow - )
\end{equation}
where $g_0 m$ is the effective bare coupling;
\begin{equation}
\label{3.20}
 \Gamma_{+-+} \sim Z_1^{-1} .
\end{equation}
If this is true, we have in Feynman gauge
\begin{equation}
\label{3.21}
\Sigma_{+-} + \Sigma_{-+} \sim \frac{Z_1^{-3}Z_2^{-1}}{(Z_1^{-2}-Z_2^{-1}
Z_3^{-1})^2} .
\end{equation}
However, $p_{\mu}\partial_{\mu}\Sigma_{--}$ has to be proportional to
$\alpha Z_1^{-1}$. This means that we have to expect
\begin{equation}
\label{3.22}
\frac{Z_1^{-3}Z_2^{-1}}{(Z_1^{-2}-Z_2^{-1}Z_3^{-1})^2} \equiv
 \frac{\alpha(p)}{\alpha_0} Z_1^{-1}
\end{equation}
i.e.
\[ Z_2 Z_1^{-2} - Z_3^{-1} = \sqrt{\frac{\alpha_0}{\alpha}Z_1^{-2}Z_2}.\]
This is our definition of $\alpha(p)$. It has all
known properties of the renormalized coupling $\alpha$.
As a result, $D$ can be written in the form
\begin{equation}
\label{3.23}
D = \sqrt{\frac{\alpha(p^2)}{\alpha_0}Z_1^2 Z_2^{1}}\{Z_2 Z_1^{-1}
 \hat{k} + C\gamma_+ + (Z_2 Z_1^{-1}\hat{k})^2\gamma_-\}\frac{1}{p^2}
 + \frac{Z_2\gamma_-}{m^2}
\end{equation}
In this approach $\alpha_0$ is not a quantity coming from the
normalization: it is the bare coupling. The theory has to be defined
as the limit $\alpha_0\rightarrow 0$. In this context the expression
(\ref{3.23}) has a very interesting property. In the limit $\alpha_0
\rightarrow 0$ the Green's function $D$ contains only $Z_1^{-1}$ and
$Z_2^{-1}$. According to (\ref{3.22}), in this limit $Z_3^{-1} = Z_1
^{-2}Z_2$. Because of this, when we regard the equations for $Z_1^{-1}$
and $Z_2^{-1}$ in the way we did for the fermionic Green's function
in QED, $\alpha_0$ has to disappear. Consequently, we have equations
for $Z_1^{-1}$ and $Z_2^{-1}$ with $\alpha(p^2)$ being arbitrary. The
equation for $Z_3^{-1}$ will not help since, due to the equality
$Z_3=Z_1^{-1}Z_2$, it has to be an identity. We will see that an
equation for $\alpha$ appears when we will consider the correction of
the order of $\sqrt{\alpha_0}$.

Before formulating the equation for the Green's function, let us
see what the Ward identity looks like in this formulation. Consider
the relation between $(p_2-p_1)_{\mu}\Gamma^{\mu}(p_2,p_1)$ and the
Green's function for the bare vertex $\Gamma^{\mu} = \hat{f}\beta^{\mu}$:
\begin{eqnarray}
\label{3.24}
 \hat{p}_{\mu}\beta_{\mu} = \hat{p}_2-\hat{p}_1 & = & m[\hat{k}_2 -
 \gamma_- -(\hat{k}_1-\gamma_-)] = \\
 = m[D^{-1}(k_2) - D^{-1}(k_1)] & + & m\zeta_0[(\hat{k}_2^2-k_2^2) -
(\hat{k}_1^2-k_1^2)]
\end{eqnarray}
At $p\rightarrow 0$
\begin{equation}
\label{3.25}
\Gamma_{\mu}^0|_{p=0} = [m\partial_{\mu}D^{-1} + m\zeta_0\partial_{\mu}
(\hat{k}^2-k^2)]\hat{f}
\end{equation}
which is, of course, the usual complication due to the Slavnov-Taylor
Ward identity. But in this formalism the additional term is equal
\[- m\partial_{\mu} \left(\zeta_0\frac{\partial}{\partial\zeta_0}D_0^{-1}\right).\]
Introducing this vertex in an arbitrary diagram we obtain the relation
\begin{equation}
\label{3.26}
\Gamma_{\mu} = m[\partial_{\mu}D^{-1}-\zeta_0\frac{\partial}{\partial
\zeta_0}\partial_{\mu}D^{-1}] .
\end{equation}
According to (\ref{3.15}), (\ref{3.18}), in Feynman gauge we can write
\begin{equation}
\label{3.27}
\Gamma_{\mu} = fm\partial_{\mu}[(Z_1^{-1}-\partial_{\zeta}Z_1^{-1})
\hat{k} - \gamma_-(Z_2^{-1}-\partial_{\zeta}Z_2^{-1}) -
\gamma_+\hat{k}^2(Z_1^{-2}Z_2 - \partial_{\zeta}(Z_1^{-2}Z_2))].
\end{equation}
The quantities $\partial_{\zeta}Z_1^{-1}$, $\partial_{\zeta}Z_2^{-1}$
must be calculated from the equations for $Z_1^{-1}$ and $Z_2^{-1}$.
But if the equations are formulated in terms of the exact Green's
functions, the dependence on $\zeta$ enters only through these Green's
functions $D$ which, according to (\ref{3.16}), (\ref{3.17}), include
$\zeta$ only through the quantity $C$. In the limit $\alpha_{0}
\rightarrow 0$, however, $C$ does not depend on $\zeta$. Consequently,
$\partial_{\zeta}Z_1^{-1}$ and $\partial_{\zeta}Z_2^{-1}$ are equal to
zero and we have the following simple relation for the vertex at zero
momentum:
\begin{equation}
\label{3.28}
\Gamma_{\mu} = \hat{f}m\partial_{\mu}\tilde{D}^{-1} .
\end{equation}
$\tilde{D}^{-1}$ is defined by (\ref{3.27}); it does not contain
gauge-fixing terms.
\begin{equation}
\label{3.29}
\tilde{D}^{-1} = Z_1^{-1}\hat{k} - Z_2^{-1}\gamma_- - \gamma_+Z_1^{-2}
Z_2\hat{k}^2 .
\end{equation}
Let us first consider the equations for $Z_1^{-1}$ and $Z_2^{-1}$ in the
same way as we did for fermions. As in the case of QED, we will use
the Feynman gauge in order to simplify the one-gluon contribution to the
equation for the Green's function.
\begin{eqnarray}
\label{3.30}
\lefteqn{\partial^2(Z_1^{-1}\hat{k}-Z_2^{-1}\gamma_-) = }\nonumber\\
 & = & \partial^2 \Big\{ \,\,\, \begin{picture}(0,0)%
\epsfig{file=fig47.pstex}%
\end{picture}%
\setlength{\unitlength}{0.00083300in}%
\begingroup\makeatletter\ifx\SetFigFont\undefined
\def\x#1#2#3#4#5#6#7\relax{\def\x{#1#2#3#4#5#6}}%
\expandafter\x\fmtname xxxxxx\relax \def\y{splain}%
\ifx\x\y   
\gdef\SetFigFont#1#2#3{%
  \ifnum #1<17\tiny\else \ifnum #1<20\small\else
  \ifnum #1<24\normalsize\else \ifnum #1<29\large\else
  \ifnum #1<34\Large\else \ifnum #1<41\LARGE\else
     \huge\fi\fi\fi\fi\fi\fi
  \csname #3\endcsname}%
\else
\gdef\SetFigFont#1#2#3{\begingroup
  \count@#1\relax \ifnum 25<\count@\count@25\fi
  \def\x{\endgroup\@setsize\SetFigFont{#2pt}}%
  \expandafter\x
    \csname \romannumeral\the\count@ pt\expandafter\endcsname
    \csname @\romannumeral\the\count@ pt\endcsname
  \csname #3\endcsname}%
\fi
\fi\endgroup
\begin{picture}(987,555)(751,-304)
\put(976,-271){\makebox(0,0)[lb]{\smash{\SetFigFont{10}{12.0}{rm}$p_1$}}}
\put(751,-136){\makebox(0,0)[lb]{\smash{\SetFigFont{10}{12.0}{rm}$p$}}}
\put(751,119){\makebox(0,0)[lb]{\smash{\SetFigFont{10}{12.0}{rm}$p\!-\!p_1$}}}
\end{picture}
 \, 
       + \,\,\, \begin{picture}(0,0)%
\epsfig{file=fig48.pstex}%
\end{picture}%
\setlength{\unitlength}{0.00083300in}%
\begingroup\makeatletter\ifx\SetFigFont\undefined
\def\x#1#2#3#4#5#6#7\relax{\def\x{#1#2#3#4#5#6}}%
\expandafter\x\fmtname xxxxxx\relax \def\y{splain}%
\ifx\x\y   
\gdef\SetFigFont#1#2#3{%
  \ifnum #1<17\tiny\else \ifnum #1<20\small\else
  \ifnum #1<24\normalsize\else \ifnum #1<29\large\else
  \ifnum #1<34\Large\else \ifnum #1<41\LARGE\else
     \huge\fi\fi\fi\fi\fi\fi
  \csname #3\endcsname}%
\else
\gdef\SetFigFont#1#2#3{\begingroup
  \count@#1\relax \ifnum 25<\count@\count@25\fi
  \def\x{\endgroup\@setsize\SetFigFont{#2pt}}%
  \expandafter\x
    \csname \romannumeral\the\count@ pt\expandafter\endcsname
    \csname @\romannumeral\the\count@ pt\endcsname
  \csname #3\endcsname}%
\fi
\fi\endgroup
\begin{picture}(987,555)(751,-304)
\put(751,-136){\makebox(0,0)[lb]{\smash{\SetFigFont{10}{12.0}{rm}$p$}}}
\put(1448,119){\makebox(0,0)[lb]{\smash{\SetFigFont{10}{12.0}{rm}$p\!-\!p_2$}}}
\put(751,119){\makebox(0,0)[lb]{\smash{\SetFigFont{10}{12.0}{rm}$p\!-\!p_1$}}}
\put(976,-271){\makebox(0,0)[lb]{\smash{\SetFigFont{10}{12.0}{rm}$p_1$}}}
\put(1445,-271){\makebox(0,0)[lb]{\smash{\SetFigFont{10}{12.0}{rm}$p_2$}}}
\end{picture}
 \, + \cdots \Big\}
\end{eqnarray}
Each line here contains the exact propagator (\ref{3.23}).
Differentiating the propagator along the upper line, we obtain the
contribution corresponding to the second derivative of one line
\begin{equation}
\label{3.31}
\partial^2 D = - 4\pi^2 i \delta^4(p) \gamma_+ Z^{\frac{1}{2}}(0) -
\frac{4p_{\mu}}{p^4}\partial_{\mu}NZ^{\frac{1}{2}} + \frac{1}{p^2}
\partial^2 NZ^{\frac{1}{2}} + \partial^2\frac{Z_2\gamma_-}{m^2}
\end{equation}
where
\begin{equation}
\label{3.32}
Z = Z_1^2 Z_2^{-1}\frac{\alpha(0)}{\alpha_0}
\end{equation}
and we introduce the notation
\begin{equation}
\label{3.33}
D = \frac{Z^{\frac{1}{2}}}{p^2}N + \frac{Z_2\gamma_-}{m^2}.
\end{equation}
The contribution of this derivative to the equation will have the form
\begin{equation}
\label{3.34}
- 4\pi^2 i Z^{\frac{1}{2}}M(p,p) + M_1 .
\end{equation}
Here $M(p,p)$ is the gluon-gluon scattering amplitude at zero momentum
of one of the gluons. The second term $M_1$ is defined diagrammatically:
\[ M_1 =  \begin{picture}(0,0)%
\epsfig{file=fig49.pstex}%
\end{picture}%
\setlength{\unitlength}{0.00083300in}%
\begingroup\makeatletter\ifx\SetFigFont\undefined
\def\x#1#2#3#4#5#6#7\relax{\def\x{#1#2#3#4#5#6}}%
\expandafter\x\fmtname xxxxxx\relax \def\y{splain}%
\ifx\x\y   
\gdef\SetFigFont#1#2#3{%
  \ifnum #1<17\tiny\else \ifnum #1<20\small\else
  \ifnum #1<24\normalsize\else \ifnum #1<29\large\else
  \ifnum #1<34\Large\else \ifnum #1<41\LARGE\else
     \huge\fi\fi\fi\fi\fi\fi
  \csname #3\endcsname}%
\else
\gdef\SetFigFont#1#2#3{\begingroup
  \count@#1\relax \ifnum 25<\count@\count@25\fi
  \def\x{\endgroup\@setsize\SetFigFont{#2pt}}%
  \expandafter\x
    \csname \romannumeral\the\count@ pt\expandafter\endcsname
    \csname @\romannumeral\the\count@ pt\endcsname
  \csname #3\endcsname}%
\fi
\fi\endgroup
\begin{picture}(924,399)(589,-373)
\end{picture}
 \,\, . \]
Taking the first derivative of two different lines, we get the
contribution
\begin{equation}
\label{3.35}
M_2 = \begin{picture}(0,0)%
\epsfig{file=fig50.pstex}%
\end{picture}%
\setlength{\unitlength}{0.00083300in}%
\begingroup\makeatletter\ifx\SetFigFont\undefined
\def\x#1#2#3#4#5#6#7\relax{\def\x{#1#2#3#4#5#6}}%
\expandafter\x\fmtname xxxxxx\relax \def\y{splain}%
\ifx\x\y   
\gdef\SetFigFont#1#2#3{%
  \ifnum #1<17\tiny\else \ifnum #1<20\small\else
  \ifnum #1<24\normalsize\else \ifnum #1<29\large\else
  \ifnum #1<34\Large\else \ifnum #1<41\LARGE\else
     \huge\fi\fi\fi\fi\fi\fi
  \csname #3\endcsname}%
\else
\gdef\SetFigFont#1#2#3{\begingroup
  \count@#1\relax \ifnum 25<\count@\count@25\fi
  \def\x{\endgroup\@setsize\SetFigFont{#2pt}}%
  \expandafter\x
    \csname \romannumeral\the\count@ pt\expandafter\endcsname
    \csname @\romannumeral\the\count@ pt\endcsname
  \csname #3\endcsname}%
\fi
\fi\endgroup
\begin{picture}(1074,346)(739,-220)
\end{picture}
 + \begin{picture}(0,0)%
\epsfig{file=fig51.pstex}%
\end{picture}%
\setlength{\unitlength}{0.00083300in}%
\begingroup\makeatletter\ifx\SetFigFont\undefined
\def\x#1#2#3#4#5#6#7\relax{\def\x{#1#2#3#4#5#6}}%
\expandafter\x\fmtname xxxxxx\relax \def\y{splain}%
\ifx\x\y   
\gdef\SetFigFont#1#2#3{%
  \ifnum #1<17\tiny\else \ifnum #1<20\small\else
  \ifnum #1<24\normalsize\else \ifnum #1<29\large\else
  \ifnum #1<34\Large\else \ifnum #1<41\LARGE\else
     \huge\fi\fi\fi\fi\fi\fi
  \csname #3\endcsname}%
\else
\gdef\SetFigFont#1#2#3{\begingroup
  \count@#1\relax \ifnum 25<\count@\count@25\fi
  \def\x{\endgroup\@setsize\SetFigFont{#2pt}}%
  \expandafter\x
    \csname \romannumeral\the\count@ pt\expandafter\endcsname
    \csname @\romannumeral\the\count@ pt\endcsname
  \csname #3\endcsname}%
\fi
\fi\endgroup
\begin{picture}(1074,349)(739,-223)
\end{picture}
  + \cdots.
\end{equation}
As a result we have
\begin{equation}
\label{3.36}
\partial^2(Z_1^{-1}\hat{k}-Z_2^{-1}\gamma_-) = \frac{\alpha_0}{\pi}
Z^{\frac{1}{2}}M(p,p) + M_1 + M_2 .
\end{equation}
We have to remember that on the right-hand side we have to take only
the matrix elements $\langle -|+\rangle$, $\langle +|-\rangle$ and
$\langle -|-\rangle$. Writing
\begin{equation}
\label{3.37}
M(p,p) = f^2\Gamma^{\mu}D\Gamma^{\mu} + \tilde{M}(p,p)
\end{equation}
and using the Ward identity (\ref{3.28}), we obtain an equation of the
same structure as for fermions:
\begin{eqnarray}
\label{3.38}
\lefteqn{\partial^2(Z_1^{-1}\hat{k} - Z_2^{-1}\gamma_-) = }\nonumber\\
 & = & f^2\frac{\alpha_0}{\pi} \frac{1}{k^2}Z^{\frac{1}{2}}(0)
 Z^{\frac{1}{2}}(p)\partial_{\mu}\tilde{D}^{-1}\left(N + \frac{Z_2
 Z^{-\frac{1}{2}}}{m^2}\gamma_-\right)\partial_{\mu}\tilde{D}^{-1}
 + L ;
\end{eqnarray}
\[ L=\tilde{M} + M_1 + M_2 \quad\mbox{,}\quad f^2 = n_c = 3 . \]
If we now make the same trick as before: replace $Z(0)$ by $Z(p)$
and redefine $L$, we will have in the limit $\alpha_0\rightarrow 0$
\begin{equation}
\label{3.39}
\partial^2 Z_1^{-1}[\gamma_-,\hat{k}] = 3\frac{\alpha(p)}{\pi}
\frac{1}{k^2}[\gamma_-, \partial_{\mu}\tilde{D}^{-1}(Z_1\hat{k} +
Z_1^2 Z_2^{-1}\gamma_+ + Z_2\hat{k}^2\gamma_-)\partial_{\mu}
\tilde{D}^{-1}] + [\gamma_-,L'] ,
\end{equation}
\begin{equation}
\label{3.40}
\partial^2 Z_2^{-1}\gamma_- = 3\frac{\alpha(p)}{\pi}\frac{1}{k^2}
\gamma_- \partial_{\mu}\tilde{D}^{-1}(Z_1\hat{k} + Z_1^2 Z_2^{-1}\gamma_+
+ Z_2\hat{k}^2\gamma_-)\partial_{\mu}\tilde{D}^{-1}\gamma_- +
\gamma_-L'\gamma_- .
\end{equation}
The equation for the fermionic Green's function in QCD will differ from
(\ref{1.21}) only by the factor $\lambda^a \lambda^a =\frac{4}{3}$
($\lambda^a$ are colour matrices). The reason is the following. The
derivation of the equation (\ref{1.21}) was based on the relation between
the fermionic Green's function and the amplitude of zero momentum photon
emission by a fermion
\begin{equation}
\label{3.41}
 \Gamma_{\mu}(q,0) = \partial_{\mu}G^{-1}(q).
\end{equation}
In the usual formulation this relation is not correct in QCD. The
simple relation (\ref{3.28}) for the amplitude of the zero momentum gluon
emission by a gluon implies, however, that the amplitude of a zero
momentum gluon emission by a quark has to be equal
\begin{equation}
\label{3.42}
  \Gamma_{\mu}^a(q,0) = \lambda^a(Z_1^{-2}Z_2)^{\frac{1}{4}}
\partial_{\mu}G^{-1}(q).
\end{equation}
Together with (\ref{3.23}) it leads to the equation (\ref{1.21}).
The equation for the colourless vertices remains also the same. The
equation for a three-gluon vertex, however, will be essentially
different. In the same way as for the fermionic case, but taking into
account the non-commutativity of the gluon coupling, we can show that
\[\Gamma_{\nu\mu\rho\sigma} =  \, _{\nu}\!(\Gamma^{\mu})_{\sigma\rho} 
  = \hat{\Gamma}^{\mu}\]
satisfies the following equation:
\begin{eqnarray}
\label{3.43}
\lefteqn{\partial^2\hat{\Gamma}^{\mu}\left(q+\frac{p}{2},p,
 q-\frac{p}{2}\right)  = }\nonumber\\
 & = & \frac{3\alpha}{\pi}\left\{A_{\nu}(p_1)\frac{\partial}{\partial
 q_{\nu}}\hat{\Gamma}^{\mu} + \frac{\partial}{\partial q_{\nu}}
\hat{\Gamma}^{\mu}\tilde{A}_{\nu}(p_3) + (A_{\nu}(p))_{\mu\mu'}\frac{
\partial}{\partial p_{\nu}}\hat{\Gamma}^{\mu'} - \right. \nonumber\\
 & - & \left. A_{\nu}(p_1) \hat{\Gamma}^{\mu}\tilde{A}_{\nu}(p_3)
  - A_{\nu}(p_1)\hat{\Gamma}^{\mu'}(\tilde{A}_{\nu}(p))_{\mu' \mu} -
(A_{\nu}(p))_{\mu\mu'}\hat{\Gamma}^{\mu'}\tilde{A}_{\nu}(p_3)\right\}
\end{eqnarray}
\[ 
A_{\nu} = \partial_{\nu}\tilde{D}^{-1}D\quad\mbox{,}\quad
\tilde{A}_{\nu} = D\partial_{\nu}\tilde{D}^{-1} \, ;\quad
 p_1 = q + \frac{p}{2} \quad , \quad p_3 = q - \frac{p}{2}
\,.
\]
The right-hand side corresponds to all possible gluon emissions from the
external lines:
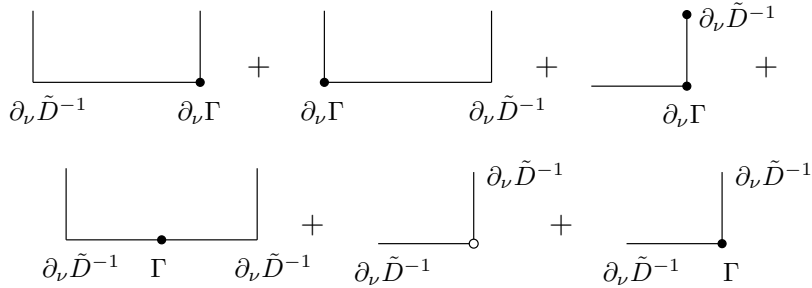
\vspace{.3cm}
\[
  \begin{picture}(0,0)%
\epsfig{file=fig52.pstex}%
\end{picture}%
\setlength{\unitlength}{0.00083300in}%
\begingroup\makeatletter\ifx\SetFigFont\undefined
\def\x#1#2#3#4#5#6#7\relax{\def\x{#1#2#3#4#5#6}}%
\expandafter\x\fmtname xxxxxx\relax \def\y{splain}%
\ifx\x\y   
\gdef\SetFigFont#1#2#3{%
  \ifnum #1<17\tiny\else \ifnum #1<20\small\else
  \ifnum #1<24\normalsize\else \ifnum #1<29\large\else
  \ifnum #1<34\Large\else \ifnum #1<41\LARGE\else
     \huge\fi\fi\fi\fi\fi\fi
  \csname #3\endcsname}%
\else
\gdef\SetFigFont#1#2#3{\begingroup
  \count@#1\relax \ifnum 25<\count@\count@25\fi
  \def\x{\endgroup\@setsize\SetFigFont{#2pt}}%
  \expandafter\x
    \csname \romannumeral\the\count@ pt\expandafter\endcsname
    \csname @\romannumeral\the\count@ pt\endcsname
  \csname #3\endcsname}%
\fi
\fi\endgroup
\begin{picture}(1233,720)(451,-619)
\put(1501,-586){\makebox(0,0)[lb]{\smash{\SetFigFont{10}{12.0}{rm}$\del_{\nu}\Gamma$}}}
\put(451,-586){\makebox(0,0)[lb]{\smash{\SetFigFont{10}{12.0}{rm}$\del_{\nu}\tilde{D}^{-1}$}}}
\end{picture}
 \hspace{.3cm} + \hspace{.3cm}
  \begin{picture}(0,0)%
\epsfig{file=fig53.pstex}%
\end{picture}%
\setlength{\unitlength}{0.00083300in}%
\begingroup\makeatletter\ifx\SetFigFont\undefined
\def\x#1#2#3#4#5#6#7\relax{\def\x{#1#2#3#4#5#6}}%
\expandafter\x\fmtname xxxxxx\relax \def\y{splain}%
\ifx\x\y   
\gdef\SetFigFont#1#2#3{%
  \ifnum #1<17\tiny\else \ifnum #1<20\small\else
  \ifnum #1<24\normalsize\else \ifnum #1<29\large\else
  \ifnum #1<34\Large\else \ifnum #1<41\LARGE\else
     \huge\fi\fi\fi\fi\fi\fi
  \csname #3\endcsname}%
\else
\gdef\SetFigFont#1#2#3{\begingroup
  \count@#1\relax \ifnum 25<\count@\count@25\fi
  \def\x{\endgroup\@setsize\SetFigFont{#2pt}}%
  \expandafter\x
    \csname \romannumeral\the\count@ pt\expandafter\endcsname
    \csname @\romannumeral\the\count@ pt\endcsname
  \csname #3\endcsname}%
\fi
\fi\endgroup
\begin{picture}(1212,720)(451,-619)
\put(1501,-586){\makebox(0,0)[lb]{\smash{\SetFigFont{10}{12.0}{rm}$\del_{\nu}\tilde{D}^{-1}$}}}
\put(451,-586){\makebox(0,0)[lb]{\smash{\SetFigFont{10}{12.0}{rm}$\del_{\nu}\Gamma$}}}
\end{picture}
 \hspace{.3cm} + \hspace{.3cm} 
  \begin{picture}(0,0)%
\epsfig{file=fig54.pstex}%
\end{picture}%
\setlength{\unitlength}{0.00083300in}%
\begingroup\makeatletter\ifx\SetFigFont\undefined
\def\x#1#2#3#4#5#6#7\relax{\def\x{#1#2#3#4#5#6}}%
\expandafter\x\fmtname xxxxxx\relax \def\y{splain}%
\ifx\x\y   
\gdef\SetFigFont#1#2#3{%
  \ifnum #1<17\tiny\else \ifnum #1<20\small\else
  \ifnum #1<24\normalsize\else \ifnum #1<29\large\else
  \ifnum #1<34\Large\else \ifnum #1<41\LARGE\else
     \huge\fi\fi\fi\fi\fi\fi
  \csname #3\endcsname}%
\else
\gdef\SetFigFont#1#2#3{\begingroup
  \count@#1\relax \ifnum 25<\count@\count@25\fi
  \def\x{\endgroup\@setsize\SetFigFont{#2pt}}%
  \expandafter\x
    \csname \romannumeral\the\count@ pt\expandafter\endcsname
    \csname @\romannumeral\the\count@ pt\endcsname
  \csname #3\endcsname}%
\fi
\fi\endgroup
\begin{picture}(687,765)(1039,-619)
\put(1501,-586){\makebox(0,0)[lb]{\smash{\SetFigFont{10}{12.0}{rm}$\del_{\nu}\Gamma$}}}
\put(1726, 14){\makebox(0,0)[lb]{\smash{\SetFigFont{10}{12.0}{rm}$\del_{\nu}\tilde{D}^{-1}$}}}
\end{picture}
 \hspace{.6cm} + 
\]
\vspace{.2cm}
\[
  \begin{picture}(0,0)%
\epsfig{file=fig55.pstex}%
\end{picture}%
\setlength{\unitlength}{0.00083300in}%
\begingroup\makeatletter\ifx\SetFigFont\undefined
\def\x#1#2#3#4#5#6#7\relax{\def\x{#1#2#3#4#5#6}}%
\expandafter\x\fmtname xxxxxx\relax \def\y{splain}%
\ifx\x\y   
\gdef\SetFigFont#1#2#3{%
  \ifnum #1<17\tiny\else \ifnum #1<20\small\else
  \ifnum #1<24\normalsize\else \ifnum #1<29\large\else
  \ifnum #1<34\Large\else \ifnum #1<41\LARGE\else
     \huge\fi\fi\fi\fi\fi\fi
  \csname #3\endcsname}%
\else
\gdef\SetFigFont#1#2#3{\begingroup
  \count@#1\relax \ifnum 25<\count@\count@25\fi
  \def\x{\endgroup\@setsize\SetFigFont{#2pt}}%
  \expandafter\x
    \csname \romannumeral\the\count@ pt\expandafter\endcsname
    \csname @\romannumeral\the\count@ pt\endcsname
  \csname #3\endcsname}%
\fi
\fi\endgroup
\begin{picture}(1362,720)(451,-619)
\put(1651,-586){\makebox(0,0)[lb]{\smash{\SetFigFont{10}{12.0}{rm}$\del_{\nu}\tilde{D}^{-1}$}}}
\put(451,-586){\makebox(0,0)[lb]{\smash{\SetFigFont{10}{12.0}{rm}$\del_{\nu}\tilde{D}^{-1}$}}}
\put(1126,-586){\makebox(0,0)[lb]{\smash{\SetFigFont{10}{12.0}{rm}$\Gamma$}}}
\end{picture}
 \hspace{.3cm} + \hspace{.3cm} 
  \begin{picture}(0,0)%
\epsfig{file=fig56.pstex}%
\end{picture}%
\setlength{\unitlength}{0.00083300in}%
\begingroup\makeatletter\ifx\SetFigFont\undefined
\def\x#1#2#3#4#5#6#7\relax{\def\x{#1#2#3#4#5#6}}%
\expandafter\x\fmtname xxxxxx\relax \def\y{splain}%
\ifx\x\y   
\gdef\SetFigFont#1#2#3{%
  \ifnum #1<17\tiny\else \ifnum #1<20\small\else
  \ifnum #1<24\normalsize\else \ifnum #1<29\large\else
  \ifnum #1<34\Large\else \ifnum #1<41\LARGE\else
     \huge\fi\fi\fi\fi\fi\fi
  \csname #3\endcsname}%
\else
\gdef\SetFigFont#1#2#3{\begingroup
  \count@#1\relax \ifnum 25<\count@\count@25\fi
  \def\x{\endgroup\@setsize\SetFigFont{#2pt}}%
  \expandafter\x
    \csname \romannumeral\the\count@ pt\expandafter\endcsname
    \csname @\romannumeral\the\count@ pt\endcsname
  \csname #3\endcsname}%
\fi
\fi\endgroup
\begin{picture}(825,765)(901,-619)
\put(1726, 14){\makebox(0,0)[lb]{\smash{\SetFigFont{10}{12.0}{rm}$\del_{\nu}\tilde{D}^{-1}$}}}
\put(901,-586){\makebox(0,0)[lb]{\smash{\SetFigFont{10}{12.0}{rm}$\del_{\nu}\tilde{D}^{-1}$}}}
\end{picture}
 \hspace{.6cm} + \hspace{.3cm} 
  \begin{picture}(0,0)%
\epsfig{file=fig57.pstex}%
\end{picture}%
\setlength{\unitlength}{0.00083300in}%
\begingroup\makeatletter\ifx\SetFigFont\undefined
\def\x#1#2#3#4#5#6#7\relax{\def\x{#1#2#3#4#5#6}}%
\expandafter\x\fmtname xxxxxx\relax \def\y{splain}%
\ifx\x\y   
\gdef\SetFigFont#1#2#3{%
  \ifnum #1<17\tiny\else \ifnum #1<20\small\else
  \ifnum #1<24\normalsize\else \ifnum #1<29\large\else
  \ifnum #1<34\Large\else \ifnum #1<41\LARGE\else
     \huge\fi\fi\fi\fi\fi\fi
  \csname #3\endcsname}%
\else
\gdef\SetFigFont#1#2#3{\begingroup
  \count@#1\relax \ifnum 25<\count@\count@25\fi
  \def\x{\endgroup\@setsize\SetFigFont{#2pt}}%
  \expandafter\x
    \csname \romannumeral\the\count@ pt\expandafter\endcsname
    \csname @\romannumeral\the\count@ pt\endcsname
  \csname #3\endcsname}%
\fi
\fi\endgroup
\begin{picture}(825,765)(901,-619)
\put(1726, 14){\makebox(0,0)[lb]{\smash{\SetFigFont{10}{12.0}{rm}$\del_{\nu}\tilde{D}^{-1}$}}}
\put(901,-586){\makebox(0,0)[lb]{\smash{\SetFigFont{10}{12.0}{rm}$\del_{\nu}\tilde{D}^{-1}$}}}
\put(1651,-586){\makebox(0,0)[lb]{\smash{\SetFigFont{10}{12.0}{rm}$\Gamma$}}}
\end{picture}

\]

\noindent
Higher order terms can be written in the same diagrammatic way. 
The third term in (\ref{3.43}) and in the corresponding diagrammatic 
expression equals zero at $\alpha = 0$ since 
$\tilde{D} \Gamma \tilde{D} = 0$ at $\alpha = 0$. 

The equation (\ref{3.43}) is, indeed, quite different from the equation
for a colourless vertex. The main difference comes from the fact that
on the right-hand side it contains derivatives of $\Gamma$ not only over
$q$ but also over $p$. Hence, in order to find $\Gamma$, we have to write
three different equations for second derivatives over three different
momenta.

A similar equation is valid for the quark-gluon vertex.

Knowing the Green's function and the vertices, all the other
amplitudes for the interactions and quarks and gluons can be
written in the usual perturbative way. These amplitudes have no
divergences and contain inside the gluon Green's function the unknown
function $\alpha(p)$. To formulate the theory in an unambiguous way,
without any references to the cutoff and the regularization, we have to
find the equation for $\alpha(p)$ as we did in QED, and learn, how to
write the higher terms more elegantly and constructively. I postpone
the investigation of this problem to another paper. What I want to
discuss now is the main difference between an asymptotically free
theory and an infrared free theory.

The equations for Green's functions of quarks and gluons are proven to
be second order integro-differential equations. To solve them, we need
boundary conditions. In an infrared theory the boundary conditions are
known: they are the conditions for the existence of free elementary
particles at small momenta. In an asymptotically free theory the
interaction is small at large momenta, and we expect to have here a
perturbative solution. However, in the region of large momenta all the
equations have two types of solutions: the perturbative solution and
solutions which decrease as some power of the momenta and therefore
contain dimensional parameters reflecting the density of different
condensates. These parameters have to be defined either by the
introduction of additional conditions of the type of conservation
laws or by the self-consistency of the solutions in the small
momentum region (or both). The next section of this paper will be
devoted, essentially, to the discussion of this problem.

\section{Spontaneous symmetry breaking in asymptotically free
theories}\label{IV}

In the present section we will consider the equation for the fermion
Green's function in QCD
\begin{equation}
\label{4.1}
\partial^2 G^{-1}(q) = g(q)\partial_{\mu}G^{-1}(q)G(q)\partial_{\mu}
 G^{-1}(q)
\end{equation}
where
\[g(q) = \frac{\alpha(q^2)}{\pi}\lambda^a\lambda^a =
  \frac{4\alpha(q^2)}{3\pi} . \]
This equation was discussed extensively [1] in connection with the
problem of quark confinement. Here we shall use the equation
for the discussion of the spontaneous breaking of chiral symmetry in
asymptotically free theories. The asymptotic freedom is reflected
in this equation by the fact that $\alpha(q^2)$ decreases when
$q^2 \rightarrow \infty$. We will assume that in the limit
$q\rightarrow 0$ $\alpha(q^2)$ approaches a finite value.
\[ \begin{picture}(0,0)%
\epsfig{file=fig58.pstex}%
\end{picture}%
\setlength{\unitlength}{0.00083300in}%
\begingroup\makeatletter\ifx\SetFigFont\undefined
\def\x#1#2#3#4#5#6#7\relax{\def\x{#1#2#3#4#5#6}}%
\expandafter\x\fmtname xxxxxx\relax \def\y{splain}%
\ifx\x\y   
\gdef\SetFigFont#1#2#3{%
  \ifnum #1<17\tiny\else \ifnum #1<20\small\else
  \ifnum #1<24\normalsize\else \ifnum #1<29\large\else
  \ifnum #1<34\Large\else \ifnum #1<41\LARGE\else
     \huge\fi\fi\fi\fi\fi\fi
  \csname #3\endcsname}%
\else
\gdef\SetFigFont#1#2#3{\begingroup
  \count@#1\relax \ifnum 25<\count@\count@25\fi
  \def\x{\endgroup\@setsize\SetFigFont{#2pt}}%
  \expandafter\x
    \csname \romannumeral\the\count@ pt\expandafter\endcsname
    \csname @\romannumeral\the\count@ pt\endcsname
  \csname #3\endcsname}%
\fi
\fi\endgroup
\begin{picture}(3232,2018)(923,-2722)
\put(1438,-824){\makebox(0,0)[lb]{\smash{\SetFigFont{11}{13.2}{rm}$g$}}}
\put(3935,-2406){\makebox(0,0)[lb]{\smash{\SetFigFont{11}{13.2}{rm}$q^2$}}}
\put(923,-1154){\makebox(0,0)[lb]{\smash{\SetFigFont{11}{13.2}{rm}$g(0)$}}}
\put(2251,-2686){\makebox(0,0)[lb]{\smash{\SetFigFont{12}{14.4}{rm}Figure 1}}}
\end{picture}
 \]
%
At $q^2 \rightarrow \infty$ the solution of the equation (\ref{4.1})
has the form
\begin{equation}
\label{4.2}
G^{-1}(q) = Z\left[(m-\hat{q}) + \frac{\nu_1^3}{q^2} + \frac{\nu_2^4
 \hat{q}}{q^4} \right] ,
\end{equation}
\[ \hat{q} = \gamma_{\mu}q_{\mu}  .\]
If $\alpha=0$, the quantities $Z$, $m$, $\nu_1$, $\nu_2$ are arbitrary
parameters. In the case when $\alpha(q^2)$ is defined by perturbation
theory, $Z,m,\nu_1,\nu_2$ are the following functions:
\begin{equation}
\label{4.3}
Z=Z_0\left(\frac{\alpha}{\alpha_0}\right)^{\gamma_Z},\quad
m=m_0\left(\frac{\alpha}{\alpha_0}\right)^{\gamma_m},\quad
\nu_{1,2}=\nu^0_{1,2}\left(\frac{\alpha}{\alpha_0}\right)^{\gamma_{1,2}}.
\end{equation}
The anomalous dimensions $\gamma_Z$, $\gamma_m$ and $\gamma_{1,2}$
can easily be found from the equation (\ref{4.1}). Generally speaking,
the solution depends on four parameters. In the limit
$q^2 \rightarrow \infty$ the chiral invariant solution can be
written as
\begin{equation}
\label{4.4}
G^{-1} = - Z\hat{q}\left(1-\frac{\nu_2^4}{q^4}\right) .
\end{equation}
The general solution (\ref{4.2}) corresponds to massive quarks. In the
solution which corresponds to spontaneously broken chiral symmetry,
$m_0=0$. In this case the mass term decreases when $q^2 \rightarrow
\infty$; the term $\nu_2$ is responsible for the violation of the
symmetry.

Multiplying (\ref{4.1}) by $G(q)G^{-1}(q)$ we obtain the equation
\begin{equation}
\label{4.5}
\partial^2 G^{-1}(q) = gA_{\mu}(q)A_{\mu}(q)G^{-1}(q)
\end{equation}
where
\begin{equation}
\label{4.6}
A_{\mu}(q) = \partial_{\mu} G^{-1}(q)G(q) .
\end{equation}
Clearly, it has a structure which corresponds to the equation for
particle propagation in the self-consistent field $gA_{\mu}A_{\mu}$.
It is easily seen that (\ref{4.1}) is equivalent to the
equation for $A_{\mu}$ of the form
\begin{equation}
\label{4.7}
\partial_{\mu}A_{\mu} = - \beta A_{\mu}A_{\mu}, \qquad \beta=1-g .
\end{equation}
The matrix $G^{-1}$ is defined by two invariant functions and can
be written as
\begin{equation}
\label{4.8}
G^{-1} = Z^{-1}(q)[m(q)-\hat{q}]\equiv \rho e^{-\hat{n}\frac{\phi}{2}}
\end{equation}
where
\[ \hat{n} = \frac{\hat{q}}{q} .\]
$A_{\mu}$ is here
\begin{equation}
\label{4.9}
A_{\mu} = \frac{\partial_{\mu}\rho}{\rho}-\frac{1}{2}\hat{n}
\partial_{\mu}\phi-\partial_{\mu}\hat{n}\sinh\frac{\phi}{2}
e^{\hat{n}\frac{\phi}{2}} .
\end{equation}
Inserting (\ref{4.8}) and (\ref{4.9}) into (\ref{4.5}) or (\ref{4.6}),
we obtain a set of non-linear equations for $\rho$ and $\phi$. We can
linearize the equation for $\rho$ by writing
\begin{equation}
\label{4.10}
\rho=\left(\frac{u}{q}\right)^{\frac{1}{\beta}}
\end{equation}
for a constant $\beta$ or
\begin{equation}
\label{4.11}
\rho = \frac{u}{q}e^{\int \frac{g}{\beta}\left(\frac{\dot{u}}{u}-1
 \right)\frac{dq}{q}}
\end{equation}
for $\beta$ which is a function of $q$. Here
\[ \dot{u}=q_{\nu}\partial_{\nu}u=\frac{\partial u}{\partial \xi},
\qquad \xi=\ln q . \]
As a result, we get for $u$ and $\phi$ the following set of equations:
\begin{equation}
\label{4.12}
\ddot{u}-u+\beta^2 \left[3\sinh^2\frac{\phi}{2} + \frac{\dot{\phi}^2}
{4}\right]u = \frac{\dot{\beta}}{\beta}(\dot{u}-u)
\end{equation}
\begin{equation}
\label{4.13}
\ddot{\phi} + 2\frac{\dot{u}}{u}\dot{\phi} - 3\sinh\phi = 0 .
\end{equation}
For a constant $\beta$ the conservation law
\[ \partial_{\xi} E = 0 \]
is fulfilled;
\begin{equation}
\label{4.14}
E = \dot{u}^2-u^2+\beta^2\left[3\sinh^2 \frac{\phi}{2} -
 \frac{\dot{\phi}^2}{4}\right] u^2 .
\end{equation}
The term $\frac{\dot{\beta}}{\beta}(\dot{u}-u)$ is of the order of
$g^2$ and it can almost always be neglected.

The asymptotic behaviour (\ref{4.2}) of the Green's function in the
limit $q^2\rightarrow \infty$ ($\beta \rightarrow 1$) corresponds to
\[ u \rightarrow Cq^2 \quad\mbox{,}\quad \phi \rightarrow i\pi .\]
The chiral invariant solution corresponds to $\phi \equiv i\pi$.
Close to the value $\phi = i\pi$ (i.e. $\phi = i\pi + \tilde{\phi}$),
at large $q^2$ we have
\begin{equation}
\label{4.15}
\frac{\dot{u}}{u} = \sqrt{1+3\beta^2}.
\end{equation}
Hence, (\ref{4.13}) is an oscillator equation with damping if
$q^2$ is growing and with acceleration if $q^2$ is decreasing:
\begin{equation}
\label{4.16}
\ddot{\tilde{\phi}}+3\tilde{\phi}=-2\sqrt{1+3\beta^2}\dot{\tilde{\phi}}.
\end{equation}
This means that the chiral invariant solution $\phi = i\pi$ is
unstable in an asymptotically free theory.

Let us consider the equation for $\phi$ at negative $q^2$ values
in detail. Due to (\ref{4.8}), $\phi=i\psi$ is in this case
purely negative and (\ref{4.13}) describes the motion of the
particle as a function of the "time" $\xi$ in a periodic field;
the damping (or the acceleration) is defined by (\ref{4.15}).
\[ \begin{picture}(0,0)%
\epsfig{file=fig59.pstex}%
\end{picture}%
\setlength{\unitlength}{0.00083300in}%
\begingroup\makeatletter\ifx\SetFigFont\undefined
\def\x#1#2#3#4#5#6#7\relax{\def\x{#1#2#3#4#5#6}}%
\expandafter\x\fmtname xxxxxx\relax \def\y{splain}%
\ifx\x\y   
\gdef\SetFigFont#1#2#3{%
  \ifnum #1<17\tiny\else \ifnum #1<20\small\else
  \ifnum #1<24\normalsize\else \ifnum #1<29\large\else
  \ifnum #1<34\Large\else \ifnum #1<41\LARGE\else
     \huge\fi\fi\fi\fi\fi\fi
  \csname #3\endcsname}%
\else
\gdef\SetFigFont#1#2#3{\begingroup
  \count@#1\relax \ifnum 25<\count@\count@25\fi
  \def\x{\endgroup\@setsize\SetFigFont{#2pt}}%
  \expandafter\x
    \csname \romannumeral\the\count@ pt\expandafter\endcsname
    \csname @\romannumeral\the\count@ pt\endcsname
  \csname #3\endcsname}%
\fi
\fi\endgroup
\begin{picture}(3429,1853)(1083,-1597)
\put(2727,-368){\makebox(0,0)[lb]{\smash{\SetFigFont{11}{13.2}{rm}$\pi$}}}
\put(3059,-678){\makebox(0,0)[lb]{\smash{\SetFigFont{11}{13.2}{rm}$2$}}}
\put(2491,-798){\makebox(0,0)[lb]{\smash{\SetFigFont{11}{13.2}{rm}$1$}}}
\put(4512,-611){\makebox(0,0)[lb]{\smash{\SetFigFont{11}{13.2}{rm}$\psi$}}}
\put(3537,-368){\makebox(0,0)[lb]{\smash{\SetFigFont{11}{13.2}{rm}$2\pi$}}}
\put(2035,124){\makebox(0,0)[lb]{\smash{\SetFigFont{11}{13.2}{rm}$\varepsilon(\psi)$}}}
\put(2551,-1561){\makebox(0,0)[lb]{\smash{\SetFigFont{12}{14.4}{rm}Figure 2}}}
\end{picture}
 \]
At $\xi \rightarrow \infty$ ($ q^2 \rightarrow \infty $) the particle is
situated at one of the minima of the potential; it accelerates as
$\xi$ decreases. The acceleration rate is defined by the parameters of
the solution (\ref{2.2}) in the limit $q^2 \rightarrow \infty$. If
$q$ is decreasing we have, generally speaking, two possible behaviours
for the solution. It goes to infinity if $\frac{\dot{u}}{u}$
remains positive all the time, or it may approach again a minimum
of the potential if $\frac{\dot{u}}{u}$ changes sign. In the latter
case it is easy to see that $G^{-1}$ has a singularity as
$q \rightarrow 0$.

There is only one possibility to avoid having a singularity in the
Euclidean region including $q=0$. We have to choose the parameters
which determine the acceleration at $q^2\rightarrow \infty$ so
that the particle appears at the maximum of the potential if
$q \rightarrow 0$ ($\xi \rightarrow -\infty$). In order to find
exceptional solutions without singularities at $q^2< 0$ it is
natural to solve the equation by fixing the solution at
$q \rightarrow 0$ ($\xi \rightarrow -\infty$). The solution of
(\ref{4.12}),(\ref{4.13}) corresponding to a maximum (e.g. $\psi=0$)
at $q\rightarrow 0$ is
\begin{equation}
\label{4.17}
i\psi = \frac{q}{m_c} \quad\mbox{,}\quad u=Z^{-1}(0)q
\end{equation}
In the case of such a solution the constant $E$ equals zero, and
\begin{equation}
\label{4.18}
\frac{\dot{u}}{u} = \sqrt{1+\beta^2\left[3\sin^2 \frac{\psi}{2} +
\frac{\dot{\psi}^2}{4}\right]}.
\end{equation}
Inserting (\ref{4.18}) into (\ref{4.13}) we obtain one non-linear
equation. It can be analysed easily for arbitrary $q^2$ values.
The solution which we are interested in contains essentially one
parameter $m_c$ which can be related to the renormalized fermion mass.
The parameter $Z^{-1}$ is irrelevant; it defines the normalization
of the Green's function at $q=0$ and can be chosen as $1$. The
solution of the equation leads to the unambiguous determination
of the asymptotic parameters of the Green's function (\ref{4.2}) by
$m_c$ and by the parameter $\lambda$ of the strong interaction which
enter $\beta(q)$.

As it was mentioned before, to the spontaneous breaking of chiral
symmetry corresponds an asymptotic behaviour of $G^{-1}$
in which $m=0$. The existence of such a solution requires a
connection between $m_c$ and $\lambda$. The renormalized fermion
mass as well as the other "condensate" parameters are then determined
by the strong interaction parameter $\lambda$.

Let us consider this solution in detail. Starting from the point
$\psi=0$ at $\xi\rightarrow -\infty$, the solution will, obviously,
reach the minimum of the potential either monotonically (if the
damping is strong enough) or in an oscillating way. In our
case the damping depends on the value of $\beta$. If $g$ is small,
$\beta$ is close to unity, the damping is strong and the solution
has a monotonic behaviour. By decreasing $\beta$ the solution may
become an oscillating one. For obtaining the value of $\beta$ at which
the oscillation begins there is no need to solve the equation at
any $q$. It will be sufficient to investigate the solution
in the region where $\psi$ becomes close to $\pi$; here
$\psi=\pi+\tilde{\psi}$ and $\tilde{\psi}$ satisfies the equation
(\ref{4.16}). The solution can be written in the form
\begin{equation}
\label{4.19}
\tilde{\psi} = e^{-p\xi}C \cos(\sqrt{2-3\beta^2}\xi + \delta),
\qquad p=\sqrt{1+3\beta^2}
\end{equation}
which oscillates if
\begin{equation}
\label{4.20}
 \beta^2<\frac{2}{3}\quad\mbox{;}\quad 1-\sqrt{\frac{2}{3}} < g <
 1 + \sqrt{\frac{2}{3}} .
\end{equation}
Oscillations in $\psi$ mean that the mass term in (\ref{4.8})
\begin{equation}
\label{4.21}
m(q) \approx i\frac{\tilde{\psi}}{2}q
\end{equation}
starts to oscillate. By solving the equation for bound states we can
check that at $g>g_c$ bound states appear with wave functions behaving
like (\ref{4.19}) which coincides with the behaviour of the solution
of the Dirac equation in the field of a point-like static charge
$Ze$ when $Z>137$. The simplest example for such bound states are
Goldstone states the wave functions of which, as we shall see it in
the next section, are proportional to $m(q)$.

We have found the oscillations and the critical value $g_c =
1-\sqrt{\frac{2}{3}}$ using the assumption that $g=const$.
In reality $g$ depends on $q$ (see Fig.1) and the situation
is somewhat more complicated. It reminds the case
of the equation for a critical charge of finite radius.
In the region of small $q$ values where $g(q)$ is close to a
constant $g(q) \approx g(0)$ we can consider $\psi$ as an
independent variable and $q^2$ as a function of $\psi$. It can
be shown that for $g(0)$ satisfying the condition (\ref{4.22}) there
are two regions $0<\psi<\pi-\psi_c$, $\pi+\psi_c<\psi<2\pi$ in the
$q^2$, $\psi$ plane (Fig.3)
\[ 
  \begin{picture}(0,0)%
\epsfig{file=fig60.pstex}%
\end{picture}%
\setlength{\unitlength}{0.00083300in}%
\begingroup\makeatletter\ifx\SetFigFont\undefined
\def\x#1#2#3#4#5#6#7\relax{\def\x{#1#2#3#4#5#6}}%
\expandafter\x\fmtname xxxxxx\relax \def\y{splain}%
\ifx\x\y   
\gdef\SetFigFont#1#2#3{%
  \ifnum #1<17\tiny\else \ifnum #1<20\small\else
  \ifnum #1<24\normalsize\else \ifnum #1<29\large\else
  \ifnum #1<34\Large\else \ifnum #1<41\LARGE\else
     \huge\fi\fi\fi\fi\fi\fi
  \csname #3\endcsname}%
\else
\gdef\SetFigFont#1#2#3{\begingroup
  \count@#1\relax \ifnum 25<\count@\count@25\fi
  \def\x{\endgroup\@setsize\SetFigFont{#2pt}}%
  \expandafter\x
    \csname \romannumeral\the\count@ pt\expandafter\endcsname
    \csname @\romannumeral\the\count@ pt\endcsname
  \csname #3\endcsname}%
\fi
\fi\endgroup
\begin{picture}(4016,2720)(451,-3226)
\put(699,-1694){\makebox(0,0)[lb]{\smash{\SetFigFont{11}{13.2}{rm}$\pi$}}}
\put(734,-2594){\makebox(0,0)[lb]{\smash{\SetFigFont{11}{13.2}{rm}$0$}}}
\put(4368,-2594){\makebox(0,0)[lb]{\smash{\SetFigFont{11}{13.2}{rm}$q^2$}}}
\put(2897,-2594){\makebox(0,0)[lb]{\smash{\SetFigFont{11}{13.2}{rm}$\lambda^2$}}}
\put(1093,-719){\makebox(0,0)[lb]{\smash{\SetFigFont{11}{13.2}{rm}$\psi$}}}
\put(451,-1439){\makebox(0,0)[lb]{\smash{\SetFigFont{11}{13.2}{rm}$\pi\!+\!\psi_c$}}}
\put(451,-1950){\makebox(0,0)[lb]{\smash{\SetFigFont{11}{13.2}{rm}$\pi\!-\!\psi_c$}}}
\put(1466,-2158){\makebox(0,0)[lb]{\smash{\SetFigFont{11}{13.2}{rm}I}}}
\put(1260,-1225){\makebox(0,0)[lb]{\smash{\SetFigFont{11}{13.2}{rm}II}}}
\put(2228,-3199){\makebox(0,0)[lb]{\smash{\SetFigFont{11}{13.2}{rm}Figure 3}}}
\put(613,-1002){\makebox(0,0)[lb]{\smash{\SetFigFont{11}{13.2}{rm}$2 \pi$}}}
\end{picture}

\]
where the solution $\psi(q)$ is a monotonic one and there is a region
$\pi-\psi_c<\psi<\pi+\psi_c$ where the solution oscillates.
The value of $\psi_c$ is determined by the equality
\begin{equation}
\label{4.22}
\sin^2 \frac{\psi_c}{2} = \left(\frac{2}{3}-\beta^2\right)
\sqrt{\frac{1+3\beta^2}{1-\beta^2}}\frac{1}{1+\sqrt{(1+3\beta^2)
 (1-\beta^2)}};
\end{equation}
it differs from zero if $\beta^2<\frac{2}{3}$. Considering $\beta$
as a function of $q^2$ in (\ref{4.22}) and taking into account that
$\beta^2 \rightarrow 1$ at $q^2 \rightarrow \infty$, we obtain a
region inside the dotted curve where the solution oscillates.
There are no oscillations if $q^2>\lambda^2$ ($\beta^2(\lambda^2)=
\frac{2}{3}$). Due to (\ref{4.2}) and (\ref{4.3}), we can write
in the region $q^2\gg \lambda^2$
\begin{equation}
\label{4.23}
\frac{i}{2}\left(\psi-\pi\right) = \frac{m_0}{q}\left(\frac{\alpha}
{\alpha_0}\right)^{\gamma_m}+\frac{\nu_1^3}{q^3}\left(\frac{\alpha}
{\alpha_0}\right)^{\gamma_1} .
\end{equation}
If $\beta^2>\frac{2}{3}$, the solution which equals $\psi=\frac{q}{m_c}$
at $q\rightarrow 0$ transforms monotonically into (\ref{4.23}) where
$m_0\neq 0$. If $\beta^2<\frac{2}{3}$, $\psi(\lambda^2)$ and
$\dot{\psi}(\lambda^2)$ start to oscillate as functions of $m_c$,
and $m_0$ can turn into zero at a certain $m_c$ value. This means that
we have a solution corresponding to broken chiral symmetry.

If $m_0=0$, there exist also a large number of solutions depending
on the parameters $\nu_1$, $\nu_2$. With the same sign of $\nu_1$
we can have different solutions corresponding to the curves I and II
which in the limit $q\rightarrow 0$ reach $\psi=0$ and $\psi=2\pi$,
respectively. The solutions $\psi(0)=2\pi$ correspond to smaller
values $\nu_1$, $m_c$.

Let us consider the solutions of types I and II in detail for complex
$q$ -- this is justified since $G^{-1}$ has to satisfy the
requirements of analiticity and unitarity. We will show that both
solutions have singularities at real positive $q^2$ values. The
solutions are chosen in such a way that they are regular as
$q \rightarrow 0$ and have no singularities at $q^2<0$. Due to the
analiticity of the equations, the behaviour of the solution for
$q^2>0$ can be found by solving the same equations (\ref{4.13}),
(\ref{4.18}) with the same boundary conditions at $q\rightarrow 0$.

If $\beta$ is fixed, the equations (\ref{4.13}), (\ref{4.18}) can be
rewritten in a simpler form. Denote
\[ \frac{\dot{u}}{u} = p(\phi) . \]
Then
\begin{equation}
\label{4.24}
\frac{\partial p}{\partial \phi} = -\beta\sqrt{p^2 + 3\sinh^2
 \frac{\phi}{2} - 1}
\end{equation}
\begin{equation}
\label{4.25}
\dot{\phi} = \frac{2}{\beta}\sqrt{p^2 + 3\sinh^2\frac{\phi}{2} - 1}
\end{equation}
with the boundary condition $p=1$ at $\phi=0$ for a I-type solution.
The phase diagram corresponding to the equation (\ref{4.24}) is
shown in Fig.4 where the solid line represents the solution of the
equation
\begin{equation}
\label{4.26}
p^2 = 1-3\sinh^2\frac{\phi}{2}.
\end{equation}
The function $p=p(\phi)$ has the shape of the dashed line in Fig.4.
\[
  \begin{picture}(0,0)%
\epsfig{file=fig61.pstex}%
\end{picture}%
\setlength{\unitlength}{0.00083300in}%
\begingroup\makeatletter\ifx\SetFigFont\undefined
\def\x#1#2#3#4#5#6#7\relax{\def\x{#1#2#3#4#5#6}}%
\expandafter\x\fmtname xxxxxx\relax \def\y{splain}%
\ifx\x\y   
\gdef\SetFigFont#1#2#3{%
  \ifnum #1<17\tiny\else \ifnum #1<20\small\else
  \ifnum #1<24\normalsize\else \ifnum #1<29\large\else
  \ifnum #1<34\Large\else \ifnum #1<41\LARGE\else
     \huge\fi\fi\fi\fi\fi\fi
  \csname #3\endcsname}%
\else
\gdef\SetFigFont#1#2#3{\begingroup
  \count@#1\relax \ifnum 25<\count@\count@25\fi
  \def\x{\endgroup\@setsize\SetFigFont{#2pt}}%
  \expandafter\x
    \csname \romannumeral\the\count@ pt\expandafter\endcsname
    \csname @\romannumeral\the\count@ pt\endcsname
  \csname #3\endcsname}%
\fi
\fi\endgroup
\begin{picture}(3324,2736)(889,-2485)
\put(2078,-2449){\makebox(0,0)[lb]{\smash{\SetFigFont{12}{14.4}{rm}Figure 4}}}
\put(4126,-1261){\makebox(0,0)[lb]{\smash{\SetFigFont{11}{13.2}{rm}$\phi$}}}
\put(2476, 89){\makebox(0,0)[lb]{\smash{\SetFigFont{11}{13.2}{rm}$p$}}}
\end{picture}

\]
The $\phi$-dependence of $p$ at $\phi \rightarrow \infty$ is different
for $\beta>\frac{1}{2}$ and $\beta<\frac{1}{2}$. In both cases $\phi$
approaches infinity at finite $\xi$ values. We have at
 $\beta>\frac{1}{2}$, $\phi\rightarrow \infty$
\begin{equation}
\label{4.27}
\frac{\partial p}{\partial \phi} = + \beta p,\qquad
p=-\frac{C}{2}e^{\beta\phi};  \qquad C>0
\end{equation}
and at $\beta<\frac{1}{2}$, $\phi\rightarrow \infty$
\begin{equation}
\label{4.28}
p=-2\beta^2 C_1 e^{\frac{\phi}{2}},\qquad
C_1 \equiv \frac{1}{2}\sqrt{\frac{3}{1-4\beta^2}}.
\end{equation}
The equation (\ref{4.25}) enables us to find $\xi$ as a function of
$\phi$.
\begin{equation}
\label{4.29}
\xi = \xi^* - \frac{\beta}{2} \int_{\phi}^{\infty} \frac{d\phi'}
 {\sqrt{p^2(\phi) + 3\sinh^2 \frac{\phi}{2}-1}}
\end{equation}
The integral in (\ref{4.29}) converges in both cases ($\beta>\frac{1}{2}$
, $\beta<\frac{1}{2}$); because of this, $\phi$ goes to infinity at a
finite $\xi=\xi^*$ ($q \rightarrow m^*$). Near $q=m^*$ at $\beta>
\frac{1}{2}$ ($g<\frac{1}{2}$) we have
\begin{equation}
\label{4.30}
u=u_0\sqrt{1-\frac{q}{m^*}}\quad\mbox{,}\quad e^{-\frac{\phi}{2}} =
\left\{C\left(1-\frac{q}{m^*}\right)\right\}^{\frac{1}{2\beta}}
\end{equation}
and, consequently,
\begin{equation}
\label{4.31}
G^{-1}(q)=Z_0^{-1}\left\{\left(1-\frac{q}{m^*}\right)^{\frac{1}{\beta}}
 \frac{1}{2}(\hat{q}+q) + \left(\frac{1}{C}\right)^{\frac{1}{\beta}}
 \frac{1}{2}(q-\hat{q})\right\} .
\end{equation}
If $\beta<\frac{1}{2}$ ($g>\frac{1}{2}$),
\begin{equation}
\label{4.32}
u=u_0\left(1-\frac{q}{m^*}\right)^{\beta^2} \quad\mbox{,}\quad
 e^{-\frac{\phi}{2}}= C_1\left(1-\frac{q}{m^*}\right),
\end{equation}
and thus
\begin{equation}
\label{4.33}
G^{-1}(q)=Z_0 \left\{\left(1-\frac{q}{m^*}\right)^2 \frac{1}{2}
 (\hat{q}+q) + \frac{1}{C_1^2}(q-\hat{q})\right\}.
\end{equation}
It can be easily shown that the relation between the position of the
singularity $q=m^*$ of the Green's function and the quantity $m_c$
can be written as
\begin{equation}
\label{4.34}
\ln\frac{m^*}{m_c} = \int_0^{\infty} d\phi \left[ \frac{\beta}{2\sqrt{
 p^2(\phi)+3\sinh^2\frac{\phi}{2}-1}} - \frac{2}{\sinh 2\phi}\right].
\end{equation}
The formulae (\ref{4.31}) and (\ref{4.33}) define the behaviour of
$G^{-1}(q)$ near the singularity for a solution of type I.
In order to obtain the behaviour of $G^{-1}$ near the singularity for
a solution of type II it is sufficient to notice that at $q\rightarrow 0$
such a solution has the form
\begin{equation}
\label{4.35}
\phi = 2\pi i - \frac{q}{m'_c} .
\end{equation}
The replacement of $\phi$ by $\phi + 2\pi i$ changes only the sign of
$G^{-1}$. Replacing $\frac{q}{m_c}$ by $-\frac{q}{m'_c}$
and solving (\ref{4.24}) and (\ref{4.25}) for $q>0$ we will
find $\phi$ to be negative, and near the singularity $\phi$ will
go to $-\infty$. This means the following. If the first
term $G_+^{-1}(q)$ of the expression
\begin{equation}
\label{4.36}
G^{-1}(q) = G_+^{-1}(q)\frac{1}{2}\left(1+\frac{\hat{q}}{q}\right) +
 G_-^{-1}(q)\frac{1}{2}\left(1-\frac{\hat{q}}{q}\right)
\end{equation}
equals zero for the solution I at $q=m^*$, then $G_-^{-1}$ is zero
for the solution II at $q=m^{'*}$.
If $q^2>m^{*2}$, both solutions become complex:
\begin{eqnarray}
\label{4.37}
\phi = - \frac{2}{\beta} \ln C \left(\frac{q}{m^*}-1\right)
 & + & \frac{i\pi}{\beta} \quad\mbox{,}\quad \beta>\frac{1}{2}
 \nonumber\\
\phi = -2 \ln C_1 \left(\frac{q}{m^*} - 1 \right)
 & + & 2i\pi \quad\mbox{,}\quad \beta<\frac{1}{2} .
\end{eqnarray}
The trajectories of $\phi(q)$ at $0<q<\infty$ in the complex plane $\phi$
are shown in Fig.5 for the solutions I,II.
\[
  \begin{picture}(0,0)%
\epsfig{file=fig62.pstex}%
\end{picture}%
\setlength{\unitlength}{0.00083300in}%
\begingroup\makeatletter\ifx\SetFigFont\undefined
\def\x#1#2#3#4#5#6#7\relax{\def\x{#1#2#3#4#5#6}}%
\expandafter\x\fmtname xxxxxx\relax \def\y{splain}%
\ifx\x\y   
\gdef\SetFigFont#1#2#3{%
  \ifnum #1<17\tiny\else \ifnum #1<20\small\else
  \ifnum #1<24\normalsize\else \ifnum #1<29\large\else
  \ifnum #1<34\Large\else \ifnum #1<41\LARGE\else
     \huge\fi\fi\fi\fi\fi\fi
  \csname #3\endcsname}%
\else
\gdef\SetFigFont#1#2#3{\begingroup
  \count@#1\relax \ifnum 25<\count@\count@25\fi
  \def\x{\endgroup\@setsize\SetFigFont{#2pt}}%
  \expandafter\x
    \csname \romannumeral\the\count@ pt\expandafter\endcsname
    \csname @\romannumeral\the\count@ pt\endcsname
  \csname #3\endcsname}%
\fi
\fi\endgroup
\begin{picture}(4074,1986)(364,-2485)
\put(2438,-736){\makebox(0,0)[lb]{\smash{\SetFigFont{11}{13.2}{rm}$2 i \pi$}}}
\put(2078,-2449){\makebox(0,0)[lb]{\smash{\SetFigFont{12}{14.4}{rm}Figure 5}}}
\put(3834,-1521){\makebox(0,0)[lb]{\smash{\SetFigFont{11}{13.2}{rm}I'}}}
\put(3924,-906){\makebox(0,0)[lb]{\smash{\SetFigFont{11}{13.2}{rm}I}}}
\put(898,-1187){\makebox(0,0)[lb]{\smash{\SetFigFont{11}{13.2}{rm}II'}}}
\put(493,-1513){\makebox(0,0)[lb]{\smash{\SetFigFont{11}{13.2}{rm}II}}}
\put(2466,-2046){\makebox(0,0)[lb]{\smash{\SetFigFont{11}{13.2}{rm}$0$}}}
\put(2435,-1336){\makebox(0,0)[lb]{\smash{\SetFigFont{11}{13.2}{rm}$i \pi$}}}
\end{picture}

\]
The solution of the type I has a remarkable feature: $\Im \phi>\pi$
for any $q>m^*$ values. Hence, taking $\phi=\phi_1+i\phi_2$, the
imaginary part of $m(q)$ in (\ref{4.8})
\begin{equation}
\label{4.38}
\Im m(q) = q\Im \frac{\cosh\frac{1}{2}(\phi_1+i\phi_2)}{\sinh
\frac{1}{2}(\phi_1+i\phi_2)} = \frac{-q\sin\phi_2}{2|\sinh\frac{1}{2}
(\phi_1+i\phi_2)|^2}
\end{equation}
turns out to be positive. At the same time $\Im m(q)$ for the solution
of the type II is an oscillating function; this can lead to a
contradiction with the unitarity condition for the Green's function.
For complex $q$ values $G^{-1}(q)$ has no singularities. Moving
in the complex plane along an arbitrary ray the trajectory of
$\phi(q)$ doesn't approach infinity (curves I' and II').

The main result of this section is the following. In the framework
of the equation for the fermion Green's function (\ref{4.1})
there exist solutions corresponding to broken symmetry
provided $g(q)$ has an asymptotically free behaviour and
$g(0)>1-\sqrt{\frac{2}{3}}$. These solutions behave at
$q^2\rightarrow \infty$ as
\begin{equation}
\label{4.39}
G^{-1}(q)=Z \left[ \frac{\nu_1^3}{q^2} - \hat{q}
 \left(1-\frac{\nu_2^4}{q^4}\right)\right].
\end{equation}
The expression (\ref{4.39}) has a mass term decreasing at infinity.

\sloppy
\section{Axial current conservation and Goldstone 
         states}\label{V}
\fussy

If a fermion Green's function corresponds to symmetry
breaking, it is natural to expect the existence of Goldstone-type
bound states. This expectation is connected with the belief that if
$m_0=0$, the axial current has to be conserved. This is, however,
not necessarily true because due to divergences in the theory a
leakage of the current is possible in the region of the ultraviolet
cutoff. A typical example for this phenomenon is the anomaly. But
even in a non-anomalous case it is not obvious whether the current
conservation is contained by the equation for the Green's function or
it is imposed as a condition on the solution of the equation. In
order to clarify this, let us consider the equation for
the bound state $\phi$ supposing that it is a pseudoscalar.
\begin{equation}
\label{5.1}
 \partial^2 \phi(p,q) =
 g(q)\{ A_{\mu}(q_2) \partial_{\mu} \phi(p,q) +
  \partial_{\mu} \phi(p,q)\tilde{A}_{\mu}(q_1) -
 A_{\mu}(q_2)\phi(p,q) A_{\mu}(q_1)\}
\end{equation}
This equation has to have a solution decreasing at large $q^2$ and
$p^2=0$. It is easy to see that indeed there exists such a solution.
At $p=0$ the equation for $\phi$ is an equation for the variation
of a function which satisfies the equation for $G^{-1}$. If this
variation is taken in the form $\phi=C\{\gamma_5,G^{-1}\}$, $\phi$
obviously satisfies the equation and decreases as $q^2 \rightarrow
\infty$
\begin{equation}
\label{5.2}
\phi \rightarrow C\frac{2\nu_1^3}{q^2}\gamma_5 .
\end{equation}
This, however, does not mean that we have particles with $p^2=0$.
Indeed, the equation (\ref{5.1}) has a solution decreasing at
$q^2\rightarrow \infty$ for any $p^2$ values. The reason for this
is that the equation is highly degenerate. It has a solution of
the form
\begin{equation}
\label{5.3}
\phi=O_1G^{-1}(q_1) + G^{-1}(q_2)O_2
\end{equation}
where $O_1$ and $O_2$ are any combinations of Dirac matrices with
coefficients not depending on $q$. This can be checked by
substituting directly (\ref{5.3}) into (\ref{5.1}) and using the
equation for $G^{-1}$.The reason for this degeneracy is the invariance
of the equation (\ref{5.1}) under Lorentz transformations, under all the
possible discrete symmetries of the Dirac equation and for a fixed
$g=const$ even scale invariance and translational invariance in
momentum space. If, for example, $O_1=O_2=\gamma_5$, then,
due to the Ward identity
\begin{equation}
\label{5.4}
p_{\mu}\Gamma^5_{\mu}(p,q) = \gamma_5 G^{-1}(q_1) + G^{-1}(q_2)\gamma_5,
\end{equation}
\begin{equation}
\label{5.5}
\phi = \gamma_5 G^{-1}(q_1) + G^{-1}(q_2)\gamma_5
\end{equation}
is the divergence of the axial current. Hence, if the equation
(\ref{5.1}) is satisfied for $p_{\mu}\Gamma_{\mu}^5$ it has to have
the solution (\ref{5.4}).

In order to show that (\ref{5.1}) has a decreasing solution at
$q^2 \rightarrow \infty$ for any $p$ value, let us notice that
in Euclidean space this equation has the structure of the
Schr\"odinger equation
\begin{equation}
\label{5.6}
(-\partial^2 + U)\Psi = \varepsilon\Psi
\end{equation}
in four dimensions at $\varepsilon=0$, with a potential depending on $q$,
spin variables and the external vector $p$. For such a potential
the total four-dimensional angular momentum is not conserved, only
its projection $\mu$ onto $p_{\nu}$. An equation of this type has always
non-singular solutions with incoming waves of given $\mu$. E.g.,
considering an incoming wave with $\mu=0$ for the pseudoscalar $\phi$,
we will have a solution
\[ \phi_0 = \gamma_5 C_0 + \quad\mbox{decreasing scattered waves}\quad \]
at $q^2 \rightarrow \infty$, or, regarding an incoming wave of the form
$\gamma_5 p_{\mu}\gamma_{\mu} \equiv \gamma_5\hat{p}$, we will have
\[ \phi_1 = \gamma_5 \hat{p} + \quad\mbox{decreasing scattered waves.}
 \quad \]
Suppose we found the solution $\phi_1$. In this case, due to the fact
that for $q \rightarrow 0$ the solution (\ref{5.4}) behaves as
\begin{equation}
\label{5.7}
\gamma_5 G^{-1}(q_1)+G^{-1}(q_2)\gamma_5 \rightarrow Z\left(\gamma_5
 \hat{p} + \frac{2\gamma_5 \nu^3}{q^2}\right),
\end{equation}
we will find that
\begin{equation}
\label{5.8}
\phi = Z\phi_1-\gamma_5 G^{-1}(q_1) - G^{-1}(q_2)\gamma_5 \rightarrow
\frac{2Z\gamma_5 \nu^3}{q^2}
\end{equation}
is decreasing with $q^2\rightarrow \infty$ at any $p$.

By stating that the equation for bound states has a solution at any
$p$ values does not mean that there are no bound states; it means
only that the mass of the bound state has to be calculated
independently. The most natural way to do this is to calculate the
self-energy of the state
\begin{equation}
\label{5.9}
    \Sigma(p) = \,\,\begin{picture}(0,0)%
\epsfig{file=fig63.pstex}%
\end{picture}%
\setlength{\unitlength}{0.00083300in}%
\begingroup\makeatletter\ifx\SetFigFont\undefined
\def\x#1#2#3#4#5#6#7\relax{\def\x{#1#2#3#4#5#6}}%
\expandafter\x\fmtname xxxxxx\relax \def\y{splain}%
\ifx\x\y   
\gdef\SetFigFont#1#2#3{%
  \ifnum #1<17\tiny\else \ifnum #1<20\small\else
  \ifnum #1<24\normalsize\else \ifnum #1<29\large\else
  \ifnum #1<34\Large\else \ifnum #1<41\LARGE\else
     \huge\fi\fi\fi\fi\fi\fi
  \csname #3\endcsname}%
\else
\gdef\SetFigFont#1#2#3{\begingroup
  \count@#1\relax \ifnum 25<\count@\count@25\fi
  \def\x{\endgroup\@setsize\SetFigFont{#2pt}}%
  \expandafter\x
    \csname \romannumeral\the\count@ pt\expandafter\endcsname
    \csname @\romannumeral\the\count@ pt\endcsname
  \csname #3\endcsname}%
\fi
\fi\endgroup
\begin{picture}(2631,516)(483,-295)
\put(1126, 89){\makebox(0,0)[lb]{\smash{\SetFigFont{11}{13.2}{rm}$\varphi$}}}
\put(2401, 89){\makebox(0,0)[lb]{\smash{\SetFigFont{11}{13.2}{rm}$\varphi$}}}
\end{picture}

\end{equation}
and to solve the equation
\begin{equation}
\label{5.10}
\Sigma(p')=0 ,
\end{equation}
or to calculate the forward Compton scattering
\begin{equation}
\label{5.11}
 \begin{picture}(0,0)%
\epsfig{file=combine3.pstex}%
\end{picture}%
\setlength{\unitlength}{0.00083300in}%
\begingroup\makeatletter\ifx\SetFigFont\undefined
\def\x#1#2#3#4#5#6#7\relax{\def\x{#1#2#3#4#5#6}}%
\expandafter\x\fmtname xxxxxx\relax \def\y{splain}%
\ifx\x\y   
\gdef\SetFigFont#1#2#3{%
  \ifnum #1<17\tiny\else \ifnum #1<20\small\else
  \ifnum #1<24\normalsize\else \ifnum #1<29\large\else
  \ifnum #1<34\Large\else \ifnum #1<41\LARGE\else
     \huge\fi\fi\fi\fi\fi\fi
  \csname #3\endcsname}%
\else
\gdef\SetFigFont#1#2#3{\begingroup
  \count@#1\relax \ifnum 25<\count@\count@25\fi
  \def\x{\endgroup\@setsize\SetFigFont{#2pt}}%
  \expandafter\x
    \csname \romannumeral\the\count@ pt\expandafter\endcsname
    \csname @\romannumeral\the\count@ pt\endcsname
  \csname #3\endcsname}%
\fi
\fi\endgroup
\begin{picture}(4728,1000)(685,-523)
\put(4801,-286){\makebox(0,0)[lb]{\smash{\SetFigFont{11}{13.2}{rm}$\varphi$}}}
\put(3751,-286){\makebox(0,0)[lb]{\smash{\SetFigFont{11}{13.2}{rm}$\varphi$}}}
\put(2251, 89){\makebox(0,0)[lb]{\smash{\SetFigFont{11}{13.2}{rm}$\varphi$}}}
\put(1201, 89){\makebox(0,0)[lb]{\smash{\SetFigFont{11}{13.2}{rm}$\varphi$}}}
\end{picture}
 
\end{equation}
of the bound state on a fermion and after that integrate over
the distribution of fermions in the vacuum. But this way we will
never get a massless Goldstone state. The reason for this is the
almost obvious fact that (\ref{5.9}) and (\ref{5.11}) contradict the
condition of axial current conservation.

Let us consider the condition for current conservation in detail. If
we introduce $\tilde{\Gamma}^5_{\mu}$ as a set of diagrams
\begin{equation}
\label{5.12}
\tilde\Gamma^5_{\mu}= \gamma_{\mu}\gamma_5 + \, 
  \begin{picture}(0,0)%
\epsfig{file=fig66.pstex}%
\end{picture}%
\setlength{\unitlength}{0.00083300in}%
\begingroup\makeatletter\ifx\SetFigFont\undefined
\def\x#1#2#3#4#5#6#7\relax{\def\x{#1#2#3#4#5#6}}%
\expandafter\x\fmtname xxxxxx\relax \def\y{splain}%
\ifx\x\y   
\gdef\SetFigFont#1#2#3{%
  \ifnum #1<17\tiny\else \ifnum #1<20\small\else
  \ifnum #1<24\normalsize\else \ifnum #1<29\large\else
  \ifnum #1<34\Large\else \ifnum #1<41\LARGE\else
     \huge\fi\fi\fi\fi\fi\fi
  \csname #3\endcsname}%
\else
\gdef\SetFigFont#1#2#3{\begingroup
  \count@#1\relax \ifnum 25<\count@\count@25\fi
  \def\x{\endgroup\@setsize\SetFigFont{#2pt}}%
  \expandafter\x
    \csname \romannumeral\the\count@ pt\expandafter\endcsname
    \csname @\romannumeral\the\count@ pt\endcsname
  \csname #3\endcsname}%
\fi
\fi\endgroup
\begin{picture}(924,441)(739,-520)
\put(1051,-211){\makebox(0,0)[lb]{\smash{\SetFigFont{10}{12.0}{rm}$\gamma_{\mu} \gamma_5$}}}
\end{picture}
 \, + \cdots = \,
  \begin{picture}(0,0)%
\epsfig{file=fig67.pstex}%
\end{picture}%
\setlength{\unitlength}{0.00083300in}%
\begingroup\makeatletter\ifx\SetFigFont\undefined
\def\x#1#2#3#4#5#6#7\relax{\def\x{#1#2#3#4#5#6}}%
\expandafter\x\fmtname xxxxxx\relax \def\y{splain}%
\ifx\x\y   
\gdef\SetFigFont#1#2#3{%
  \ifnum #1<17\tiny\else \ifnum #1<20\small\else
  \ifnum #1<24\normalsize\else \ifnum #1<29\large\else
  \ifnum #1<34\Large\else \ifnum #1<41\LARGE\else
     \huge\fi\fi\fi\fi\fi\fi
  \csname #3\endcsname}%
\else
\gdef\SetFigFont#1#2#3{\begingroup
  \count@#1\relax \ifnum 25<\count@\count@25\fi
  \def\x{\endgroup\@setsize\SetFigFont{#2pt}}%
  \expandafter\x
    \csname \romannumeral\the\count@ pt\expandafter\endcsname
    \csname @\romannumeral\the\count@ pt\endcsname
  \csname #3\endcsname}%
\fi
\fi\endgroup
\begin{picture}(474,624)(964,-298)
\end{picture}

\end{equation}
with a massive fermionic Green's function, it will not satisfy the
Ward identity. However, the Ward identity will be satisfied by the
sum of $\tilde{\Gamma}^5_{\mu}$ and of the Goldstone contribution
\begin{equation}
\label{5.13}
\Gamma^5_{\mu}= \,\begin{picture}(0,0)%
\epsfig{file=fig68.pstex}%
\end{picture}%
\setlength{\unitlength}{0.00083300in}%
\begingroup\makeatletter\ifx\SetFigFont\undefined
\def\x#1#2#3#4#5#6#7\relax{\def\x{#1#2#3#4#5#6}}%
\expandafter\x\fmtname xxxxxx\relax \def\y{splain}%
\ifx\x\y   
\gdef\SetFigFont#1#2#3{%
  \ifnum #1<17\tiny\else \ifnum #1<20\small\else
  \ifnum #1<24\normalsize\else \ifnum #1<29\large\else
  \ifnum #1<34\Large\else \ifnum #1<41\LARGE\else
     \huge\fi\fi\fi\fi\fi\fi
  \csname #3\endcsname}%
\else
\gdef\SetFigFont#1#2#3{\begingroup
  \count@#1\relax \ifnum 25<\count@\count@25\fi
  \def\x{\endgroup\@setsize\SetFigFont{#2pt}}%
  \expandafter\x
    \csname \romannumeral\the\count@ pt\expandafter\endcsname
    \csname @\romannumeral\the\count@ pt\endcsname
  \csname #3\endcsname}%
\fi
\fi\endgroup
\begin{picture}(774,945)(814,-544)
\put(1351,-511){\makebox(0,0)[lb]{\smash{\SetFigFont{11}{13.2}{rm}$q_1$}}}
\put(976,-511){\makebox(0,0)[lb]{\smash{\SetFigFont{11}{13.2}{rm}$q_2$}}}
\put(1276,239){\makebox(0,0)[lb]{\smash{\SetFigFont{11}{13.2}{rm}$p$}}}
\end{picture}
  \, 
  + \, \begin{picture}(0,0)%
\epsfig{file=fig69.pstex}%
\end{picture}%
\setlength{\unitlength}{0.00083300in}%
\begingroup\makeatletter\ifx\SetFigFont\undefined
\def\x#1#2#3#4#5#6#7\relax{\def\x{#1#2#3#4#5#6}}%
\expandafter\x\fmtname xxxxxx\relax \def\y{splain}%
\ifx\x\y   
\gdef\SetFigFont#1#2#3{%
  \ifnum #1<17\tiny\else \ifnum #1<20\small\else
  \ifnum #1<24\normalsize\else \ifnum #1<29\large\else
  \ifnum #1<34\Large\else \ifnum #1<41\LARGE\else
     \huge\fi\fi\fi\fi\fi\fi
  \csname #3\endcsname}%
\else
\gdef\SetFigFont#1#2#3{\begingroup
  \count@#1\relax \ifnum 25<\count@\count@25\fi
  \def\x{\endgroup\@setsize\SetFigFont{#2pt}}%
  \expandafter\x
    \csname \romannumeral\the\count@ pt\expandafter\endcsname
    \csname @\romannumeral\the\count@ pt\endcsname
  \csname #3\endcsname}%
\fi
\fi\endgroup
\begin{picture}(624,1459)(1189,-882)
\put(1576,145){\makebox(0,0)[lb]{\smash{\SetFigFont{11}{13.2}{rm}$f p_\mu$}}}
\put(1335,145){\makebox(0,0)[lb]{\smash{\SetFigFont{11}{13.2}{rm}$i$}}}
\put(1471,-702){\makebox(0,0)[lb]{\smash{\SetFigFont{11}{13.2}{rm}$g$}}}
\put(1651,-849){\makebox(0,0)[lb]{\smash{\SetFigFont{11}{13.2}{rm}$q_1$}}}
\put(1228,-849){\makebox(0,0)[lb]{\smash{\SetFigFont{11}{13.2}{rm}$q_2$}}}
\end{picture}
 \,\,\,\,\,:
\end{equation}
\vspace{.2cm}
\begin{equation}
\label{5.14}
p_{\mu}\Gamma^5_{\mu}= p_{\mu}\tilde{\Gamma}^5_{\mu} - if\hat{g} =
\gamma_5 G^{-1}(q_1) + G^{-1}(q_2)\gamma_5 .
\end{equation}
Here $g$ is the Yukawa coupling of a Goldstone boson to a fermion:
\begin{equation}
\label{5.15}
 ifp_{\mu} = \gamma_{\mu}\gamma_5\begin{picture}(0,0)%
\epsfig{file=fig70.pstex}%
\end{picture}%
\setlength{\unitlength}{0.00083300in}%
\begingroup\makeatletter\ifx\SetFigFont\undefined
\def\x#1#2#3#4#5#6#7\relax{\def\x{#1#2#3#4#5#6}}%
\expandafter\x\fmtname xxxxxx\relax \def\y{splain}%
\ifx\x\y   
\gdef\SetFigFont#1#2#3{%
  \ifnum #1<17\tiny\else \ifnum #1<20\small\else
  \ifnum #1<24\normalsize\else \ifnum #1<29\large\else
  \ifnum #1<34\Large\else \ifnum #1<41\LARGE\else
     \huge\fi\fi\fi\fi\fi\fi
  \csname #3\endcsname}%
\else
\gdef\SetFigFont#1#2#3{\begingroup
  \count@#1\relax \ifnum 25<\count@\count@25\fi
  \def\x{\endgroup\@setsize\SetFigFont{#2pt}}%
  \expandafter\x
    \csname \romannumeral\the\count@ pt\expandafter\endcsname
    \csname @\romannumeral\the\count@ pt\endcsname
  \csname #3\endcsname}%
\fi
\fi\endgroup
\begin{picture}(941,537)(593,-629)
\end{picture}
g.
\end{equation}
Knowing $p_{\mu}\tilde\Gamma^5_{\mu}$, we can define the Yukawa
coupling $g$ by (\ref{5.14}), (\ref{5.15}). Let us apply the operator
$-\partial^2+U$ to (\ref{5.14}). We find that $gf$ satisfies
the equation (\ref{5.1}) since $p_{\mu}\tilde{\Gamma}^5_{\mu}$
and the right-hand-side of (\ref{5.14}) satisfy the same equation. But
the existence of (\ref{5.13}) implies that the mass of the Goldstone
boson has to be zero.

In order to clarify the situation with Compton scattering, let us
consider the Ward identity for the amplitude
\begin{equation}
\label{5.16}
\Gamma_{\mu}^5(k',q_2,q_1,k) = \,\,\begin{picture}(0,0)%
\epsfig{file=fig71.pstex}%
\end{picture}%
\setlength{\unitlength}{0.00083300in}%
\begingroup\makeatletter\ifx\SetFigFont\undefined
\def\x#1#2#3#4#5#6#7\relax{\def\x{#1#2#3#4#5#6}}%
\expandafter\x\fmtname xxxxxx\relax \def\y{splain}%
\ifx\x\y   
\gdef\SetFigFont#1#2#3{%
  \ifnum #1<17\tiny\else \ifnum #1<20\small\else
  \ifnum #1<24\normalsize\else \ifnum #1<29\large\else
  \ifnum #1<34\Large\else \ifnum #1<41\LARGE\else
     \huge\fi\fi\fi\fi\fi\fi
  \csname #3\endcsname}%
\else
\gdef\SetFigFont#1#2#3{\begingroup
  \count@#1\relax \ifnum 25<\count@\count@25\fi
  \def\x{\endgroup\@setsize\SetFigFont{#2pt}}%
  \expandafter\x
    \csname \romannumeral\the\count@ pt\expandafter\endcsname
    \csname @\romannumeral\the\count@ pt\endcsname
  \csname #3\endcsname}%
\fi
\fi\endgroup
\begin{picture}(1250,1085)(95,-1123)
\put( 95,-928){\makebox(0,0)[lb]{\smash{\SetFigFont{11}{13.2}{rm}$q_2$}}}
\put(1070,-472){\makebox(0,0)[lb]{\smash{\SetFigFont{11}{13.2}{rm}$k$}}}
\put(751,-999){\makebox(0,0)[lb]{\smash{\SetFigFont{11}{13.2}{rm}$q_1$}}}
\put(151,-286){\makebox(0,0)[lb]{\smash{\SetFigFont{11}{13.2}{rm}$\mu$}}}
\put(376,-211){\makebox(0,0)[lb]{\smash{\SetFigFont{11}{13.2}{rm}$k'$}}}
\end{picture}

\end{equation}
where the dotted line with the momentum $k'$ corresponds to the axial
current, the wavy line with $k$ corresponds to the Goldstone state and
the other two lines to the fermions. The Ward identity for this
amplitude is
\begin{equation}
\label{5.17}
 k'_{\mu}\Gamma_{\mu}^5(q_2,q_1,k) =  \,\,\begin{picture}(0,0)%
\epsfig{file=fig72.pstex}%
\end{picture}%
\setlength{\unitlength}{0.00083300in}%
\begingroup\makeatletter\ifx\SetFigFont\undefined
\def\x#1#2#3#4#5#6#7\relax{\def\x{#1#2#3#4#5#6}}%
\expandafter\x\fmtname xxxxxx\relax \def\y{splain}%
\ifx\x\y   
\gdef\SetFigFont#1#2#3{%
  \ifnum #1<17\tiny\else \ifnum #1<20\small\else
  \ifnum #1<24\normalsize\else \ifnum #1<29\large\else
  \ifnum #1<34\Large\else \ifnum #1<41\LARGE\else
     \huge\fi\fi\fi\fi\fi\fi
  \csname #3\endcsname}%
\else
\gdef\SetFigFont#1#2#3{\begingroup
  \count@#1\relax \ifnum 25<\count@\count@25\fi
  \def\x{\endgroup\@setsize\SetFigFont{#2pt}}%
  \expandafter\x
    \csname \romannumeral\the\count@ pt\expandafter\endcsname
    \csname @\romannumeral\the\count@ pt\endcsname
  \csname #3\endcsname}%
\fi
\fi\endgroup
\begin{picture}(1654,1176)(1438,-863)
\put(2780,-830){\makebox(0,0)[lb]{\smash{\SetFigFont{11}{13.2}{rm}$q_1$}}}
\put(3003, 70){\makebox(0,0)[lb]{\smash{\SetFigFont{11}{13.2}{rm}$k$}}}
\put(1565,161){\makebox(0,0)[lb]{\smash{\SetFigFont{11}{13.2}{rm}$k'$}}}
\put(1438,-560){\makebox(0,0)[lb]{\smash{\SetFigFont{11}{13.2}{rm}$\gamma_5$}}}
\put(2177,-560){\makebox(0,0)[lb]{\smash{\SetFigFont{11}{13.2}{rm}$g(k,q_1)$}}}
\end{picture}
 \, 
  + \,\,\begin{picture}(0,0)%
\epsfig{file=fig73.pstex}%
\end{picture}%
\setlength{\unitlength}{0.00083300in}%
\begingroup\makeatletter\ifx\SetFigFont\undefined
\def\x#1#2#3#4#5#6#7\relax{\def\x{#1#2#3#4#5#6}}%
\expandafter\x\fmtname xxxxxx\relax \def\y{splain}%
\ifx\x\y   
\gdef\SetFigFont#1#2#3{%
  \ifnum #1<17\tiny\else \ifnum #1<20\small\else
  \ifnum #1<24\normalsize\else \ifnum #1<29\large\else
  \ifnum #1<34\Large\else \ifnum #1<41\LARGE\else
     \huge\fi\fi\fi\fi\fi\fi
  \csname #3\endcsname}%
\else
\gdef\SetFigFont#1#2#3{\begingroup
  \count@#1\relax \ifnum 25<\count@\count@25\fi
  \def\x{\endgroup\@setsize\SetFigFont{#2pt}}%
  \expandafter\x
    \csname \romannumeral\the\count@ pt\expandafter\endcsname
    \csname @\romannumeral\the\count@ pt\endcsname
  \csname #3\endcsname}%
\fi
\fi\endgroup
\begin{picture}(1680,1183)(1076,-572)
\put(1529,-262){\makebox(0,0)[lb]{\smash{\SetFigFont{11}{13.2}{rm}$g(q_2,k)$}}}
\put(2756,459){\makebox(0,0)[lb]{\smash{\SetFigFont{11}{13.2}{rm}$k'$}}}
\put(2666,-262){\makebox(0,0)[lb]{\smash{\SetFigFont{11}{13.2}{rm}$\gamma_5$}}}
\end{picture}
 \,\,\,\,\,.
\end{equation}
To fulfil (\ref{5.17}), $\Gamma_{\mu}^5(q_2,q_1,k)$ has to be equal
\begin{equation}
\label{5.18}
\Gamma_{\mu}^5(q_2,q_1,k) = \,\begin{picture}(0,0)%
\epsfig{file=fig74.pstex}%
\end{picture}%
\setlength{\unitlength}{0.00083300in}%
\begingroup\makeatletter\ifx\SetFigFont\undefined
\def\x#1#2#3#4#5#6#7\relax{\def\x{#1#2#3#4#5#6}}%
\expandafter\x\fmtname xxxxxx\relax \def\y{splain}%
\ifx\x\y   
\gdef\SetFigFont#1#2#3{%
  \ifnum #1<17\tiny\else \ifnum #1<20\small\else
  \ifnum #1<24\normalsize\else \ifnum #1<29\large\else
  \ifnum #1<34\Large\else \ifnum #1<41\LARGE\else
     \huge\fi\fi\fi\fi\fi\fi
  \csname #3\endcsname}%
\else
\gdef\SetFigFont#1#2#3{\begingroup
  \count@#1\relax \ifnum 25<\count@\count@25\fi
  \def\x{\endgroup\@setsize\SetFigFont{#2pt}}%
  \expandafter\x
    \csname \romannumeral\the\count@ pt\expandafter\endcsname
    \csname @\romannumeral\the\count@ pt\endcsname
  \csname #3\endcsname}%
\fi
\fi\endgroup
\begin{picture}(1374,950)(1264,-822)
\put(1877,-560){\makebox(0,0)[lb]{\smash{\SetFigFont{11}{13.2}{rm}$g(q_1,k)$}}}
\put(2488, 11){\makebox(0,0)[lb]{\smash{\SetFigFont{11}{13.2}{rm}$k$}}}
\put(2354,-789){\makebox(0,0)[lb]{\smash{\SetFigFont{11}{13.2}{rm}$q_1$}}}
\put(1394,-789){\makebox(0,0)[lb]{\smash{\SetFigFont{11}{13.2}{rm}$q_2$}}}
\put(1569,-271){\makebox(0,0)[lb]{\smash{\SetFigFont{11}{13.2}{rm}$\Gamma^5_\mu$}}}
\put(1565, 11){\makebox(0,0)[lb]{\smash{\SetFigFont{11}{13.2}{rm}$k'$}}}
\end{picture}
\, 
  + \,\begin{picture}(0,0)%
\epsfig{file=fig75.pstex}%
\end{picture}%
\setlength{\unitlength}{0.00083300in}%
\begingroup\makeatletter\ifx\SetFigFont\undefined
\def\x#1#2#3#4#5#6#7\relax{\def\x{#1#2#3#4#5#6}}%
\expandafter\x\fmtname xxxxxx\relax \def\y{splain}%
\ifx\x\y   
\gdef\SetFigFont#1#2#3{%
  \ifnum #1<17\tiny\else \ifnum #1<20\small\else
  \ifnum #1<24\normalsize\else \ifnum #1<29\large\else
  \ifnum #1<34\Large\else \ifnum #1<41\LARGE\else
     \huge\fi\fi\fi\fi\fi\fi
  \csname #3\endcsname}%
\else
\gdef\SetFigFont#1#2#3{\begingroup
  \count@#1\relax \ifnum 25<\count@\count@25\fi
  \def\x{\endgroup\@setsize\SetFigFont{#2pt}}%
  \expandafter\x
    \csname \romannumeral\the\count@ pt\expandafter\endcsname
    \csname @\romannumeral\the\count@ pt\endcsname
  \csname #3\endcsname}%
\fi
\fi\endgroup
\begin{picture}(1374,941)(1264,-822)
\put(2354,-789){\makebox(0,0)[lb]{\smash{\SetFigFont{11}{13.2}{rm}$q_1$}}}
\put(2147,-267){\makebox(0,0)[lb]{\smash{\SetFigFont{11}{13.2}{rm}$\Gamma^5_\mu$}}}
\put(1565, 11){\makebox(0,0)[lb]{\smash{\SetFigFont{11}{13.2}{rm}$k$}}}
\put(1394,-789){\makebox(0,0)[lb]{\smash{\SetFigFont{11}{13.2}{rm}$q_2$}}}
\put(1528,-560){\makebox(0,0)[lb]{\smash{\SetFigFont{11}{13.2}{rm}$g(q_2,k)$}}}
\put(2488, 11){\makebox(0,0)[lb]{\smash{\SetFigFont{11}{13.2}{rm}$k'$}}}
\end{picture}
\, + \,\begin{picture}(0,0)%
\epsfig{file=fig76.pstex}%
\end{picture}%
\setlength{\unitlength}{0.00083300in}%
\begingroup\makeatletter\ifx\SetFigFont\undefined
\def\x#1#2#3#4#5#6#7\relax{\def\x{#1#2#3#4#5#6}}%
\expandafter\x\fmtname xxxxxx\relax \def\y{splain}%
\ifx\x\y   
\gdef\SetFigFont#1#2#3{%
  \ifnum #1<17\tiny\else \ifnum #1<20\small\else
  \ifnum #1<24\normalsize\else \ifnum #1<29\large\else
  \ifnum #1<34\Large\else \ifnum #1<41\LARGE\else
     \huge\fi\fi\fi\fi\fi\fi
  \csname #3\endcsname}%
\else
\gdef\SetFigFont#1#2#3{\begingroup
  \count@#1\relax \ifnum 25<\count@\count@25\fi
  \def\x{\endgroup\@setsize\SetFigFont{#2pt}}%
  \expandafter\x
    \csname \romannumeral\the\count@ pt\expandafter\endcsname
    \csname @\romannumeral\the\count@ pt\endcsname
  \csname #3\endcsname}%
\fi
\fi\endgroup
\begin{picture}(1270,1250)(495,-1348)
\put(1426,-1036){\makebox(0,0)[lb]{\smash{\SetFigFont{11}{13.2}{rm}$\Lambda(q_2q_1,k',k)$}}}
\put(1745,-472){\makebox(0,0)[lb]{\smash{\SetFigFont{11}{13.2}{rm}$k$}}}
\put(630,-215){\makebox(0,0)[lb]{\smash{\SetFigFont{11}{13.2}{rm}$k'$}}}
\end{picture}
\,\,\,;
\end{equation}
here $\Gamma_{\mu}^5$ is defined by (\ref{5.13}). Using (\ref{5.13})
and (\ref{5.17}) we obtain
\begin{equation}
\label{5.19}
if\Lambda = \gamma_5 g(k,q_1) + g(q_2,k)\gamma_5.
\end{equation}
At large $q_1^2$, $q_2^2$ values we will have
\begin{equation}
\label{5.20}
g=i\frac{2\gamma_5}{f}\frac{\nu^3}{q^2} \quad\mbox{;}\quad
  \Lambda = -\frac{4\nu^3}{f^2 q^2}.
\end{equation}
This means that similarly to the case of asymptotically non-free
theories the Goldstone -- fermion scattering amplitude does not depend
on the momentum of the Goldstone boson; it decreases only as a function
of fermion virtuality. Under these circumstances it is obvious that even
in an asymptotically free theory the Goldstone boson has a point-like
structure.

Amplitudes for the interactions of many Goldstones with fermions
can be found in an analogous way and have the same properties.
The self-energy of the Goldstone state is now different from (\ref{5.9}).
It contains two terms
\begin{equation}
\label{5.21}
\Sigma(p) = \, \begin{picture}(0,0)%
\epsfig{file=fig77.pstex}%
\end{picture}%
\setlength{\unitlength}{0.00083300in}%
\begingroup\makeatletter\ifx\SetFigFont\undefined
\def\x#1#2#3#4#5#6#7\relax{\def\x{#1#2#3#4#5#6}}%
\expandafter\x\fmtname xxxxxx\relax \def\y{splain}%
\ifx\x\y   
\gdef\SetFigFont#1#2#3{%
  \ifnum #1<17\tiny\else \ifnum #1<20\small\else
  \ifnum #1<24\normalsize\else \ifnum #1<29\large\else
  \ifnum #1<34\Large\else \ifnum #1<41\LARGE\else
     \huge\fi\fi\fi\fi\fi\fi
  \csname #3\endcsname}%
\else
\gdef\SetFigFont#1#2#3{\begingroup
  \count@#1\relax \ifnum 25<\count@\count@25\fi
  \def\x{\endgroup\@setsize\SetFigFont{#2pt}}%
  \expandafter\x
    \csname \romannumeral\the\count@ pt\expandafter\endcsname
    \csname @\romannumeral\the\count@ pt\endcsname
  \csname #3\endcsname}%
\fi
\fi\endgroup
\begin{picture}(1412,285)(420,-530)
\put(852,-500){\makebox(0,0)[lb]{\smash{\SetFigFont{11}{13.2}{rm}$g$}}}
\put(1336,-500){\makebox(0,0)[lb]{\smash{\SetFigFont{11}{13.2}{rm}$g$}}}
\end{picture}
 \,+\,\begin{picture}(0,0)%
\epsfig{file=fig78.pstex}%
\end{picture}%
\setlength{\unitlength}{0.00083300in}%
\begingroup\makeatletter\ifx\SetFigFont\undefined
\def\x#1#2#3#4#5#6#7\relax{\def\x{#1#2#3#4#5#6}}%
\expandafter\x\fmtname xxxxxx\relax \def\y{splain}%
\ifx\x\y   
\gdef\SetFigFont#1#2#3{%
  \ifnum #1<17\tiny\else \ifnum #1<20\small\else
  \ifnum #1<24\normalsize\else \ifnum #1<29\large\else
  \ifnum #1<34\Large\else \ifnum #1<41\LARGE\else
     \huge\fi\fi\fi\fi\fi\fi
  \csname #3\endcsname}%
\else
\gdef\SetFigFont#1#2#3{\begingroup
  \count@#1\relax \ifnum 25<\count@\count@25\fi
  \def\x{\endgroup\@setsize\SetFigFont{#2pt}}%
  \expandafter\x
    \csname \romannumeral\the\count@ pt\expandafter\endcsname
    \csname @\romannumeral\the\count@ pt\endcsname
  \csname #3\endcsname}%
\fi
\fi\endgroup
\begin{picture}(962,660)(420,-563)
\put(863,-530){\makebox(0,0)[lb]{\smash{\SetFigFont{11}{13.2}{rm}$\phi$}}}
\end{picture}
 
\end{equation}
and equals $p^2$ at small $p^2$.

As we already have said, in this approach the Ward identity becomes
the definition of the Yukawa coupling $g$ (the wave function of the
Goldstone boson) through $p_{\mu}\Gamma_{\mu}^5$ which has a clear
diagrammatic meaning. The equation contains the amplitude $f$ of the
Goldstone -- axial current transition. The expression (\ref{5.15})
for $fp_{\mu}$ is highly symbolic, because it contains overlapping
divergences. In order to write a sensible expression we can use the
same procedure as we did in section \ref{I} when we calculated
the polarization operator of the photon. Applying the Ward identity,
we can write instead of (\ref{5.15})
\begin{equation}
\label{5.22}
 f^2 p_{\mu} = \gamma_{\mu}\gamma_5 
               \begin{picture}(0,0)%
\epsfig{file=fig79.pstex}%
\end{picture}%
\setlength{\unitlength}{0.00083300in}%
\begingroup\makeatletter\ifx\SetFigFont\undefined
\def\x#1#2#3#4#5#6#7\relax{\def\x{#1#2#3#4#5#6}}%
\expandafter\x\fmtname xxxxxx\relax \def\y{splain}%
\ifx\x\y   
\gdef\SetFigFont#1#2#3{%
  \ifnum #1<17\tiny\else \ifnum #1<20\small\else
  \ifnum #1<24\normalsize\else \ifnum #1<29\large\else
  \ifnum #1<34\Large\else \ifnum #1<41\LARGE\else
     \huge\fi\fi\fi\fi\fi\fi
  \csname #3\endcsname}%
\else
\gdef\SetFigFont#1#2#3{\begingroup
  \count@#1\relax \ifnum 25<\count@\count@25\fi
  \def\x{\endgroup\@setsize\SetFigFont{#2pt}}%
  \expandafter\x
    \csname \romannumeral\the\count@ pt\expandafter\endcsname
    \csname @\romannumeral\the\count@ pt\endcsname
  \csname #3\endcsname}%
\fi
\fi\endgroup
\begin{picture}(956,537)(593,-629)
\end{picture}
\Gamma_{\mu}^5 p_{\mu} \,.
\end{equation}
Differentiating (\ref{5.22}) over $p$ we will get for $p=0$
\begin{equation}
\label{5.23}
 f^2 \delta_{\mu\nu} = \gamma_{\mu}\gamma_5 \begin{picture}(0,0)%
\epsfig{file=fig80.pstex}%
\end{picture}%
\setlength{\unitlength}{0.00083300in}%
\begingroup\makeatletter\ifx\SetFigFont\undefined
\def\x#1#2#3#4#5#6#7\relax{\def\x{#1#2#3#4#5#6}}%
\expandafter\x\fmtname xxxxxx\relax \def\y{splain}%
\ifx\x\y   
\gdef\SetFigFont#1#2#3{%
  \ifnum #1<17\tiny\else \ifnum #1<20\small\else
  \ifnum #1<24\normalsize\else \ifnum #1<29\large\else
  \ifnum #1<34\Large\else \ifnum #1<41\LARGE\else
     \huge\fi\fi\fi\fi\fi\fi
  \csname #3\endcsname}%
\else
\gdef\SetFigFont#1#2#3{\begingroup
  \count@#1\relax \ifnum 25<\count@\count@25\fi
  \def\x{\endgroup\@setsize\SetFigFont{#2pt}}%
  \expandafter\x
    \csname \romannumeral\the\count@ pt\expandafter\endcsname
    \csname @\romannumeral\the\count@ pt\endcsname
  \csname #3\endcsname}%
\fi
\fi\endgroup
\begin{picture}(916,536)(593,-628)
\end{picture}

  \gamma_{\nu}\gamma_5 \,.
\end{equation}
If we want to get rid of the $\gamma_5$-s we have to commute $\gamma_5$
with the Green's functions and the interaction vertices along one
of the fermionic lines. Doing this, we obtain
\begin{equation}
\label{5.24}
 f^2 \delta_{\mu\nu} = \gamma_{\mu}\begin{picture}(0,0)%
\epsfig{file=fig81.pstex}%
\end{picture}%
\setlength{\unitlength}{0.00083300in}%
\begingroup\makeatletter\ifx\SetFigFont\undefined
\def\x#1#2#3#4#5#6#7\relax{\def\x{#1#2#3#4#5#6}}%
\expandafter\x\fmtname xxxxxx\relax \def\y{splain}%
\ifx\x\y   
\gdef\SetFigFont#1#2#3{%
  \ifnum #1<17\tiny\else \ifnum #1<20\small\else
  \ifnum #1<24\normalsize\else \ifnum #1<29\large\else
  \ifnum #1<34\Large\else \ifnum #1<41\LARGE\else
     \huge\fi\fi\fi\fi\fi\fi
  \csname #3\endcsname}%
\else
\gdef\SetFigFont#1#2#3{\begingroup
  \count@#1\relax \ifnum 25<\count@\count@25\fi
  \def\x{\endgroup\@setsize\SetFigFont{#2pt}}%
  \expandafter\x
    \csname \romannumeral\the\count@ pt\expandafter\endcsname
    \csname @\romannumeral\the\count@ pt\endcsname
  \csname #3\endcsname}%
\fi
\fi\endgroup
\begin{picture}(916,468)(593,-595)
\end{picture}
\gamma_{\nu} +
\gamma_{\mu}\begin{picture}(0,0)%
\epsfig{file=fig82.pstex}%
\end{picture}%
\setlength{\unitlength}{0.00083300in}%
\begingroup\makeatletter\ifx\SetFigFont\undefined
\def\x#1#2#3#4#5#6#7\relax{\def\x{#1#2#3#4#5#6}}%
\expandafter\x\fmtname xxxxxx\relax \def\y{splain}%
\ifx\x\y   
\gdef\SetFigFont#1#2#3{%
  \ifnum #1<17\tiny\else \ifnum #1<20\small\else
  \ifnum #1<24\normalsize\else \ifnum #1<29\large\else
  \ifnum #1<34\Large\else \ifnum #1<41\LARGE\else
     \huge\fi\fi\fi\fi\fi\fi
  \csname #3\endcsname}%
\else
\gdef\SetFigFont#1#2#3{\begingroup
  \count@#1\relax \ifnum 25<\count@\count@25\fi
  \def\x{\endgroup\@setsize\SetFigFont{#2pt}}%
  \expandafter\x
    \csname \romannumeral\the\count@ pt\expandafter\endcsname
    \csname @\romannumeral\the\count@ pt\endcsname
  \csname #3\endcsname}%
\fi
\fi\endgroup
\begin{picture}(1134,1202)(600,-662)
\put(894,318){\makebox(0,0)[lb]{\smash{\SetFigFont{10}{12.0}{rm}$\{\gamma_5,G^{-1}\}\gamma_5$}}}
\end{picture}
\gamma_{\nu} +
\gamma_{\mu}\begin{picture}(0,0)%
\epsfig{file=fig83.pstex}%
\end{picture}%
\setlength{\unitlength}{0.00083300in}%
\begingroup\makeatletter\ifx\SetFigFont\undefined
\def\x#1#2#3#4#5#6#7\relax{\def\x{#1#2#3#4#5#6}}%
\expandafter\x\fmtname xxxxxx\relax \def\y{splain}%
\ifx\x\y   
\gdef\SetFigFont#1#2#3{%
  \ifnum #1<17\tiny\else \ifnum #1<20\small\else
  \ifnum #1<24\normalsize\else \ifnum #1<29\large\else
  \ifnum #1<34\Large\else \ifnum #1<41\LARGE\else
     \huge\fi\fi\fi\fi\fi\fi
  \csname #3\endcsname}%
\else
\gdef\SetFigFont#1#2#3{\begingroup
  \count@#1\relax \ifnum 25<\count@\count@25\fi
  \def\x{\endgroup\@setsize\SetFigFont{#2pt}}%
  \expandafter\x
    \csname \romannumeral\the\count@ pt\expandafter\endcsname
    \csname @\romannumeral\the\count@ pt\endcsname
  \csname #3\endcsname}%
\fi
\fi\endgroup
\begin{picture}(1185,1202)(549,-662)
\put(1452,314){\makebox(0,0)[lb]{\smash{\SetFigFont{10}{12.0}{rm}$\{\gamma_5,G^{-1}\}$}}}
\put(549,316){\makebox(0,0)[lb]{\smash{\SetFigFont{10}{12.0}{rm}$\{\gamma_5,G^{-1}\}$}}}
\end{picture}
\gamma_{\nu} .
\end{equation}
Due to the conservation of the vector current, the first two terms in
(\ref{5.24}) are zero at $p=0$. The first one is just the photon
polarization operator at $p=0$, the second one is the amplitude for
the decay of a zero momentum scalar into two zero momentum photons.
The last term looks like the zero momentum pseudoscalar -- photon
scattering amplitude which also has to be zero. This, however, is not
true, because it does not contain all the necessary diagrams.
It does not involve overlapping divergences. As a result, we can write
\begin{eqnarray}
4 f^2 &=& \partial_{\mu}G^{-1}\begin{picture}(0,0)%
\epsfig{file=fig84.pstex}%
\end{picture}%
\setlength{\unitlength}{0.00083300in}%
\begingroup\makeatletter\ifx\SetFigFont\undefined
\def\x#1#2#3#4#5#6#7\relax{\def\x{#1#2#3#4#5#6}}%
\expandafter\x\fmtname xxxxxx\relax \def\y{splain}%
\ifx\x\y   
\gdef\SetFigFont#1#2#3{%
  \ifnum #1<17\tiny\else \ifnum #1<20\small\else
  \ifnum #1<24\normalsize\else \ifnum #1<29\large\else
  \ifnum #1<34\Large\else \ifnum #1<41\LARGE\else
     \huge\fi\fi\fi\fi\fi\fi
  \csname #3\endcsname}%
\else
\gdef\SetFigFont#1#2#3{\begingroup
  \count@#1\relax \ifnum 25<\count@\count@25\fi
  \def\x{\endgroup\@setsize\SetFigFont{#2pt}}%
  \expandafter\x
    \csname \romannumeral\the\count@ pt\expandafter\endcsname
    \csname @\romannumeral\the\count@ pt\endcsname
  \csname #3\endcsname}%
\fi
\fi\endgroup
\begin{picture}(1215,752)(519,-437)
\put(519,168){\makebox(0,0)[lb]{\smash{\SetFigFont{10}{12.0}{rm}$\{\gamma_5,G^{-1}\}$}}}
\put(1419,168){\makebox(0,0)[lb]{\smash{\SetFigFont{10}{12.0}{rm}$\{\gamma_5,G^{-1}\}$}}}
\end{picture}
 \,\partial_{\mu}G^{-1}
+ \partial_{\mu}G^{-1}\begin{picture}(0,0)%
\epsfig{file=fig85.pstex}%
\end{picture}%
\setlength{\unitlength}{0.00083300in}%
\begingroup\makeatletter\ifx\SetFigFont\undefined
\def\x#1#2#3#4#5#6#7\relax{\def\x{#1#2#3#4#5#6}}%
\expandafter\x\fmtname xxxxxx\relax \def\y{splain}%
\ifx\x\y   
\gdef\SetFigFont#1#2#3{%
  \ifnum #1<17\tiny\else \ifnum #1<20\small\else
  \ifnum #1<24\normalsize\else \ifnum #1<29\large\else
  \ifnum #1<34\Large\else \ifnum #1<41\LARGE\else
     \huge\fi\fi\fi\fi\fi\fi
  \csname #3\endcsname}%
\else
\gdef\SetFigFont#1#2#3{\begingroup
  \count@#1\relax \ifnum 25<\count@\count@25\fi
  \def\x{\endgroup\@setsize\SetFigFont{#2pt}}%
  \expandafter\x
    \csname \romannumeral\the\count@ pt\expandafter\endcsname
    \csname @\romannumeral\the\count@ pt\endcsname
  \csname #3\endcsname}%
\fi
\fi\endgroup
\begin{picture}(1215,752)(519,-437)
\put(519,168){\makebox(0,0)[lb]{\smash{\SetFigFont{10}{12.0}{rm}$\{\gamma_5,G^{-1}\}$}}}
\put(1419,168){\makebox(0,0)[lb]{\smash{\SetFigFont{10}{12.0}{rm}$\{\gamma_5,G^{-1}\}$}}}
\end{picture}
\partial_{\mu}G^{-1} +  
\nonumber \\
& & + \,\partial_{\mu}G^{-1}\begin{picture}(0,0)%
\epsfig{file=fig86.pstex}%
\end{picture}%
\setlength{\unitlength}{0.00083300in}%
\begingroup\makeatletter\ifx\SetFigFont\undefined
\def\x#1#2#3#4#5#6#7\relax{\def\x{#1#2#3#4#5#6}}%
\expandafter\x\fmtname xxxxxx\relax \def\y{splain}%
\ifx\x\y   
\gdef\SetFigFont#1#2#3{%
  \ifnum #1<17\tiny\else \ifnum #1<20\small\else
  \ifnum #1<24\normalsize\else \ifnum #1<29\large\else
  \ifnum #1<34\Large\else \ifnum #1<41\LARGE\else
     \huge\fi\fi\fi\fi\fi\fi
  \csname #3\endcsname}%
\else
\gdef\SetFigFont#1#2#3{\begingroup
  \count@#1\relax \ifnum 25<\count@\count@25\fi
  \def\x{\endgroup\@setsize\SetFigFont{#2pt}}%
  \expandafter\x
    \csname \romannumeral\the\count@ pt\expandafter\endcsname
    \csname @\romannumeral\the\count@ pt\endcsname
  \csname #3\endcsname}%
\fi
\fi\endgroup
\begin{picture}(1215,752)(519,-437)
\put(519,168){\makebox(0,0)[lb]{\smash{\SetFigFont{10}{12.0}{rm}$\{\gamma_5,G^{-1}\}$}}}
\put(1419,168){\makebox(0,0)[lb]{\smash{\SetFigFont{10}{12.0}{rm}$\{\gamma_5,G^{-1}\}$}}}
\end{picture}
\partial_{\mu}G^{-1} + \cdots\; .
\label{5.25}
\end{eqnarray}
In the zeroth order of $\frac{\alpha}{\pi}$
\begin{equation}
\label{5.26}
f^2 = \frac{1}{4} \int \frac{d^4 q}{(2\pi)^4i} Tr \{\gamma_5,G^{-1}\}G
\{\gamma_5,G^{-1}\}G A_{\mu}(q)A_{\mu}(q) .
\end{equation}

\section{Flavour singlet and flavour non-singlet Goldstones states.
The $U(1)$ problem}\label{VI}

Up to now we have discussed the Goldstone states in a relatively
abstract way without fixing the concrete asymptotically free
theory. In real QCD we have quarks with different flavours and
there is a difference between flavour singlet and flavour non-singlet
states. In order to clarify the picture it will be useful to
describe the Goldstone state in a different way.

The previous discussion shows that the Goldstone states in
asymptotically free and non-free theories are rather similar.
Therefore it is natural to try to introduce the Goldstone boson in
the usual way as a point-like state and to see how this state will
interact. In the usual discussion of a Goldstone particle we
suppose that there is a point-like pseudoscalar interaction
between this particle and a fermion with the pseudovector coupling
\begin{equation}
\label{6.1}
\begin{picture}(0,0)%
\epsfig{file=fig87.pstex}%
\end{picture}%
\setlength{\unitlength}{0.00083300in}%
\begingroup\makeatletter\ifx\SetFigFont\undefined
\def\x#1#2#3#4#5#6#7\relax{\def\x{#1#2#3#4#5#6}}%
\expandafter\x\fmtname xxxxxx\relax \def\y{splain}%
\ifx\x\y   
\gdef\SetFigFont#1#2#3{%
  \ifnum #1<17\tiny\else \ifnum #1<20\small\else
  \ifnum #1<24\normalsize\else \ifnum #1<29\large\else
  \ifnum #1<34\Large\else \ifnum #1<41\LARGE\else
     \huge\fi\fi\fi\fi\fi\fi
  \csname #3\endcsname}%
\else
\gdef\SetFigFont#1#2#3{\begingroup
  \count@#1\relax \ifnum 25<\count@\count@25\fi
  \def\x{\endgroup\@setsize\SetFigFont{#2pt}}%
  \expandafter\x
    \csname \romannumeral\the\count@ pt\expandafter\endcsname
    \csname @\romannumeral\the\count@ pt\endcsname
  \csname #3\endcsname}%
\fi
\fi\endgroup
\begin{picture}(774,530)(814,-99)
\put(1291,288){\makebox(0,0)[lb]{\smash{\SetFigFont{11}{13.2}{rm}$p$}}}
\end{picture}
 = \hat{p}\gamma_5\frac{1}{f_0}.
\end{equation}
This interaction induces the radiative correction to the propagator
$D(p^2)$ of the pseudoscalar
\begin{equation}
\label{6.2}
D(p^2) = \, \begin{picture}(0,0)%
\epsfig{file=fig88.pstex}%
\end{picture}%
\setlength{\unitlength}{0.00083300in}%
\begingroup\makeatletter\ifx\SetFigFont\undefined
\def\x#1#2#3#4#5#6#7\relax{\def\x{#1#2#3#4#5#6}}%
\expandafter\x\fmtname xxxxxx\relax \def\y{splain}%
\ifx\x\y   
\gdef\SetFigFont#1#2#3{%
  \ifnum #1<17\tiny\else \ifnum #1<20\small\else
  \ifnum #1<24\normalsize\else \ifnum #1<29\large\else
  \ifnum #1<34\Large\else \ifnum #1<41\LARGE\else
     \huge\fi\fi\fi\fi\fi\fi
  \csname #3\endcsname}%
\else
\gdef\SetFigFont#1#2#3{\begingroup
  \count@#1\relax \ifnum 25<\count@\count@25\fi
  \def\x{\endgroup\@setsize\SetFigFont{#2pt}}%
  \expandafter\x
    \csname \romannumeral\the\count@ pt\expandafter\endcsname
    \csname @\romannumeral\the\count@ pt\endcsname
  \csname #3\endcsname}%
\fi
\fi\endgroup
\begin{picture}(961,110)(465,-633)
\end{picture}
 \,+\, \begin{picture}(0,0)%
\epsfig{file=fig89.pstex}%
\end{picture}%
\setlength{\unitlength}{0.00083300in}%
\begingroup\makeatletter\ifx\SetFigFont\undefined
\def\x#1#2#3#4#5#6#7\relax{\def\x{#1#2#3#4#5#6}}%
\expandafter\x\fmtname xxxxxx\relax \def\y{splain}%
\ifx\x\y   
\gdef\SetFigFont#1#2#3{%
  \ifnum #1<17\tiny\else \ifnum #1<20\small\else
  \ifnum #1<24\normalsize\else \ifnum #1<29\large\else
  \ifnum #1<34\Large\else \ifnum #1<41\LARGE\else
     \huge\fi\fi\fi\fi\fi\fi
  \csname #3\endcsname}%
\else
\gdef\SetFigFont#1#2#3{\begingroup
  \count@#1\relax \ifnum 25<\count@\count@25\fi
  \def\x{\endgroup\@setsize\SetFigFont{#2pt}}%
  \expandafter\x
    \csname \romannumeral\the\count@ pt\expandafter\endcsname
    \csname @\romannumeral\the\count@ pt\endcsname
  \csname #3\endcsname}%
\fi
\fi\endgroup
\begin{picture}(1387,387)(359,-554)
\end{picture}

  \,+ \cdots \hspace{.5cm} ;
\end{equation}
$D_0$ is the bare pseudoscalar Green's function.
If the fermion is massless and the axial current is conserved,
this pseudoscalar will not interact; its self-energy
\begin{equation}
\label{6.3}
\Sigma(p^2) = \frac{1}{f_0^2}p_{\mu}p_{\nu}\gamma_{\mu}\gamma_5 
              \tilde{\Gamma}_{\nu}^5
\end{equation}
is equal to zero. If, due to symmetry breaking, the fermion becomes
massive, it starts to interact and acquires a self-energy different
from zero. By using the diagrammatic definition of $\Gamma_{\mu}^5$
we will find
\begin{equation}
\label{6.4}
\Sigma(p^2) = \frac{p^2 f^2}{f_0^2},
\end{equation}
where $f$ is the same amplitude for the Goldstone -- current transition
as what was discussed in the previous section.
Hence,
\begin{equation}
\label{6.5}
D(p^2) = \frac{f_0^2}{D_0^{-1}f_0^2 - p^2 f^2}.
\end{equation}
In the limit $f_0 \rightarrow 0$ we will have
\begin{equation}
\label{6.6}
D(p^2) = - \frac{f_0^2}{p^2 f^2};
\end{equation}
the pseudovector Goldstone -- fermion interaction is $\frac{1}{f}p_{\mu}
\tilde{\Gamma}_{\mu}^5$ with a pseudovector coupling $\frac{1}{f}$
defined by the fermion mass. The limiting procedure $f_0 \rightarrow 0$
can be understood if we accept that the interaction responsible for
symmetry breaking changes the fermion vacuum fluctuations not only
at finite momenta but also near the ultraviolet cutoff. This change
in the fermion vacuum fluctuations is responsible for the leakage
of the axial current in the region of finite momenta; it can produce
the driving term for the Goldstone state, recovering the current
conservation.

In general, the pseudovector coupling has a disadvantage compared to
the pseudoscalar coupling which we have discussed before: it looks
unrenormalizable. But in the case of flavour non-singlet states
it can always be replaced by the pseudoscalar coupling with the help
of the trivial relation
\begin{equation}
\label{6.7}
\hat{p}\gamma_5 = \big(q_2^2-m(q_2)\big)\gamma_5 + \gamma_5\big(
q_1^2-m(q_1)\big) + m(q_2)\gamma_5 + \gamma_5 m(q_1).
\end{equation}
which leads to the Ward identity (\ref{5.14}) for pseudoscalar coupling.
If we include the emission and the absorption of Goldstone bosons
inside the diagram, then in the process of this replacement
a point-like amplitude appears which corresponds to the quark
interaction with many Goldstone bosons. Nevertheless it is possible to
prove that, due to the decrease of this amplitude as the function
of quark virtuality, the theory is renormalizable.

Due to the anomaly, the case of a flavour singlet current is very
different even on the level of the Goldstone Green's function. In
this case the corresponding polarization operator $\Sigma$ will
contain not only the quark loop which we have discussed but also a
gluonic contribution
\begin{equation}
\label{6.8}
\Sigma(p^2) = \begin{picture}(0,0)%
\epsfig{file=fig90.pstex}%
\end{picture}%
\setlength{\unitlength}{0.00083300in}%
\begingroup\makeatletter\ifx\SetFigFont\undefined
\def\x#1#2#3#4#5#6#7\relax{\def\x{#1#2#3#4#5#6}}%
\expandafter\x\fmtname xxxxxx\relax \def\y{splain}%
\ifx\x\y   
\gdef\SetFigFont#1#2#3{%
  \ifnum #1<17\tiny\else \ifnum #1<20\small\else
  \ifnum #1<24\normalsize\else \ifnum #1<29\large\else
  \ifnum #1<34\Large\else \ifnum #1<41\LARGE\else
     \huge\fi\fi\fi\fi\fi\fi
  \csname #3\endcsname}%
\else
\gdef\SetFigFont#1#2#3{\begingroup
  \count@#1\relax \ifnum 25<\count@\count@25\fi
  \def\x{\endgroup\@setsize\SetFigFont{#2pt}}%
  \expandafter\x
    \csname \romannumeral\the\count@ pt\expandafter\endcsname
    \csname @\romannumeral\the\count@ pt\endcsname
  \csname #3\endcsname}%
\fi
\fi\endgroup
\begin{picture}(1412,318)(359,-520)
\end{picture}
  \,+\, \begin{picture}(0,0)%
\epsfig{file=fig91.pstex}%
\end{picture}%
\setlength{\unitlength}{0.00083300in}%
\begingroup\makeatletter\ifx\SetFigFont\undefined
\def\x#1#2#3#4#5#6#7\relax{\def\x{#1#2#3#4#5#6}}%
\expandafter\x\fmtname xxxxxx\relax \def\y{splain}%
\ifx\x\y   
\gdef\SetFigFont#1#2#3{%
  \ifnum #1<17\tiny\else \ifnum #1<20\small\else
  \ifnum #1<24\normalsize\else \ifnum #1<29\large\else
  \ifnum #1<34\Large\else \ifnum #1<41\LARGE\else
     \huge\fi\fi\fi\fi\fi\fi
  \csname #3\endcsname}%
\else
\gdef\SetFigFont#1#2#3{\begingroup
  \count@#1\relax \ifnum 25<\count@\count@25\fi
  \def\x{\endgroup\@setsize\SetFigFont{#2pt}}%
  \expandafter\x
    \csname \romannumeral\the\count@ pt\expandafter\endcsname
    \csname @\romannumeral\the\count@ pt\endcsname
  \csname #3\endcsname}%
\fi
\fi\endgroup
\begin{picture}(1922,495)(242,-619)
\put(301,-586){\makebox(0,0)[lb]{\smash{\SetFigFont{11}{13.2}{rm}$p$}}}
\end{picture}
 \,\,\,.
\end{equation}
The triangle diagram $f_{\mu\nu}$ included in (\ref{6.8}) was calculated
many years ago by Adler, Bell and Jackiw [3]
\begin{equation}
\label{6.9}
f_{\mu\nu} = \frac{\alpha}{\pi} \varepsilon_{\mu\nu\rho\sigma}k_{1\rho}
 k_{2\sigma} .
\end{equation}
With this expression for $f_{\mu\nu}$, $\Sigma(p^2)$ still has the form
(\ref{6.4}) but it will be quadratically divergent and
\begin{equation}
\label{6.10}
 f^2 = f^2_{q\bar{q}} + \left(\frac{\alpha}{\pi}\right)^2 \Lambda^2
 \rightarrow \infty
\end{equation}
where $\Lambda$ is the ultraviolet cutoff. This means that Goldstone
particles exist in the anomalous case but they are decoupled from any
physical state. At the same time the Ward identity still makes sense
because the product $gf$ does not depend on $f$. Nevertheless, the
concrete form of the Ward identity will change. The reason for this
is again the Adler-Bell-Jackiw anomaly [3]. For the triangle diagram
\begin{equation}
\label{6.11}
\Delta_{\rho\sigma}^{\mu} = \begin{picture}(0,0)%
\epsfig{file=fig92.pstex}%
\end{picture}%
\setlength{\unitlength}{0.00083300in}%
\begingroup\makeatletter\ifx\SetFigFont\undefined
\def\x#1#2#3#4#5#6#7\relax{\def\x{#1#2#3#4#5#6}}%
\expandafter\x\fmtname xxxxxx\relax \def\y{splain}%
\ifx\x\y   
\gdef\SetFigFont#1#2#3{%
  \ifnum #1<17\tiny\else \ifnum #1<20\small\else
  \ifnum #1<24\normalsize\else \ifnum #1<29\large\else
  \ifnum #1<34\Large\else \ifnum #1<41\LARGE\else
     \huge\fi\fi\fi\fi\fi\fi
  \csname #3\endcsname}%
\else
\gdef\SetFigFont#1#2#3{\begingroup
  \count@#1\relax \ifnum 25<\count@\count@25\fi
  \def\x{\endgroup\@setsize\SetFigFont{#2pt}}%
  \expandafter\x
    \csname \romannumeral\the\count@ pt\expandafter\endcsname
    \csname @\romannumeral\the\count@ pt\endcsname
  \csname #3\endcsname}%
\fi
\fi\endgroup
\begin{picture}(1374,1369)(664,-991)
\put(1456, 89){\makebox(0,0)[lb]{\smash{\SetFigFont{11}{13.2}{rm}$\gamma_\mu \gamma_5$}}}
\put(1726,-961){\makebox(0,0)[lb]{\smash{\SetFigFont{11}{13.2}{rm}$\sigma$}}}
\put(901,-961){\makebox(0,0)[lb]{\smash{\SetFigFont{11}{13.2}{rm}$\rho$}}}
\end{picture}

\end{equation}
the replacement of the pseudovector coupling by the pseudoscalar
coupling gives an incorrect result: instead of the correct expression
\begin{equation}
\label{6.12}
p_{\mu}\Delta_{\rho\sigma}^{\mu} = \begin{picture}(0,0)%
\epsfig{file=fig93.pstex}%
\end{picture}%
\setlength{\unitlength}{0.00083300in}%
\begingroup\makeatletter\ifx\SetFigFont\undefined
\def\x#1#2#3#4#5#6#7\relax{\def\x{#1#2#3#4#5#6}}%
\expandafter\x\fmtname xxxxxx\relax \def\y{splain}%
\ifx\x\y   
\gdef\SetFigFont#1#2#3{%
  \ifnum #1<17\tiny\else \ifnum #1<20\small\else
  \ifnum #1<24\normalsize\else \ifnum #1<29\large\else
  \ifnum #1<34\Large\else \ifnum #1<41\LARGE\else
     \huge\fi\fi\fi\fi\fi\fi
  \csname #3\endcsname}%
\else
\gdef\SetFigFont#1#2#3{\begingroup
  \count@#1\relax \ifnum 25<\count@\count@25\fi
  \def\x{\endgroup\@setsize\SetFigFont{#2pt}}%
  \expandafter\x
    \csname \romannumeral\the\count@ pt\expandafter\endcsname
    \csname @\romannumeral\the\count@ pt\endcsname
  \csname #3\endcsname}%
\fi
\fi\endgroup
\begin{picture}(1224,1219)(739,-994)
\put(1456,-256){\makebox(0,0)[lb]{\smash{\SetFigFont{11}{13.2}{rm}$2 m \gamma_5$}}}
\put(1618,-961){\makebox(0,0)[lb]{\smash{\SetFigFont{11}{13.2}{rm}$k_1$}}}
\put(976,-961){\makebox(0,0)[lb]{\smash{\SetFigFont{11}{13.2}{rm}$k_2$}}}
\end{picture}
 
  + \frac{\alpha}{\pi} 
  \varepsilon_{\rho\sigma\delta\gamma}k_{2\delta}k_{1\rho}
\end{equation}
which was obtained in [3], we get just the first term. We can
try to write the Ward identity using (\ref{6.12}), but this turns
out not to be necessary. The reason is that in this approach
the axial current $\Gamma^{\mu}_{\rho\sigma}$ between gluonic states,
defined symbolically by the relation
\begin{equation}
\label{6.13}
\Gamma^{\mu}_{\rho\sigma} = \tilde{\Gamma}^{\mu}_{\rho\sigma} +
  \begin{picture}(0,0)%
\epsfig{file=fig94.pstex}%
\end{picture}%
\setlength{\unitlength}{0.00083300in}%
\begingroup\makeatletter\ifx\SetFigFont\undefined
\def\x#1#2#3#4#5#6#7\relax{\def\x{#1#2#3#4#5#6}}%
\expandafter\x\fmtname xxxxxx\relax \def\y{splain}%
\ifx\x\y   
\gdef\SetFigFont#1#2#3{%
  \ifnum #1<17\tiny\else \ifnum #1<20\small\else
  \ifnum #1<24\normalsize\else \ifnum #1<29\large\else
  \ifnum #1<34\Large\else \ifnum #1<41\LARGE\else
     \huge\fi\fi\fi\fi\fi\fi
  \csname #3\endcsname}%
\else
\gdef\SetFigFont#1#2#3{\begingroup
  \count@#1\relax \ifnum 25<\count@\count@25\fi
  \def\x{\endgroup\@setsize\SetFigFont{#2pt}}%
  \expandafter\x
    \csname \romannumeral\the\count@ pt\expandafter\endcsname
    \csname @\romannumeral\the\count@ pt\endcsname
  \csname #3\endcsname}%
\fi
\fi\endgroup
\begin{picture}(774,1183)(1114,-823)
\put(1576,-72){\makebox(0,0)[lb]{\smash{\SetFigFont{11}{13.2}{rm}$f p_\mu$}}}
\put(1335,-72){\makebox(0,0)[lb]{\smash{\SetFigFont{11}{13.2}{rm}$i$}}}
\put(1621,-526){\makebox(0,0)[lb]{\smash{\SetFigFont{11}{13.2}{rm}$g_{\rho \sigma}$}}}
\end{picture}

\end{equation}
(where the term $\tilde{\Gamma}^{\mu}_{\rho\sigma}$ is defined
diagrammatically and $g_{\rho\sigma}=\frac{p_{\mu}}{f}
\tilde{\Gamma}^{\mu}_{\rho\sigma}$) is just the transverse part of
$\tilde{\Gamma}^{\mu}_{\rho\sigma}$:
\begin{equation}
\label{6.14}
\Gamma^{\mu}_{\rho\sigma} = \tilde{\Gamma}^{\mu}_{\rho\sigma} -
 \frac{p_{\mu}p_{\nu}}{p^2}\tilde{\Gamma}^{\nu}_{\rho\sigma} .
\end{equation}
For the axial current between quark states $\tilde{\Gamma}^{\mu}$
can be written as
\begin{equation}
\label{6.15}
\tilde{\Gamma}^{\mu} = \tilde{\tilde{\Gamma}}^{\mu}(q_2,q_1) +
\tilde{\Gamma}^{\mu}_g(q_2,q_1).
\end{equation}
Here
\begin{equation}
\label{6.16}
\tilde{\tilde{\Gamma}} = \begin{picture}(0,0)%
\epsfig{file=fig95.pstex}%
\end{picture}%
\setlength{\unitlength}{0.00083300in}%
\begingroup\makeatletter\ifx\SetFigFont\undefined
\def\x#1#2#3#4#5#6#7\relax{\def\x{#1#2#3#4#5#6}}%
\expandafter\x\fmtname xxxxxx\relax \def\y{splain}%
\ifx\x\y   
\gdef\SetFigFont#1#2#3{%
  \ifnum #1<17\tiny\else \ifnum #1<20\small\else
  \ifnum #1<24\normalsize\else \ifnum #1<29\large\else
  \ifnum #1<34\Large\else \ifnum #1<41\LARGE\else
     \huge\fi\fi\fi\fi\fi\fi
  \csname #3\endcsname}%
\else
\gdef\SetFigFont#1#2#3{\begingroup
  \count@#1\relax \ifnum 25<\count@\count@25\fi
  \def\x{\endgroup\@setsize\SetFigFont{#2pt}}%
  \expandafter\x
    \csname \romannumeral\the\count@ pt\expandafter\endcsname
    \csname @\romannumeral\the\count@ pt\endcsname
  \csname #3\endcsname}%
\fi
\fi\endgroup
\begin{picture}(1224,846)(589,-595)
\end{picture}

\end{equation}
is the same set of diagrams as in the non-singlet case and
$\tilde{\Gamma}^{\mu}_g$ is the "axial current of gluons"
\begin{equation}
\label{6.17}
\tilde{\Gamma}^{\mu}_g = \begin{picture}(0,0)%
\epsfig{file=fig96.pstex}%
\end{picture}%
\setlength{\unitlength}{0.00083300in}%
\begingroup\makeatletter\ifx\SetFigFont\undefined
\def\x#1#2#3#4#5#6#7\relax{\def\x{#1#2#3#4#5#6}}%
\expandafter\x\fmtname xxxxxx\relax \def\y{splain}%
\ifx\x\y   
\gdef\SetFigFont#1#2#3{%
  \ifnum #1<17\tiny\else \ifnum #1<20\small\else
  \ifnum #1<24\normalsize\else \ifnum #1<29\large\else
  \ifnum #1<34\Large\else \ifnum #1<41\LARGE\else
     \huge\fi\fi\fi\fi\fi\fi
  \csname #3\endcsname}%
\else
\gdef\SetFigFont#1#2#3{\begingroup
  \count@#1\relax \ifnum 25<\count@\count@25\fi
  \def\x{\endgroup\@setsize\SetFigFont{#2pt}}%
  \expandafter\x
    \csname \romannumeral\the\count@ pt\expandafter\endcsname
    \csname @\romannumeral\the\count@ pt\endcsname
  \csname #3\endcsname}%
\fi
\fi\endgroup
\begin{picture}(924,921)(589,-520)
\put(1100,186){\makebox(0,0)[lb]{\smash{\SetFigFont{11}{13.2}{rm}$p$}}}
\end{picture}
 + \begin{picture}(0,0)%
\epsfig{file=fig97.pstex}%
\end{picture}%
\setlength{\unitlength}{0.00083300in}%
\begingroup\makeatletter\ifx\SetFigFont\undefined
\def\x#1#2#3#4#5#6#7\relax{\def\x{#1#2#3#4#5#6}}%
\expandafter\x\fmtname xxxxxx\relax \def\y{splain}%
\ifx\x\y   
\gdef\SetFigFont#1#2#3{%
  \ifnum #1<17\tiny\else \ifnum #1<20\small\else
  \ifnum #1<24\normalsize\else \ifnum #1<29\large\else
  \ifnum #1<34\Large\else \ifnum #1<41\LARGE\else
     \huge\fi\fi\fi\fi\fi\fi
  \csname #3\endcsname}%
\else
\gdef\SetFigFont#1#2#3{\begingroup
  \count@#1\relax \ifnum 25<\count@\count@25\fi
  \def\x{\endgroup\@setsize\SetFigFont{#2pt}}%
  \expandafter\x
    \csname \romannumeral\the\count@ pt\expandafter\endcsname
    \csname @\romannumeral\the\count@ pt\endcsname
  \csname #3\endcsname}%
\fi
\fi\endgroup
\begin{picture}(924,921)(589,-520)
\put(1100,186){\makebox(0,0)[lb]{\smash{\SetFigFont{11}{13.2}{rm}$p$}}}
\end{picture}
 
  + \cdots \hspace{1cm}.
\end{equation}
In the same way the Goldstone boson -- quark interaction can also be
divided into two parts. In the first part we can replace the
pseudovector coupling by the pseudoscalar coupling. The second part
is the longitudinal part of $\tilde{\Gamma}^{\mu}_g$. Consequently,
\begin{equation}
\label{6.18}
\Gamma^{\mu} = \tilde{\tilde{\Gamma}}^{\mu} + \begin{picture}(0,0)%
\epsfig{file=fig98.pstex}%
\end{picture}%
\setlength{\unitlength}{0.00083300in}%
\begingroup\makeatletter\ifx\SetFigFont\undefined
\def\x#1#2#3#4#5#6#7\relax{\def\x{#1#2#3#4#5#6}}%
\expandafter\x\fmtname xxxxxx\relax \def\y{splain}%
\ifx\x\y   
\gdef\SetFigFont#1#2#3{%
  \ifnum #1<17\tiny\else \ifnum #1<20\small\else
  \ifnum #1<24\normalsize\else \ifnum #1<29\large\else
  \ifnum #1<34\Large\else \ifnum #1<41\LARGE\else
     \huge\fi\fi\fi\fi\fi\fi
  \csname #3\endcsname}%
\else
\gdef\SetFigFont#1#2#3{\begingroup
  \count@#1\relax \ifnum 25<\count@\count@25\fi
  \def\x{\endgroup\@setsize\SetFigFont{#2pt}}%
  \expandafter\x
    \csname \romannumeral\the\count@ pt\expandafter\endcsname
    \csname @\romannumeral\the\count@ pt\endcsname
  \csname #3\endcsname}%
\fi
\fi\endgroup
\begin{picture}(774,1183)(1114,-823)
\put(1576,-72){\makebox(0,0)[lb]{\smash{\SetFigFont{11}{13.2}{rm}$f p_\mu$}}}
\put(1335,-72){\makebox(0,0)[lb]{\smash{\SetFigFont{11}{13.2}{rm}$i$}}}
\put(1464,-714){\makebox(0,0)[lb]{\smash{\SetFigFont{11}{13.2}{rm}$g$}}}
\end{picture}
 +
\tilde{\Gamma}^{\mu} - \frac{p_{\mu}p_{\nu}}{p^2}\tilde{\Gamma}^{\nu},
\end{equation}
and the Ward identity can be written in the form
\begin{equation}
\label{6.19}
p_{\mu}\tilde{\tilde{\Gamma}}^{\mu} - ifg = \gamma_5 G^{-1}(q_1) +
 G^{-1}(q_2)\gamma_5 .
\end{equation}
This expression enables us to answer the question, what happens with
particles like $\eta'$ which were Goldstone states if we would not
take into account that they can decay into two gluons.
Let us consider the contribution of the massive pseudoscalar flavour
singlet particle $\eta'$ to the Ward identity (\ref{6.19}).
$p_{\mu}\tilde{\tilde{\Gamma^{\mu}}}$ has a pole corresponding to
$\eta'$:
\begin{equation}
\label{6.20}
p_{\mu}\tilde{\tilde{\Gamma}}^{\mu} = \begin{picture}(0,0)%
\epsfig{file=fig99.pstex}%
\end{picture}%
\setlength{\unitlength}{0.00083300in}%
\begingroup\makeatletter\ifx\SetFigFont\undefined
\def\x#1#2#3#4#5#6#7\relax{\def\x{#1#2#3#4#5#6}}%
\expandafter\x\fmtname xxxxxx\relax \def\y{splain}%
\ifx\x\y   
\gdef\SetFigFont#1#2#3{%
  \ifnum #1<17\tiny\else \ifnum #1<20\small\else
  \ifnum #1<24\normalsize\else \ifnum #1<29\large\else
  \ifnum #1<34\Large\else \ifnum #1<41\LARGE\else
     \huge\fi\fi\fi\fi\fi\fi
  \csname #3\endcsname}%
\else
\gdef\SetFigFont#1#2#3{\begingroup
  \count@#1\relax \ifnum 25<\count@\count@25\fi
  \def\x{\endgroup\@setsize\SetFigFont{#2pt}}%
  \expandafter\x
    \csname \romannumeral\the\count@ pt\expandafter\endcsname
    \csname @\romannumeral\the\count@ pt\endcsname
  \csname #3\endcsname}%
\fi
\fi\endgroup
\begin{picture}(774,1532)(1114,-1273)
\put(1434,-1164){\makebox(0,0)[lb]{\smash{\SetFigFont{11}{13.2}{rm}$g_{\eta'}$}}}
\put(1572,  3){\makebox(0,0)[lb]{\smash{\SetFigFont{11}{13.2}{rm}$p$}}}
\put(1594,-331){\makebox(0,0)[lb]{\smash{\SetFigFont{11}{13.2}{rm}$g_{\eta'}$}}}
\end{picture}
  =
p^2 f_{\eta'}\frac{1}{\mu^2 - p^2}g_{\eta'}\,.
\end{equation}
The Yukawa coupling to the Goldstone state has also a pole:
\begin{equation}
\label{6.21}
-fg = \begin{picture}(0,0)%
\epsfig{file=fig100.pstex}%
\end{picture}%
\setlength{\unitlength}{0.00083300in}%
\begingroup\makeatletter\ifx\SetFigFont\undefined
\def\x#1#2#3#4#5#6#7\relax{\def\x{#1#2#3#4#5#6}}%
\expandafter\x\fmtname xxxxxx\relax \def\y{splain}%
\ifx\x\y   
\gdef\SetFigFont#1#2#3{%
  \ifnum #1<17\tiny\else \ifnum #1<20\small\else
  \ifnum #1<24\normalsize\else \ifnum #1<29\large\else
  \ifnum #1<34\Large\else \ifnum #1<41\LARGE\else
     \huge\fi\fi\fi\fi\fi\fi
  \csname #3\endcsname}%
\else
\gdef\SetFigFont#1#2#3{\begingroup
  \count@#1\relax \ifnum 25<\count@\count@25\fi
  \def\x{\endgroup\@setsize\SetFigFont{#2pt}}%
  \expandafter\x
    \csname \romannumeral\the\count@ pt\expandafter\endcsname
    \csname @\romannumeral\the\count@ pt\endcsname
  \csname #3\endcsname}%
\fi
\fi\endgroup
\begin{picture}(624,1047)(1189,-942)
\put(1594,-331){\makebox(0,0)[lb]{\smash{\SetFigFont{11}{13.2}{rm}$g_{\eta'}$}}}
\put(1434,-909){\makebox(0,0)[lb]{\smash{\SetFigFont{11}{13.2}{rm}$g_{\eta'}$}}}
\put(1456,-27){\makebox(0,0)[lb]{\smash{\SetFigFont{11}{13.2}{rm}$\{\gamma_5,G^{-1}\}$}}}
\end{picture}
  
    = \{\gamma_5,G^{-1}(q)\} \begin{picture}(0,0)%
\epsfig{file=fig101.pstex}%
\end{picture}%
\setlength{\unitlength}{0.00083300in}%
\begingroup\makeatletter\ifx\SetFigFont\undefined
\def\x#1#2#3#4#5#6#7\relax{\def\x{#1#2#3#4#5#6}}%
\expandafter\x\fmtname xxxxxx\relax \def\y{splain}%
\ifx\x\y   
\gdef\SetFigFont#1#2#3{%
  \ifnum #1<17\tiny\else \ifnum #1<20\small\else
  \ifnum #1<24\normalsize\else \ifnum #1<29\large\else
  \ifnum #1<34\Large\else \ifnum #1<41\LARGE\else
     \huge\fi\fi\fi\fi\fi\fi
  \csname #3\endcsname}%
\else
\gdef\SetFigFont#1#2#3{\begingroup
  \count@#1\relax \ifnum 25<\count@\count@25\fi
  \def\x{\endgroup\@setsize\SetFigFont{#2pt}}%
  \expandafter\x
    \csname \romannumeral\the\count@ pt\expandafter\endcsname
    \csname @\romannumeral\the\count@ pt\endcsname
  \csname #3\endcsname}%
\fi
\fi\endgroup
\begin{picture}(541,169)(893,-145)
\end{picture}
 g_{\eta'}\;
 \frac{1}{\mu^2-p^2}g_{\eta'} .
\end{equation}
The right-hand side of (\ref{6.19}) has, however, no poles. This
condition can be satisfied if
\begin{equation}
\label{6.22}
\mu^2 f_{\eta'} = 
  \{\gamma_5,G^{-1}(q)\}\;\begin{picture}(0,0)%
\epsfig{file=fig101.pstex}%
\end{picture}%
\setlength{\unitlength}{0.00083300in}%
\begingroup\makeatletter\ifx\SetFigFont\undefined
\def\x#1#2#3#4#5#6#7\relax{\def\x{#1#2#3#4#5#6}}%
\expandafter\x\fmtname xxxxxx\relax \def\y{splain}%
\ifx\x\y   
\gdef\SetFigFont#1#2#3{%
  \ifnum #1<17\tiny\else \ifnum #1<20\small\else
  \ifnum #1<24\normalsize\else \ifnum #1<29\large\else
  \ifnum #1<34\Large\else \ifnum #1<41\LARGE\else
     \huge\fi\fi\fi\fi\fi\fi
  \csname #3\endcsname}%
\else
\gdef\SetFigFont#1#2#3{\begingroup
  \count@#1\relax \ifnum 25<\count@\count@25\fi
  \def\x{\endgroup\@setsize\SetFigFont{#2pt}}%
  \expandafter\x
    \csname \romannumeral\the\count@ pt\expandafter\endcsname
    \csname @\romannumeral\the\count@ pt\endcsname
  \csname #3\endcsname}%
\fi
\fi\endgroup
\begin{picture}(541,169)(893,-145)
\end{picture}
 g_{\eta'}\,.
\end{equation}
The same Ward identity (\ref{6.19}) gives at $p^2=0$
\begin{equation}
\label{6.23}
f_{\eta'}g_{\eta'} = \{i\gamma_5,G^{-1}(q)\}
\end{equation}
from what it follows that $\mu^2$ for $\eta'$ is equal
\begin{equation}
\label{6.24}
\mu_{\eta'}^2 = \frac{\{\gamma_5,G^{-1}(q)\} \begin{picture}(0,0)%
\epsfig{file=fig102.pstex}%
\end{picture}%
\setlength{\unitlength}{0.00083300in}%
\begingroup\makeatletter\ifx\SetFigFont\undefined
\def\x#1#2#3#4#5#6#7\relax{\def\x{#1#2#3#4#5#6}}%
\expandafter\x\fmtname xxxxxx\relax \def\y{splain}%
\ifx\x\y   
\gdef\SetFigFont#1#2#3{%
  \ifnum #1<17\tiny\else \ifnum #1<20\small\else
  \ifnum #1<24\normalsize\else \ifnum #1<29\large\else
  \ifnum #1<34\Large\else \ifnum #1<41\LARGE\else
     \huge\fi\fi\fi\fi\fi\fi
  \csname #3\endcsname}%
\else
\gdef\SetFigFont#1#2#3{\begingroup
  \count@#1\relax \ifnum 25<\count@\count@25\fi
  \def\x{\endgroup\@setsize\SetFigFont{#2pt}}%
  \expandafter\x
    \csname \romannumeral\the\count@ pt\expandafter\endcsname
    \csname @\romannumeral\the\count@ pt\endcsname
  \csname #3\endcsname}%
\fi
\fi\endgroup
\begin{picture}(616,251)(893,-486)
\end{picture}

 \{\gamma_5,G^{-1}(q)\}}{f_{\eta'}^2}.
\end{equation}
This means that $\eta'$ acquired a mass due to the transition into
a Goldstone boson which itself is decoupled.
It is interesting to notice that at relatively small $\mu^2$ when
the comparison between Ward identities for different values of
$p^2$ ($p^2=\mu^2$, $p^2=0$) makes sense, (\ref{6.24}) gives us the
same result as the equation (\ref{5.9}) (without the additional
point-like term which is present in the $\pi$-meson case (\ref{5.21})).
Indeed, in (\ref{5.9}) we can write
\[ g\begin{picture}(0,0)%
\epsfig{file=fig103.pstex}%
\end{picture}%
\setlength{\unitlength}{0.00083300in}%
\begingroup\makeatletter\ifx\SetFigFont\undefined
\def\x#1#2#3#4#5#6#7\relax{\def\x{#1#2#3#4#5#6}}%
\expandafter\x\fmtname xxxxxx\relax \def\y{splain}%
\ifx\x\y   
\gdef\SetFigFont#1#2#3{%
  \ifnum #1<17\tiny\else \ifnum #1<20\small\else
  \ifnum #1<24\normalsize\else \ifnum #1<29\large\else
  \ifnum #1<34\Large\else \ifnum #1<41\LARGE\else
     \huge\fi\fi\fi\fi\fi\fi
  \csname #3\endcsname}%
\else
\gdef\SetFigFont#1#2#3{\begingroup
  \count@#1\relax \ifnum 25<\count@\count@25\fi
  \def\x{\endgroup\@setsize\SetFigFont{#2pt}}%
  \expandafter\x
    \csname \romannumeral\the\count@ pt\expandafter\endcsname
    \csname @\romannumeral\the\count@ pt\endcsname
  \csname #3\endcsname}%
\fi
\fi\endgroup
\begin{picture}(616,620)(893,-662)
\put(1092,-174){\makebox(0,0)[lb]{\smash{\SetFigFont{11}{13.2}{rm}$\Sigma(p)$}}}
\end{picture}
g = \, \begin{picture}(0,0)%
\epsfig{file=fig104.pstex}%
\end{picture}%
\setlength{\unitlength}{0.00083300in}%
\begingroup\makeatletter\ifx\SetFigFont\undefined
\def\x#1#2#3#4#5#6#7\relax{\def\x{#1#2#3#4#5#6}}%
\expandafter\x\fmtname xxxxxx\relax \def\y{splain}%
\ifx\x\y   
\gdef\SetFigFont#1#2#3{%
  \ifnum #1<17\tiny\else \ifnum #1<20\small\else
  \ifnum #1<24\normalsize\else \ifnum #1<29\large\else
  \ifnum #1<34\Large\else \ifnum #1<41\LARGE\else
     \huge\fi\fi\fi\fi\fi\fi
  \csname #3\endcsname}%
\else
\gdef\SetFigFont#1#2#3{\begingroup
  \count@#1\relax \ifnum 25<\count@\count@25\fi
  \def\x{\endgroup\@setsize\SetFigFont{#2pt}}%
  \expandafter\x
    \csname \romannumeral\the\count@ pt\expandafter\endcsname
    \csname @\romannumeral\the\count@ pt\endcsname
  \csname #3\endcsname}%
\fi
\fi\endgroup
\begin{picture}(1236,318)(359,-520)
\end{picture}
 \,
  + \begin{picture}(0,0)%
\epsfig{file=fig105.pstex}%
\end{picture}%
\setlength{\unitlength}{0.00083300in}%
\begingroup\makeatletter\ifx\SetFigFont\undefined
\def\x#1#2#3#4#5#6#7\relax{\def\x{#1#2#3#4#5#6}}%
\expandafter\x\fmtname xxxxxx\relax \def\y{splain}%
\ifx\x\y   
\gdef\SetFigFont#1#2#3{%
  \ifnum #1<17\tiny\else \ifnum #1<20\small\else
  \ifnum #1<24\normalsize\else \ifnum #1<29\large\else
  \ifnum #1<34\Large\else \ifnum #1<41\LARGE\else
     \huge\fi\fi\fi\fi\fi\fi
  \csname #3\endcsname}%
\else
\gdef\SetFigFont#1#2#3{\begingroup
  \count@#1\relax \ifnum 25<\count@\count@25\fi
  \def\x{\endgroup\@setsize\SetFigFont{#2pt}}%
  \expandafter\x
    \csname \romannumeral\the\count@ pt\expandafter\endcsname
    \csname @\romannumeral\the\count@ pt\endcsname
  \csname #3\endcsname}%
\fi
\fi\endgroup
\begin{picture}(694,512)(553,-415)
\end{picture}
 - \begin{picture}(0,0)%
\epsfig{file=fig105.pstex}%
\end{picture}%
\setlength{\unitlength}{0.00083300in}%
\begingroup\makeatletter\ifx\SetFigFont\undefined
\def\x#1#2#3#4#5#6#7\relax{\def\x{#1#2#3#4#5#6}}%
\expandafter\x\fmtname xxxxxx\relax \def\y{splain}%
\ifx\x\y   
\gdef\SetFigFont#1#2#3{%
  \ifnum #1<17\tiny\else \ifnum #1<20\small\else
  \ifnum #1<24\normalsize\else \ifnum #1<29\large\else
  \ifnum #1<34\Large\else \ifnum #1<41\LARGE\else
     \huge\fi\fi\fi\fi\fi\fi
  \csname #3\endcsname}%
\else
\gdef\SetFigFont#1#2#3{\begingroup
  \count@#1\relax \ifnum 25<\count@\count@25\fi
  \def\x{\endgroup\@setsize\SetFigFont{#2pt}}%
  \expandafter\x
    \csname \romannumeral\the\count@ pt\expandafter\endcsname
    \csname @\romannumeral\the\count@ pt\endcsname
  \csname #3\endcsname}%
\fi
\fi\endgroup
\begin{picture}(694,512)(553,-415)
\end{picture}
 = 0\, . \]
The first two terms are equal to $p^2$ (as in (\ref{5.21})) and therefore
\begin{equation}
\label{6.25}
p^2 = \begin{picture}(0,0)%
\epsfig{file=fig105.pstex}%
\end{picture}%
\setlength{\unitlength}{0.00083300in}%
\begingroup\makeatletter\ifx\SetFigFont\undefined
\def\x#1#2#3#4#5#6#7\relax{\def\x{#1#2#3#4#5#6}}%
\expandafter\x\fmtname xxxxxx\relax \def\y{splain}%
\ifx\x\y   
\gdef\SetFigFont#1#2#3{%
  \ifnum #1<17\tiny\else \ifnum #1<20\small\else
  \ifnum #1<24\normalsize\else \ifnum #1<29\large\else
  \ifnum #1<34\Large\else \ifnum #1<41\LARGE\else
     \huge\fi\fi\fi\fi\fi\fi
  \csname #3\endcsname}%
\else
\gdef\SetFigFont#1#2#3{\begingroup
  \count@#1\relax \ifnum 25<\count@\count@25\fi
  \def\x{\endgroup\@setsize\SetFigFont{#2pt}}%
  \expandafter\x
    \csname \romannumeral\the\count@ pt\expandafter\endcsname
    \csname @\romannumeral\the\count@ pt\endcsname
  \csname #3\endcsname}%
\fi
\fi\endgroup
\begin{picture}(694,512)(553,-415)
\end{picture}
 \approx -g\begin{picture}(0,0)%
\epsfig{file=fig101.pstex}%
\end{picture}%
\setlength{\unitlength}{0.00083300in}%
\begingroup\makeatletter\ifx\SetFigFont\undefined
\def\x#1#2#3#4#5#6#7\relax{\def\x{#1#2#3#4#5#6}}%
\expandafter\x\fmtname xxxxxx\relax \def\y{splain}%
\ifx\x\y   
\gdef\SetFigFont#1#2#3{%
  \ifnum #1<17\tiny\else \ifnum #1<20\small\else
  \ifnum #1<24\normalsize\else \ifnum #1<29\large\else
  \ifnum #1<34\Large\else \ifnum #1<41\LARGE\else
     \huge\fi\fi\fi\fi\fi\fi
  \csname #3\endcsname}%
\else
\gdef\SetFigFont#1#2#3{\begingroup
  \count@#1\relax \ifnum 25<\count@\count@25\fi
  \def\x{\endgroup\@setsize\SetFigFont{#2pt}}%
  \expandafter\x
    \csname \romannumeral\the\count@ pt\expandafter\endcsname
    \csname @\romannumeral\the\count@ pt\endcsname
  \csname #3\endcsname}%
\fi
\fi\endgroup
\begin{picture}(541,169)(893,-145)
\end{picture}
g|_{p=0} \,,
\end{equation}
in agreement with (\ref{6.24}). We see that the subtraction term in
the $\pi$-meson self energy, reflecting the quasi-local structure
of the pion, disappears in the case of $\eta'$. In this sense $\eta'$
is a normal bound state without a point-like structure.

The approach we presented here for the resolution of the $U(1)$-problem
is technically very close to the approach developed by Veneziano [4]
but the underlying physics differs essentially.
In Veneziano's approach big long-range fluctuations are responsible
for the $\eta'$ mass. In our approach $\eta'$ is a normal $q\bar{q}$
bound state with no local structure which would be responsible for its
small mass if it were a Goldstone state. This local structure is
destroyed by the decay on hard gluons.

\section{QCD with massive quarks}\label{VII}

We have discussed in detail an asymptotically free theory with
massless fermions. We came to the conclusion that in order to
obtain the correct spectrum of Goldstone particles, axial
current conservation has to be imposed on the theory. QCD, however,
contains massive quarks and the same spectrum of massive
quasi-Goldstone particles (the pseudo-scalar octet) as the theory with
massless quarks. The problem is, how to impose the condition of
axial current conservation on a theory which obviously does not
conserve the axial current.

In general, I don't know how to do this. For our real world, however,
there is a natural possibility to solve the problem.

In the real world QCD is part of the standard model describing strong,
electromagnetic and weak interactions. In the standard model all
particles are supposed to be intrinsically massless and their masses
appear as the result of symmetry breaking due to some kind of Higgs
mechanism with or without elementary Higgs particles. The possibility of
such a mechanism is guaranteed by the conservation of the left-handed
$SU(2)$ current $j_{\mu}^a $. For the matrix element $\Gamma^{\mu}_a$ of
this current between any two quarks with momenta $q_2$, $q_1$ we have
the Ward identity
\begin{equation}
\label{7.1}
p_{\mu}\Gamma^{\mu}_a(q_2,q_1) = \frac{1}{2}\tau_a\frac{1}{2}(1-\gamma_5)
 G^{-1}(q_1) - G^{-1}(q_2)\frac{1}{2}\tau_a\frac{1}{2}(1+\gamma_5).
\end{equation}
In the case of massive fermions $p_{\mu}\Gamma^{\mu}_a$ contains
the contribution of three Goldstone bosons responsible for the masses
of $W^{\pm}$ and $Z^0$
\begin{equation}
\label{7.2}
p_{\mu}\Gamma^{\mu}_a = p_{\mu}\tilde{\Gamma}^{\mu}_a - fg.
\end{equation}
For large $q^2$ values $G^{-1}$ contains a massive term $Z^{-1}m_0$.
Hence, at large $q^2$ we have
\begin{equation}
\label{7.3}
(fg)_0 = - \frac{1}{4}\{\tau_a\gamma_5,m_0 Z^{-1}\} + \frac{1}{4}[\tau_a,
m_0 Z^{-1}] .
\end{equation}
At small $q^2$ around the QCD scale $fg$ is defined by the total
quark mass $m(q)$ which is for the light quark much larger than $m_0$.
Using the relations (\ref{7.1}) - (\ref{7.3}) we can calculate the
the masses of the pseudoscalar octet in the same way as we did for
$\eta'$. Suppose there is a bound state of light quarks which is a
pseudoscalar particle (the $\pi$-meson) with a finite mass $\mu$. The
pole corresponding to this particle will contribute to both terms
in (\ref{7.2}). With this pole contribution, $p_{\mu}\tilde{\Gamma}^{\mu}
_a$ and $fg$ can be written in the form
\begin{equation}
\label{7.4}
p_{\mu}\tilde{\Gamma}^{\mu}_a = \begin{picture}(0,0)%
\epsfig{file=fig106.pstex}%
\end{picture}%
\setlength{\unitlength}{0.00083300in}%
\begingroup\makeatletter\ifx\SetFigFont\undefined
\def\x#1#2#3#4#5#6#7\relax{\def\x{#1#2#3#4#5#6}}%
\expandafter\x\fmtname xxxxxx\relax \def\y{splain}%
\ifx\x\y   
\gdef\SetFigFont#1#2#3{%
  \ifnum #1<17\tiny\else \ifnum #1<20\small\else
  \ifnum #1<24\normalsize\else \ifnum #1<29\large\else
  \ifnum #1<34\Large\else \ifnum #1<41\LARGE\else
     \huge\fi\fi\fi\fi\fi\fi
  \csname #3\endcsname}%
\else
\gdef\SetFigFont#1#2#3{\begingroup
  \count@#1\relax \ifnum 25<\count@\count@25\fi
  \def\x{\endgroup\@setsize\SetFigFont{#2pt}}%
  \expandafter\x
    \csname \romannumeral\the\count@ pt\expandafter\endcsname
    \csname @\romannumeral\the\count@ pt\endcsname
  \csname #3\endcsname}%
\fi
\fi\endgroup
\begin{picture}(774,1569)(1114,-1273)
\put(1434,-1164){\makebox(0,0)[lb]{\smash{\SetFigFont{11}{13.2}{rm}$g_\pi$}}}
\put(1594,-331){\makebox(0,0)[lb]{\smash{\SetFigFont{11}{13.2}{rm}$g_\pi$}}}
\put(1594,164){\makebox(0,0)[lb]{\smash{\SetFigFont{11}{13.2}{rm}$p_\mu \Gamma^\mu$}}}
\end{picture}
 
  + (p_{\mu}\tilde{\Gamma}^{\mu}_a)_{n,p} 
  = \frac{f_{\pi}p^2\frac{\tau_a}{2}}{\mu^2-p^2} +
  (p_{\mu}\tilde{\Gamma}^{\mu}_a)_{n,p} ,
\end{equation}
\begin{equation}
\label{7.5}
fg =  \begin{picture}(0,0)%
\epsfig{file=fig107.pstex}%
\end{picture}%
\setlength{\unitlength}{0.00083300in}%
\begingroup\makeatletter\ifx\SetFigFont\undefined
\def\x#1#2#3#4#5#6#7\relax{\def\x{#1#2#3#4#5#6}}%
\expandafter\x\fmtname xxxxxx\relax \def\y{splain}%
\ifx\x\y   
\gdef\SetFigFont#1#2#3{%
  \ifnum #1<17\tiny\else \ifnum #1<20\small\else
  \ifnum #1<24\normalsize\else \ifnum #1<29\large\else
  \ifnum #1<34\Large\else \ifnum #1<41\LARGE\else
     \huge\fi\fi\fi\fi\fi\fi
  \csname #3\endcsname}%
\else
\gdef\SetFigFont#1#2#3{\begingroup
  \count@#1\relax \ifnum 25<\count@\count@25\fi
  \def\x{\endgroup\@setsize\SetFigFont{#2pt}}%
  \expandafter\x
    \csname \romannumeral\the\count@ pt\expandafter\endcsname
    \csname @\romannumeral\the\count@ pt\endcsname
  \csname #3\endcsname}%
\fi
\fi\endgroup
\begin{picture}(774,1523)(1114,-1134)
\put(1434,-1025){\makebox(0,0)[lb]{\smash{\SetFigFont{11}{13.2}{rm}$g_\pi$}}}
\put(1362,257){\makebox(0,0)[lb]{\smash{\SetFigFont{11}{13.2}{rm}$(fg)_0$}}}
\put(1594,-331){\makebox(0,0)[lb]{\smash{\SetFigFont{11}{13.2}{rm}$g_\pi$}}}
\end{picture}
 + (fg)_{n,p} = 
 \frac{(fg)_0\begin{picture}(0,0)%
\epsfig{file=fig101.pstex}%
\end{picture}%
\setlength{\unitlength}{0.00083300in}%
\begingroup\makeatletter\ifx\SetFigFont\undefined
\def\x#1#2#3#4#5#6#7\relax{\def\x{#1#2#3#4#5#6}}%
\expandafter\x\fmtname xxxxxx\relax \def\y{splain}%
\ifx\x\y   
\gdef\SetFigFont#1#2#3{%
  \ifnum #1<17\tiny\else \ifnum #1<20\small\else
  \ifnum #1<24\normalsize\else \ifnum #1<29\large\else
  \ifnum #1<34\Large\else \ifnum #1<41\LARGE\else
     \huge\fi\fi\fi\fi\fi\fi
  \csname #3\endcsname}%
\else
\gdef\SetFigFont#1#2#3{\begingroup
  \count@#1\relax \ifnum 25<\count@\count@25\fi
  \def\x{\endgroup\@setsize\SetFigFont{#2pt}}%
  \expandafter\x
    \csname \romannumeral\the\count@ pt\expandafter\endcsname
    \csname @\romannumeral\the\count@ pt\endcsname
  \csname #3\endcsname}%
\fi
\fi\endgroup
\begin{picture}(541,169)(893,-145)
\end{picture}
g_{\pi}}{\mu^2-p^2} + (fg)_{n,p} .
\end{equation}
As in the case of $\eta'$, the condition for the pole cancellation
gives us the value of $\mu^2$:
\begin{equation}
\label{7.6}
 \mu^2 = \frac{1}{f_{\pi}} (fg)_0\begin{picture}(0,0)%
\epsfig{file=fig108.pstex}%
\end{picture}%
\setlength{\unitlength}{0.00083300in}%
\begingroup\makeatletter\ifx\SetFigFont\undefined
\def\x#1#2#3#4#5#6#7\relax{\def\x{#1#2#3#4#5#6}}%
\expandafter\x\fmtname xxxxxx\relax \def\y{splain}%
\ifx\x\y   
\gdef\SetFigFont#1#2#3{%
  \ifnum #1<17\tiny\else \ifnum #1<20\small\else
  \ifnum #1<24\normalsize\else \ifnum #1<29\large\else
  \ifnum #1<34\Large\else \ifnum #1<41\LARGE\else
     \huge\fi\fi\fi\fi\fi\fi
  \csname #3\endcsname}%
\else
\gdef\SetFigFont#1#2#3{\begingroup
  \count@#1\relax \ifnum 25<\count@\count@25\fi
  \def\x{\endgroup\@setsize\SetFigFont{#2pt}}%
  \expandafter\x
    \csname \romannumeral\the\count@ pt\expandafter\endcsname
    \csname @\romannumeral\the\count@ pt\endcsname
  \csname #3\endcsname}%
\fi
\fi\endgroup
\begin{picture}(541,229)(893,-174)
\end{picture}
g_{\pi} .
\end{equation}
At $p^2=0$ we have
\begin{equation}
\label{7.7}
f_{\pi}g_{\pi} + (fg)_{n,p} = \frac{1}{4}\left\{\tau_a\gamma_5,\hat{m}(q)
 Z^{-1}\right\} + \frac{1}{4}\left[\tau_a,\hat{m}Z^{-1}\right].
\end{equation}
The Yukawa coupling $g_{\pi}$ of the bound state has to decrease at
large $q^2$. Therefore we have to identify $(fg)_{n,p}$ with $(fg)_0$.
Then
\begin{equation}
\label{7.8}
f_{\pi}g_{\pi} = \frac{1}{2}\tau_a\gamma_5 m_s(q)Z^{-1}
\end{equation}
where $m_s(q)$ is the isotopically invariant part of the quark mass
term behaving at $q \rightarrow \infty$ as $\nu^3/q^2$. Hence,
$(fg)_0 \sim m_0$ is defined by weak interactions with Goldstone
states, and $f_{\pi}g_{\pi} \sim m_s$ is determined by pion exchange.
The conclusion of these considerations is that in the case of
massive quarks the conservation of the left-handed $SU(2)$ current
can play the same role for the calculation of the coupling and the
masses of the flavour non-singlet pseudoscalar particles as in the
massless case the conservation of the axial current. The result is that
the masses become different from zero while the Yukawa coupling remains
the same at least for small $m_0$ values.

The expression (\ref{7.6}) for $\mu^2$ can be also obtained without
discussing the current conservation, just by calculating $\Sigma(p)$
as it is defined by (\ref{5.21}) for the quark mass $m_s+m_0$, keeping
only the term linear in $m_0$. It is in agreement with the old idea
that $m_0$ influences the mass of the Goldstone state but not its
wave function.

Unfortunately, if we calculate $\mu^2$ by using (\ref{7.6}) or
(\ref{5.21}), we get a logarithmic divergence. For the $\pi$-meson
mass the main divergent part equals
\begin{equation}
\label{7.9}
m_{\pi}^2 = \frac{3}{4\pi^2}\frac{1}{f_{\pi}^2} \int_{\nu_0^2}^{\infty}
 \frac{d(q^2)}{q^2} m_0(q)\nu^3(q)
\end{equation}
(here $\nu_0$ is of the order of $\lambda_{QCD}$).

In the standard model the behaviour of $m_0$ and $\nu^3$ can be
calculated in a fantastically large region of $q^2$: from 1 GeV up to
the scale where one of the couplings of the standard model ($g_1$ --
the $U(1)$-coupling, $h$ -- the Yukawa-coupling or $\lambda$ -- the
coupling of the self-interaction of the Higgs particles) becomes of
the order of unity. In the reasonable case when the $U(1)$-coupling
$g_1$ is the first to become unity this scale is equal to
$\Lambda = 10^{38}$ GeV.

At $q^2>\Lambda^2$ the behaviour of $m_0$ and $\nu^3$
is unknown. Because of this, $m_{\pi}$ is not calculable in principle
and it has to be considered as an arbitrary parameter. If we, however,
assume that at $q^2$ larger than $\Lambda$ $m_0(q)$ and $\nu^3(q)$
become equal to zero, we will be able to calculate $m_{\pi}$ and,
surprisingly, this calculation gives a reasonable value for $m_{\pi}$
with $\Lambda \approx 10^{38}$ GeV [2]. The expression (\ref{7.9})
for $m_{\pi}^2$ is also in agreement with the naive expression
\[ m_{\pi}^2 = \frac{2m_0}{f_{\pi}^2} \langle \bar{\Psi}\Psi \rangle \]
with the important difference that now $\langle \bar{\Psi}\Psi \rangle$
is defined not only by strong but also by weak interactions and it
goes to infinity if the weak interaction is removed.

\sloppy
\section{The pion contribution to the equation for light quark
Green's functions}\label{VIII}
\fussy

From the previous discussion it is obvious that the small mass pion
contribution has to be included in the equation for the light quark
Green's function. Fortunately this is very easy to do. Having in mind
that now the diagrams contain not only the gluon contribution
but also the emission and the absorption of pions we will find that
as before, the main contribution to $\partial^2 G^{-1}$ comes from
the simplest diagram \begin{picture}(0,0)%
\epsfig{file=fig109.pstex}%
\end{picture}%
\setlength{\unitlength}{0.00083300in}%
\begingroup\makeatletter\ifx\SetFigFont\undefined
\def\x#1#2#3#4#5#6#7\relax{\def\x{#1#2#3#4#5#6}}%
\expandafter\x\fmtname xxxxxx\relax \def\y{splain}%
\ifx\x\y   
\gdef\SetFigFont#1#2#3{%
  \ifnum #1<17\tiny\else \ifnum #1<20\small\else
  \ifnum #1<24\normalsize\else \ifnum #1<29\large\else
  \ifnum #1<34\Large\else \ifnum #1<41\LARGE\else
     \huge\fi\fi\fi\fi\fi\fi
  \csname #3\endcsname}%
\else
\gdef\SetFigFont#1#2#3{\begingroup
  \count@#1\relax \ifnum 25<\count@\count@25\fi
  \def\x{\endgroup\@setsize\SetFigFont{#2pt}}%
  \expandafter\x
    \csname \romannumeral\the\count@ pt\expandafter\endcsname
    \csname @\romannumeral\the\count@ pt\endcsname
  \csname #3\endcsname}%
\fi
\fi\endgroup
\begin{picture}(624,198)(589,-673)
\end{picture}
 
with the coupling $\{i\gamma_5,G^{-1}\}$
at zero momentum instead of the gluon coupling $\partial_{\mu}G^{-1}$.
It leads to the following equation for the Green's function:
\begin{equation}
\label{8.1}
\partial^2G^{-1}(q) = g\partial_{\mu}G^{-1}(q)G(q)\partial_{\mu}G^{-1}(q) -
\{i\gamma_5,G^{-1}\}G(q)\{i\gamma_5,G^{-1}\}\frac{3}{16\pi^2 f_{\pi}^2}
\end{equation}
The equation for bound states (\ref{2.6}) has also to be changed.
The correction comes from the diagrams
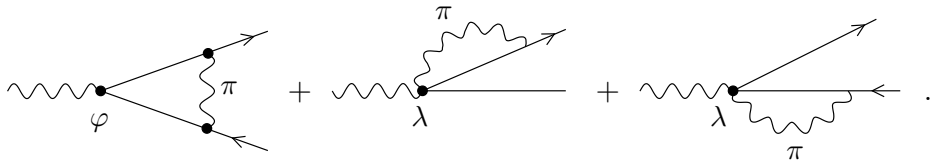
\[ \begin{picture}(0,0)%
\epsfig{file=fig110.pstex}%
\end{picture}%
\setlength{\unitlength}{0.00083300in}%
\begingroup\makeatletter\ifx\SetFigFont\undefined
\def\x#1#2#3#4#5#6#7\relax{\def\x{#1#2#3#4#5#6}}%
\expandafter\x\fmtname xxxxxx\relax \def\y{splain}%
\ifx\x\y   
\gdef\SetFigFont#1#2#3{%
  \ifnum #1<17\tiny\else \ifnum #1<20\small\else
  \ifnum #1<24\normalsize\else \ifnum #1<29\large\else
  \ifnum #1<34\Large\else \ifnum #1<41\LARGE\else
     \huge\fi\fi\fi\fi\fi\fi
  \csname #3\endcsname}%
\else
\gdef\SetFigFont#1#2#3{\begingroup
  \count@#1\relax \ifnum 25<\count@\count@25\fi
  \def\x{\endgroup\@setsize\SetFigFont{#2pt}}%
  \expandafter\x
    \csname \romannumeral\the\count@ pt\expandafter\endcsname
    \csname @\romannumeral\the\count@ pt\endcsname
  \csname #3\endcsname}%
\fi
\fi\endgroup
\begin{picture}(1655,774)(458,-748)
\put(1801,-372){\makebox(0,0)[lb]{\smash{\SetFigFont{11}{13.2}{rm}$\pi$}}}
\put(987,-571){\makebox(0,0)[lb]{\smash{\SetFigFont{11}{13.2}{rm}$\varphi$}}}
\end{picture}
 + \,\, \begin{picture}(0,0)%
\epsfig{file=fig111.pstex}%
\end{picture}%
\setlength{\unitlength}{0.00083300in}%
\begingroup\makeatletter\ifx\SetFigFont\undefined
\def\x#1#2#3#4#5#6#7\relax{\def\x{#1#2#3#4#5#6}}%
\expandafter\x\fmtname xxxxxx\relax \def\y{splain}%
\ifx\x\y   
\gdef\SetFigFont#1#2#3{%
  \ifnum #1<17\tiny\else \ifnum #1<20\small\else
  \ifnum #1<24\normalsize\else \ifnum #1<29\large\else
  \ifnum #1<34\Large\else \ifnum #1<41\LARGE\else
     \huge\fi\fi\fi\fi\fi\fi
  \csname #3\endcsname}%
\else
\gdef\SetFigFont#1#2#3{\begingroup
  \count@#1\relax \ifnum 25<\count@\count@25\fi
  \def\x{\endgroup\@setsize\SetFigFont{#2pt}}%
  \expandafter\x
    \csname \romannumeral\the\count@ pt\expandafter\endcsname
    \csname @\romannumeral\the\count@ pt\endcsname
  \csname #3\endcsname}%
\fi
\fi\endgroup
\begin{picture}(1551,1202)(476,-962)
\put(1133, 97){\makebox(0,0)[lb]{\smash{\SetFigFont{11}{13.2}{rm}$\pi$}}}
\put(987,-571){\makebox(0,0)[lb]{\smash{\SetFigFont{11}{13.2}{rm}$\lambda$}}}
\end{picture}
 + 
  \,\, \begin{picture}(0,0)%
\epsfig{file=fig112.pstex}%
\end{picture}%
\setlength{\unitlength}{0.00083300in}%
\begingroup\makeatletter\ifx\SetFigFont\undefined
\def\x#1#2#3#4#5#6#7\relax{\def\x{#1#2#3#4#5#6}}%
\expandafter\x\fmtname xxxxxx\relax \def\y{splain}%
\ifx\x\y   
\gdef\SetFigFont#1#2#3{%
  \ifnum #1<17\tiny\else \ifnum #1<20\small\else
  \ifnum #1<24\normalsize\else \ifnum #1<29\large\else
  \ifnum #1<34\Large\else \ifnum #1<41\LARGE\else
     \huge\fi\fi\fi\fi\fi\fi
  \csname #3\endcsname}%
\else
\gdef\SetFigFont#1#2#3{\begingroup
  \count@#1\relax \ifnum 25<\count@\count@25\fi
  \def\x{\endgroup\@setsize\SetFigFont{#2pt}}%
  \expandafter\x
    \csname \romannumeral\the\count@ pt\expandafter\endcsname
    \csname @\romannumeral\the\count@ pt\endcsname
  \csname #3\endcsname}%
\fi
\fi\endgroup
\begin{picture}(1655,1202)(458,-962)
\put(1388,-773){\makebox(0,0)[lb]{\smash{\SetFigFont{11}{13.2}{rm}$\pi$}}}
\put(912,-571){\makebox(0,0)[lb]{\smash{\SetFigFont{11}{13.2}{rm}$\lambda$}}}
\end{picture}
 \,\,. \]
Instead of (\ref{2.6}) we will have
\begin{eqnarray}
\label{8.2}
\lefteqn{ \partial^2 \phi(p,q) = } \nonumber\\
 & = & g(q)\{ A_{\nu}(q_2) \partial_{\nu} \phi(p,q) +
  \partial_{\nu} \phi(p,q)\tilde{A}_{\nu}(q_1) -
 A_{\nu}(q_2)\phi(p,q) A_{\nu}(q_1)\} + \nonumber\\
 & + & \frac{1}{4\pi^2 f_{\pi}^2} \left[
 \{i\gamma_5,G^{-1}(q_2)\}G(q_2)\frac{\tau_a}{2}\phi\frac{\tau_a}{2}
G(q_1)\{i\gamma_5,G^{-1}(q_1)\}\right. - \nonumber\\
 & - & \left.\lambda G(q_1)\{i\gamma_5,G^{-1}(q_1)\}
\frac{\tau_a}{2} - \{i\gamma_5,G^{-1}(q_2)\}G(q_2)\frac{\tau_a}{2}\lambda
\right].
\end{eqnarray}
Here $\lambda$ is the emission amplitude of the zero momentum pion
in the transition of the bound state to the $q\bar{q}$-pair. This
amplitude has to be defined by the axial current conservation.
There is another important quantity in this equation, namely: $f_{\pi}$.
In section \ref{V} we have written an explicit expression for
$f_{\pi}^2$ including only the gauge field contribution and ignoring
the pion contribution. Now we include the pion contribution in the
equation for the Green's function and, to be self-consistent,
we have to do the same for $f_{\pi}^2$. I was not able to carry out
this in any order of the Yukawa coupling, but in the first order of
$g^2$ the equation (\ref{5.25}) is correct if one adds the diagrams
\begin{equation}
\label{8.3}
\{\gamma_5,G^{-1}\} \begin{picture}(0,0)%
\epsfig{file=fig113.pstex}%
\end{picture}%
\setlength{\unitlength}{0.00083300in}%
\begingroup\makeatletter\ifx\SetFigFont\undefined
\def\x#1#2#3#4#5#6#7\relax{\def\x{#1#2#3#4#5#6}}%
\expandafter\x\fmtname xxxxxx\relax \def\y{splain}%
\ifx\x\y   
\gdef\SetFigFont#1#2#3{%
  \ifnum #1<17\tiny\else \ifnum #1<20\small\else
  \ifnum #1<24\normalsize\else \ifnum #1<29\large\else
  \ifnum #1<34\Large\else \ifnum #1<41\LARGE\else
     \huge\fi\fi\fi\fi\fi\fi
  \csname #3\endcsname}%
\else
\gdef\SetFigFont#1#2#3{\begingroup
  \count@#1\relax \ifnum 25<\count@\count@25\fi
  \def\x{\endgroup\@setsize\SetFigFont{#2pt}}%
  \expandafter\x
    \csname \romannumeral\the\count@ pt\expandafter\endcsname
    \csname @\romannumeral\the\count@ pt\endcsname
  \csname #3\endcsname}%
\fi
\fi\endgroup
\begin{picture}(1216,902)(593,-512)
\put(766,239){\makebox(0,0)[lb]{\smash{\SetFigFont{10}{12.0}{rm}$\partial_{\mu} G^{-1}$}}}
\put(1569,-61){\makebox(0,0)[lb]{\smash{\SetFigFont{10}{12.0}{rm}$g$}}}
\put(1351,239){\makebox(0,0)[lb]{\smash{\SetFigFont{10}{12.0}{rm}$\partial_{\mu} G^{-1}$}}}
\put(762,-62){\makebox(0,0)[lb]{\smash{\SetFigFont{10}{12.0}{rm}$g$}}}
\end{picture}
 \{\gamma_5,G^{-1}\} +
\{\gamma_5,G^{-1}\} \begin{picture}(0,0)%
\epsfig{file=fig114.pstex}%
\end{picture}%
\setlength{\unitlength}{0.00083300in}%
\begingroup\makeatletter\ifx\SetFigFont\undefined
\def\x#1#2#3#4#5#6#7\relax{\def\x{#1#2#3#4#5#6}}%
\expandafter\x\fmtname xxxxxx\relax \def\y{splain}%
\ifx\x\y   
\gdef\SetFigFont#1#2#3{%
  \ifnum #1<17\tiny\else \ifnum #1<20\small\else
  \ifnum #1<24\normalsize\else \ifnum #1<29\large\else
  \ifnum #1<34\Large\else \ifnum #1<41\LARGE\else
     \huge\fi\fi\fi\fi\fi\fi
  \csname #3\endcsname}%
\else
\gdef\SetFigFont#1#2#3{\begingroup
  \count@#1\relax \ifnum 25<\count@\count@25\fi
  \def\x{\endgroup\@setsize\SetFigFont{#2pt}}%
  \expandafter\x
    \csname \romannumeral\the\count@ pt\expandafter\endcsname
    \csname @\romannumeral\the\count@ pt\endcsname
  \csname #3\endcsname}%
\fi
\fi\endgroup
\begin{picture}(1216,902)(893,-662)
\put(1801, 51){\makebox(0,0)[lb]{\smash{\SetFigFont{10}{12.0}{rm}$\partial_{\mu} G^{-1}$}}}
\put(976, 51){\makebox(0,0)[lb]{\smash{\SetFigFont{10}{12.0}{rm}$\partial_{\mu} G^{-1}$}}}
\end{picture}
 \{\gamma_5,G^{-1}\}.
\end{equation}
As we see, the gluonic correction of the order of $\frac{\alpha}{\pi}$
and the pionic correction of the order of $g^2$ have the same
diagrammatic structure. In order to estimate the value of these
corrections, let us take just the contributions of zero momentum
gluons and pions to them. It can be shown that the contribution
of zero momentum gluons cancels in the last two
diagrams of (\ref{5.25}). Zero momentum pion contribution comes
only from the first diagram of (\ref{8.3}). Transferring the
differentiation from the fermionic line to the pionic line in this
diagram we obtain the contribution of zero momentum pions in the
form
\begin{equation}
\label{8.4}
\{\gamma_5,G^{-1}\} \begin{picture}(0,0)%
\epsfig{file=fig115.pstex}%
\end{picture}%
\setlength{\unitlength}{0.00083300in}%
\begingroup\makeatletter\ifx\SetFigFont\undefined
\def\x#1#2#3#4#5#6#7\relax{\def\x{#1#2#3#4#5#6}}%
\expandafter\x\fmtname xxxxxx\relax \def\y{splain}%
\ifx\x\y   
\gdef\SetFigFont#1#2#3{%
  \ifnum #1<17\tiny\else \ifnum #1<20\small\else
  \ifnum #1<24\normalsize\else \ifnum #1<29\large\else
  \ifnum #1<34\Large\else \ifnum #1<41\LARGE\else
     \huge\fi\fi\fi\fi\fi\fi
  \csname #3\endcsname}%
\else
\gdef\SetFigFont#1#2#3{\begingroup
  \count@#1\relax \ifnum 25<\count@\count@25\fi
  \def\x{\endgroup\@setsize\SetFigFont{#2pt}}%
  \expandafter\x
    \csname \romannumeral\the\count@ pt\expandafter\endcsname
    \csname @\romannumeral\the\count@ pt\endcsname
  \csname #3\endcsname}%
\fi
\fi\endgroup
\begin{picture}(1216,468)(893,-1045)
\put(1726,-811){\makebox(0,0)[lb]{\smash{\SetFigFont{11}{13.2}{rm}$g$}}}
\put(1201,-811){\makebox(0,0)[lb]{\smash{\SetFigFont{11}{13.2}{rm}$g$}}}
\end{picture}
  
\{\gamma_5,G^{-1}\}\frac{1}{8\pi^2}\,.
\end{equation}
The expression for $f_{\pi}^2$ which includes the zero momentum pion
contribution is
\begin{eqnarray}
\label{8.5}
 8 f_{\pi}^2 &=& \int \frac{d^4 q}{(2\pi)^4i} Tr \{\gamma_5,G^{-1}\}G
\{\gamma_5,G^{-1}\}G \partial_{\mu}A\partial_{\mu}A + \nonumber\\
 & & + \frac{1}{8\pi^2 f_{\pi}^2}\int \frac{d^4 q}{(2\pi)^4i} Tr \big(
\{\gamma_5,G^{-1}\}G\big)^4 .
\end{eqnarray}
It gives us an understanding of the scale of possible pion contributions.
It is interesting to note, that (\ref{8.5}) is not an expression
in terms of the Green's functions but an equation for $f_{\pi}$.

In the next paper we will show that the pion contribution changes
essentially the structure of the equation for the Green's function.
The new equation has a solution corresponding to the confined quark.
At the same time the symmetry breaking solution will not necessarily
survive (at least if $g(0)$ is large).

\section*{Acknowledgments}

I would like to thank Yu.\ Dokshitzer, C.\ Ewerz, B.\ Metsch, J.\ Nyiri,
H.\ Petry, D.\ Sch\"utte and A.\ Vainshtein for many helpful discussions.
I am very grateful to all the members of the Institute for Theoretical
Nuclear Physics for their kind hospitality and the warm and inspiring
atmosphere I experienced during my stay in Bonn. I express my gratitude
to the Humboldt foundation for making this stay possible.

\section*{References}

\begin{description}
\item [1] V.\,N.\ Gribov, Lund preprint LU-TP 91-7 (1991)
\item [2] V.\,N.\ Gribov, Bonn preprint TK-95-35 (1995), hep-ph/9512352
\item [3] S.\ Adler, Phys.\ Rev.\ {\bf 177} (1969), 2426; 
 J.\,S.\ Bell, R.\ Jackiw, Nuov.\ Cimen.\ {\bf 60A} (1969), 47 
\item [4] G.\ Veneziano, Nucl.\ Phys.\ {\bf B159} (1979), 357; 
          Phys.\ Lett.\ {\bf 95B} (1980), 90
\end{description}

\end{document}